\documentclass[a4paper,fleqn,usenatbib]{mnras}
\usepackage[T1]{fontenc}
\usepackage{ae,aecompl}

\usepackage{graphicx}
\usepackage{color}
\usepackage{times}
\usepackage{natbib}
\usepackage{setspace}
\usepackage{hyperref}
\usepackage{amsmath}
\usepackage{amsbsy}
\usepackage{gensymb}
\usepackage{txfonts}

\newif\ifAMStwofonts
\AMStwofontstrue
\definecolor{red}{rgb}{1,0.,0.}

\newcommand{\hmsun}{\,h^{-1}\,{\rm M_\odot}}

\newcommand{\arepo}{{\sc arepo}}

\newcommand{\kpc}{{\rm kpc}}
\newcommand{\Mpc}{{\rm Mpc}}
\newcommand{\erg}{{\rm erg}}
\newcommand{\kms}{{\rm km~s^{-1}}}
\newcommand{\cm}{{\rm cm}}

\newcommand{\gsim}{\,\lower.7ex\hbox{$\;\stackrel{\textstyle>}{\sim}\;$}}
\newcommand{\lsim}{\,\lower.7ex\hbox{$\;\stackrel{\textstyle<}{\sim}\;$}}

\newcommand{\muG}{{\rm \mu G}}
\newcommand{\nG}{{\rm nG}}

\newcommand{\FM}[1]{#1}
\newcommand{\FMR}[1]{#1}
\newcommand{\FMRR}[1]{#1}

\defcitealias{Marinacci2014b}{b} 
\defcitealias{Planck2016}{Planck Collaboration XIII} 
\defcitealias{Neronov2010}{Neronov \& Vovk 2010}

\newcommand{\planckpappar}{\citetalias{Planck2016} \citeyear{Planck2016}}

\ifpdf
  \DeclareGraphicsExtensions{.pdf}
\else
  \DeclareGraphicsExtensions{.eps}
\fi

\title[Radio haloes and B fields in IllustrisTNG] 
{First results from the IllustrisTNG simulations: radio haloes and magnetic 
fields}

\author[F.~Marinacci et al.]
{Federico Marinacci${^1}$\thanks{E-mail: fmarinac@mit.edu}, Mark 
Vogelsberger${^1}$, R\"udiger~Pakmor${^2}$, Paul Torrey${^1}$,
\newauthor Volker Springel$^{2,3}$, Lars Hernquist${^4}$, Dylan Nelson${^5}$, 
Rainer Weinberger${^2}$,
\newauthor Annalisa Pillepich${^6}$, Jill Naiman${^4}$, Shy Genel$^{7,8}$
\vspace*{0.2cm}\\
  $^1$Kavli Institute for Astrophysics and Space Research, 
  Massachusetts Institute of Technology, Cambridge, MA 02139, USA\\
  $^2$Heidelberger Institut f\"{u}r Theoretische Studien,
  Schloss-Wolfsbrunnenweg 35, D-69118 Heidelberg, Germany\\
  $^3$Zentrum f\"ur Astronomie der Universit\"at Heidelberg,
  Astronomisches Recheninstitut, M\"{o}nchhofstr. 12-14, D-69120
  Heidelberg, Germany\\
  $^4$Harvard--Smithsonian Center for Astrophysics, 
  60 Garden Street, Cambridge, MA 02138, USA\\
  $^5$Max-Planck-Institut f\"ur Astrophysik, Karl-Schwarzschild-Stra{\ss}e 1, 
  D-85741 Garching bei M\"unchen, Germany\\
  $^6$Max-Planck-Institut f\"ur Astronomie, K\"onigstuhl 17, 
  D-69117, Heidelberg, Germany\\
  $^7$Center  for  Computational  Astrophysics,  Flatiron  Institute, 162 Fifth 
Avenue, New York, NY 10010, USA\\
  $^8$Columbia Astrophysics Laboratory, Columbia University, 550 West 120th 
Street, New York, NY 10027, USA
 }
  
\date{Accepted 2018 August 10. Received 2018 August 08; in original form 2017 July 12.}

\begin{document}

\pagerange{\pageref{firstpage}--\pageref{lastpage}}
\pubyear{2018}

\maketitle

\label{firstpage}

\begin{abstract}
We introduce the IllustrisTNG project, a new suite of cosmological 
magnetohydrodynamical simulations performed with the moving-mesh code \arepo\ 
employing an updated Illustris galaxy formation model. Here we focus on the 
general properties of magnetic fields and the diffuse radio emission in galaxy 
clusters. Magnetic fields are prevalent in galaxies, and their build-up is 
closely linked to structure formation. We find that structure formation 
amplifies the initial seed fields ($10^{-14}$ comoving Gauss) to the values 
observed in low-redshift galaxies ($1-10\,\muG$). The magnetic field topology is 
closely connected to galaxy morphology such that irregular fields are hosted by 
early-type galaxies, while large-scale, ordered fields are present in disc 
galaxies. Using \FMR{two simple models} for the energy distribution of relativistic 
electrons we predict the diffuse radio emission of $280$ clusters with a 
baryonic mass resolution of $1.1\times 10^{7}\,{\rm M_{\odot}}$, and generate 
mock observations for VLA, LOFAR, ASKAP and SKA. Our simulated clusters show 
extended radio emission, whose detectability correlates with their virial mass.  
We reproduce the observed scaling relations between total radio power and X-ray 
emission, $M_{500}$, and the Sunyaev-Zel'dovich $Y_{\rm 500}$ parameter. The 
radio emission surface brightness profiles of our most massive clusters are in 
reasonable agreement with VLA measurements of Coma and Perseus. Finally, we 
discuss the fraction of detected extended radio haloes as a function of virial 
mass and source count functions for different instruments. Overall our results 
agree encouragingly well with observations, but a refined analysis requires a 
more sophisticated treatment of relativistic particles in large-scale galaxy 
formation simulations. 
\end{abstract}

\begin{keywords}
magnetic fields -- MHD -- methods: numerical -- cosmology: theory --
radio continuum: general -- galaxies: clusters: general
\end{keywords}

\section{Introduction} \label{sec:intro}

The ubiquitous presence of magnetic fields in the Universe is an established 
observational fact \citep{Beck2013c}. The progress of observational techniques 
has allowed for the measurement of the present-day field intensities and the 
mapping of their orientations with an increasing degree of accuracy.  The 
situation is far more uncertain in low-density environments, such as 
cosmological filaments and voids, where a robust detection is still lacking. 
Indeed, the considerably lower field strength expected in those regions ($\lsim 
1\,\nG$) makes a direct detection very challenging.  This lack of detections on 
large scales is particularly unfortunate because it leaves us with very little 
information about the origin of magnetic fields and their subsequent 
amplification during structure formation. 

Magnetic fields are also an essential component of many astrophysical phenomena 
occurring at all scales. It is now generally known that clusters are permeated 
by magnetic fields that extend into the intergalactic medium. The existence of 
those large-scale ($\sim 1\,{\rm Mpc}$) fields is inferred through Faraday 
rotation signals of polarized radio galaxies \citep{Murgia2004, Govoni2006, 
Guidetti2008, Bonafede2010} and extended synchrotron radio emission 
\citep[e.g.][]{Feretti2012}, which cannot be directly associated to individual 
galaxies hosting an active galactic nucleus (AGN). This radio emission can be 
broadly divided in \FM{three} categories: (i) radio relics, which are 
predominantly found in the outskirts of clusters and trace mergers and shocks, 
(ii) \FM{(giant)} radio haloes, that show a more regular morphology and whose 
emission is predominantly unpolarized, \FM{appears to be approximately centred 
on the X-ray emission of the cluster and extends on scales of $\sim 1\,{\rm 
Mpc}$, and (iii) mini radio haloes associated to relaxed clusters and extending 
on smaller ($\sim 50-300\,\kpc$) scales}. Diffuse radio emission from clusters 
has been extensively studied since it is an important tool to unveil the complex 
interplay among physical processes occurring in the intracluster medium (ICM). 
Indeed, these diffuse radio sources are very sensitive to the microphysics of 
the ICM, such as turbulence level and shock structures, and thus they represent 
an important probe to determine its properties. 

Despite extensive observational campaigns, many aspects of the diffuse radio 
emission in clusters are still poorly understood. For example, there is no 
definitive model to explain their origin. Given the short cooling time of 
electrons emitting synchrotron radiation, an efficient mechanism that 
re-accelerates them or injects fresh electrons at relativistic speeds is needed 
to maintain radio emission on Mpc scales. Two main models have been proposed. In 
one scenario, relativistic electrons are produced by hadronic interactions of 
cosmic ray protons with the ICM \citep[e.g.][]{Blasi1999, Pfrommer2008, 
Ensslin2011}. A mechanism to accelerate protons at relativistic speed is thus 
needed, and several of them, like accretion shocks or feedback from AGN and 
galactic winds, might be at work. In the second scenario, local re-acceleration 
of electrons occurs through interaction with plasma waves driven by turbulence 
in the ICM, possibly caused by cluster mergers/assembly \citep{Giovannini1993, 
Brunetti2009, Donnert2013}. Each of these models has its difficulties. \FM{For 
instance, the classic hadronic scenario predicts in galaxy clusters a level of 
gamma-ray emission from inelastic collision between the cosmic ray protons and 
thermal protons in the ICM that is in tension with the persistent non-detections 
coming from the {\it Fermi} satellite \citep[][]{Ackermann2016}. While this can 
be somewhat mitigated by fast streaming of cosmic ray protons 
\citep{Ensslin2011}, it is unclear whether this mechanism can work when  physics 
of the ICM is fully considered \citep[see e.g.][]{Wiener2013,Pinzke2017}. 
Conversely, even the re-acceleration scenario needs seed electrons, which are 
usually originated by hadronic interactions \citep[see e.g.][]{Brunetti2017}, to 
operate.} Moreover, radio haloes are not present in all clusters. Observations 
suggest that the detection probability increases with the X-ray luminosity of 
the host cluster. At $z\lsim 0.2$, radio haloes are found in about $30-35$ per 
cent of galaxy clusters with an X-ray luminosity above $\simeq 
1.3\times10^{45}\,{\rm erg\,s^{-1}}$ \citep{Giovannini1999}. It is unclear 
whether this is the result of insufficient sensitivity of current radio 
observations, or reflects a more profound reason linked to the physics of the 
ICM. 

Indeed, only about 50 objects are known to host radio haloes (see 
\citealt{Feretti2012}, for a review and \citealt{Giacintucci2017} for a 
reasonably updated census of mini haloes). All-sky surveys have been 
instrumental to the study of diffuse radio emission from clusters, as for 
instance the VLA NVSS survey \citep{Condon1998}. This survey  mapped all radio 
sky above $-40$ deg declination (almost 82 per cent of the sky) at 1.4 GHz, 
with a resolution of $45$ arcsec (FWHM) and an rms sensitivity of $\simeq 
0.45\,{\rm mJy\,beam^{-1}}$. Although it was not primarily designed to detect 
radio haloes in galaxy clusters, it laid the foundation to their study 
\citep{Giovannini1999} by triggering follow-up campaigns, also employing other 
radio-telescopes, trying to extend the known sample both in number and in 
redshift \citep{Venturi2007, Venturi2008, Giacintucci2011}.

In the near future, the number of radio haloes in galaxy clusters is expected to 
increase thanks to the next generation of radio instruments, with observation 
targeted at detecting such diffuse structures. For example, one of the main 
science goals of the LOFAR Survey Key Project is to detect thousands of diffuse 
radio sources in clusters out to $z\sim 0.8$, with more than 100 at redshift 
above $z\gsim 0.6$, other than studying already detected radio haloes in 
clusters in more detail \citep{Rottgering2010}. This should be possible thanks 
to the increased sensitivity of $0.25\,{\rm mJy\,beam^{-1}}$ and a resolution of 
$25$ arcsec full width at half-maximum (FWHM) for an observing frequency of $120\,{\rm MHz}$ (Tier 1 Large 
Area Survey). Similar performance is expected for future SKA observations, with 
the predicted discovery of about $\sim 2500$ radio haloes up to $z \sim 0.6$, 
and probing a cluster mass range down to $10^{14}\,{\rm M}_\odot$, which is not 
explored with current observations \citep{Cassano2015}. The sensitivity 
($0.02\,{\rm mJy\,beam^{-1}}$) and resolution ($10$ arcsec FWHM) at $120\,{\rm 
MHz}$ are superior to those of the current generation of radio instruments.

The challenge for any successful picture trying to explain galaxy formation,
and in particular the assembly of galaxy clusters, is therefore to properly
model and predict the salient features of this extended radio emission.  To
this end, reproducing the scaling relations that link the radio halo power with
structural properties of the halo, such as the virial mass \citep{Cassano2013},
or the ICM, such as the total X-ray power \citep{Giovannini2009}, give
important insights into the diffuse radio emission phenomenology in clusters,
other than being an important test bed for any galaxy formation and galaxy
cluster physics theory.

\begin{figure*}
\centering
\includegraphics[width=0.97\textwidth]{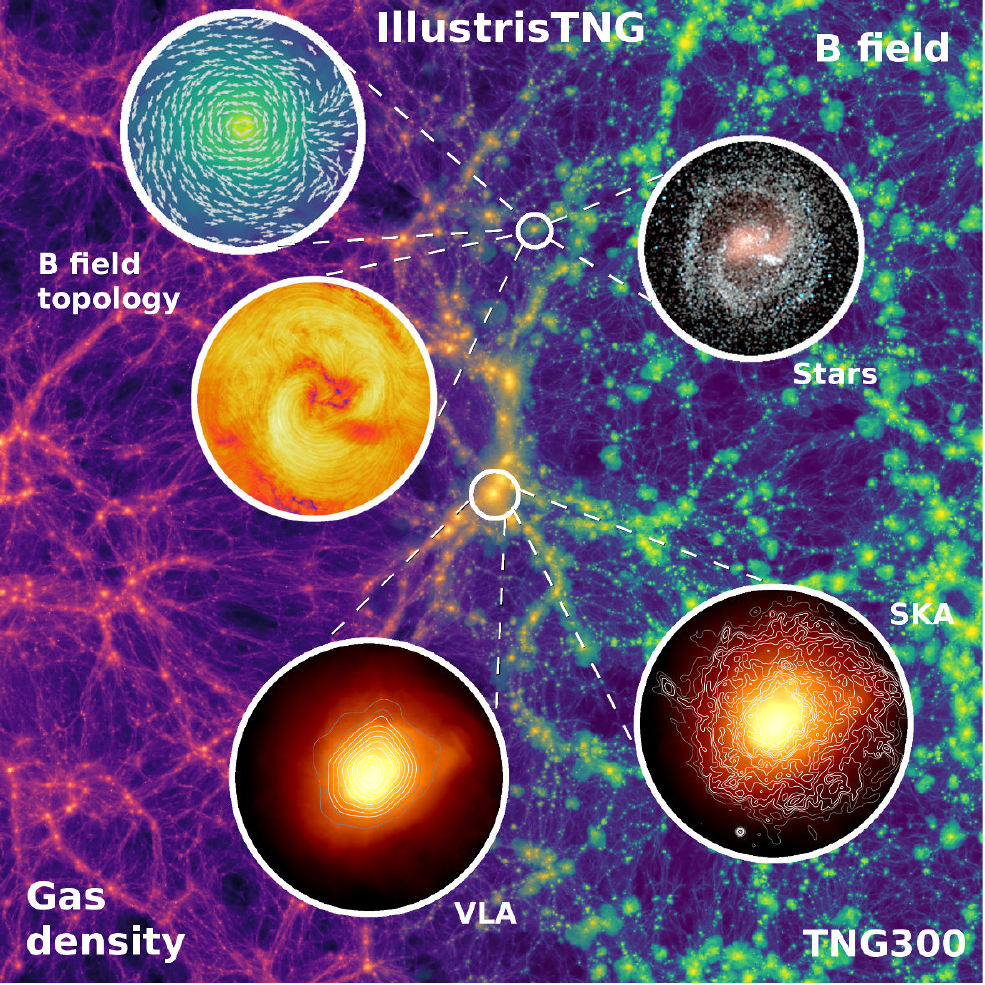}
\caption{Gas column density map (left-hand side) and density-weighted projection of the 
magnetic field intensity (right-hand side) for the TNG300 box at $z = 0$. The projection 
extends for the full size of the TNG300 box ($\simeq 300$ Mpc) for a thickness 
of $\simeq 22.13$ Mpc and is centred on the most massive galaxy cluster. The 
zoomed panels on this structure show the X-ray emission (colours) overlayed with 
the extended ($\sim 1\,\Mpc$) radio emission as seen with two different 
instruments. The zoomed panels on the small object illustrate the link between 
the magnetic field topology within a galaxy (on scales $\sim 10\,\kpc$) and the 
galaxy morphology.} \label{fig:overview}
\end{figure*}

Cosmological magnetohydrodynamics (MHD) simulations represent the most complete 
approach to investigate the complex interplay between magnetic fields and 
structures in the Universe. Performing those calculations from first principles, 
although desirable, still appears to be a remote goal given the tremendous 
dynamic range of spatial scales to be simultaneously modelled and the current 
uncertainties in our knowledge of baryon processes relevant for galaxy 
formation. Notwithstanding the advancement in the field 
\citep[e.g.][]{Dolag2016, Marinacci2015, Marinacci2016}, many of the 
cosmological MHD calculations performed so far do not include baryonic processes 
that are currently thought to play an essential role in galaxy formation and/or 
study the large-scale distribution of the field \citep{Vazza2014,Vazza2015a}. 

Although there exist examples of simulations studying diffuse radio emission in 
galaxy clusters with different techniques to model relativistic particles 
\citep[see e.g.][]{Dolag2000, Xu2012}, they are limited by several factors. Most 
importantly the great majority of the baryonic physics relevant for galaxy 
formation\FM{, and which has a non-negligible impact in setting the present-day 
strength of magnetic fields \citep[see e.g.][]{Marinacci2015},} is largely 
ignored. Another limitation is that to reach high resolution, these calculations 
use the so-called zoom-in technique \citep{Hahn2011}, simulating one object at a 
time.  For instance, \citet{ZuHone2015} simulate the emergence of a radio mini 
halo by following the evolution and transport of cosmic rays, through passive 
tracer particles, in a simulation of cluster formation. While they include a 
detailed model for the cosmic rays evolution, the cosmological context is 
missing, together with physics of galaxy formation. \citet{Donnert2013} studied 
radio halo emission in an idealized MHD simulation of a merging galaxy cluster 
with cosmic ray transport, again with the same limitations. Moreover, most of 
the theoretical work has focused recently on studying the acceleration of 
particles to relativistic speeds in the ICM, an essential ingredient in 
synchrotron radio emission, with different levels of sophistication 
\citep{Pinzke2013, Vazza2014b, Vazza2016, Wittor2016}.  Therefore, in the study 
of radio haloes the superior statistical power that uniformly sampled 
\FM{magnetohydrodynamical cosmological simulations, which evolve magnetic 
fields (the other key ingredient for synchrotron emission) self-consistently and 
include the most important baryonic physics processes in galaxy formation,} can 
offer is paramount to make decisive steps forward in our theoretical 
understanding. 

\begin{table*}
\begin{tabular}{llcrrcccl}
\hline
Simulation name & Realization & $L$   & $N_{\rm DM}$  & $N_{\rm gas}$ & $m_{\rm 
DM}$ & $m_{\rm b}$ & $\epsilon_{\rm DM,*}$ & $\epsilon_{\rm gas}$ \\
               &             &   $[h^{-1}{\rm Mpc}]$  &                      &   
               & $[h^{-1}{\rm M}_\odot]$ & $[h^{-1}{\rm M}_\odot]$ &  
$[h^{-1}{\rm kpc}]$ & $[h^{-1}{\rm kpc}]$ \\
\hline
{\bf TNG300} &  TNG300(-1)  & 205  & $2500^3$ & $2500^3$ & $3.98\times 10^7$  & 
$7.44\times 10^6$  & 1.0 & 0.25\\
             &  TNG300-2  & 205  & $1250^3$ & $1250^3$ & $3.19\times 10^8$  & 
$5.95\times 10^7$  & 2.0 & 0.5 \\
             &  TNG300-3  & 205  & $625^3$  & $625^3$  & $2.55\times 10^9$  & 
$4.76\times 10^8$  & 4.0 & 1.0 \\
             &  TNG300-DM-1  & 205 & $2500^3$ &  & $4.73\times 10^7$ & & 1.0 \\
     &  TNG300-DM-2  & 205 & $1250^3$ & &   $3.78\times 10^8$ & & 2.0 \\
     &  TNG300-DM-3  & 205 & $625^3$ & &   $3.03\times 10^9$ & & 4.0 \\
\hline
{\bf TNG100} &  TNG100(-1)  & 75 & $1820^3$ & $1820^3$ & $5.06\times 10^6$ & 
$9.44\times 10^5$  & 0.5 & 0.125\\
             &  TNG100-2  & 75 & $910^3$ & $910^3$ & $4.04\times 10^7$ & 
$7.55\times 10^6$  & 1.0 & 0.25\\
             &  TNG100-3  & 75 & $455^3$ & $455^3$ & $3.24\times 10^8$ & 
$6.04\times 10^7$  & 2.0 & 0.5\\
             &  TNG100-DM-1  & 75 & $1820^3$ & &   $6.00\times 10^6$ & & 0.5 \\
     &  TNG100-DM-2  & 75 & $910^3$ & &   $4.80\times 10^7$ & & 1.0 \\
     &  TNG100-DM-3  & 75 & $455^3$ & &   $3.84\times 10^8$ & & 2.0 \\
\hline
\end{tabular}
\caption{IllustrisTNG contains three simulations covering three volumes, roughly 
$\sim 50^3, 100^3, 300^3\,{\rm Mpc}^3$: {\bf TNG50} ($N_{\rm gas} + N_{\rm DM} = 
2 \times 2160^3$, $m_{b}=5.74\times 10^4\hmsun$), {\bf TNG100} ($N_{\rm gas} + 
N_{\rm DM} = 2 \times 1820^3$, $m_{\rm b} = 9.44 \times 10^5\hmsun$), and {\bf 
TNG300} ($N_{\rm gas} + N_{\rm DM} = 2 \times 2500^3$, $m_{b}= 7.44 \times 
10^6\hmsun$). Here we use TNG100 and TNG300, ideally suited for cluster studies, 
and list the numerical parameters in the table. For each simulated box, we have 
performed simulations at three different resolution levels indicated by the 
numbers appended to the realization name (larger numbers denote progressively 
coarser resolution). Each level is spaced by a factor of $8$ in mass resolution 
and a factor of $2$ in softening length. For all runs we indicate the box side 
length ($L$), the number of collisionless DM particles ($N_{\rm DM}$), and the 
initial number of gas cells ($N_{\rm gas}$) employed. We also report the masses 
of the DM particles ($m_{\rm DM}$) and of the baryonic component ($m_{\rm b}$). 
The gravitational softening lengths $\epsilon_{\rm DM,*}$ indicate the maximum 
physical softening length (reached at $z < 1$) of dark matter and star 
particles; $\epsilon_{\rm gas}$ is the minimum comoving value for the 
gravitational softening length of gas cells. 
\label{tab:sims} 
} 
\end{table*}

In this paper we extend previous work based on cosmological simulations by 
analysing the general magnetic field properties and the diffuse radio halo 
emission in galaxy clusters in the IllustrisTNG project, a set of cosmological 
magnetohydrodynamics simulations run with the moving-mesh code \arepo\ 
\citep{Arepo} that include a comprehensive module for galaxy formation physics. 
The main and novel aspect of our work is the analysis of the diffuse radio 
emission resulting from radio haloes in galaxy clusters \citep{Feretti1996, 
Feretti2012, Murgia2009, Vacca2011}. We investigate radio emission from clusters 
by a detailed comparison with observations, trying to match the current 
observational constraints and to make predictions for the upcoming radio surveys 
that will be performed with the new generation of radio instruments such as SKA 
and LOFAR. The analysis of simulated radio haloes gives us a complementary view 
on the spatial extent and energy content of magnetic fields in galaxy clusters, 
since the radio emission is proportional to their strength.  As such, the study 
of radio halo scaling relations \citep{Giovannini2009, Cassano2013, 
Zandanel2014} with the total X-ray power and halo mass may yield important 
information about the amplification mechanisms of magnetic fields in clusters 
and the level of turbulence in the ICM. The modelling of radio emission makes it 
also possible to study the  transport of charged particles and their 
re-acceleration to relativistic speeds, and it constrains the probability of 
detecting extended radio-emitting structures in a statistical sample of 
realistic simulated clusters. The comparison of the simulated radio emission 
with actual observations might also be employed as a useful check for the 
implementation of the galaxy formation physics modules used to perform the 
simulations, although our modelling of relativistic particles is rather 
preliminary and might have a non-negligible impact on the results.

Figure \ref{fig:overview} gives a visual overview of the analysis performed in 
this paper. In the background it shows projected maps of gas density (left-hand side) and 
magnetic field intensity (right-hand side) in the TNG300 box (see Section 
\ref{sec:methods} for details on this simulation) centred on the most massive 
galaxy cluster. The correlation between the large-scale magnetic field and the 
large-scale structure of the Universe appears evident. The zoom panels 
illustrate the aspects that we are going to analyse on individual haloes, namely 
the connection between magnetic fields and radio emission in galaxy clusters (on 
scales of $\sim 1\,\Mpc$) and the link between the properties of the magnetic 
field (on scales of $\sim 10\,\kpc$) and those of the hosting galaxies.

\begin{figure*}
\centering
\includegraphics[width=0.245\textwidth]{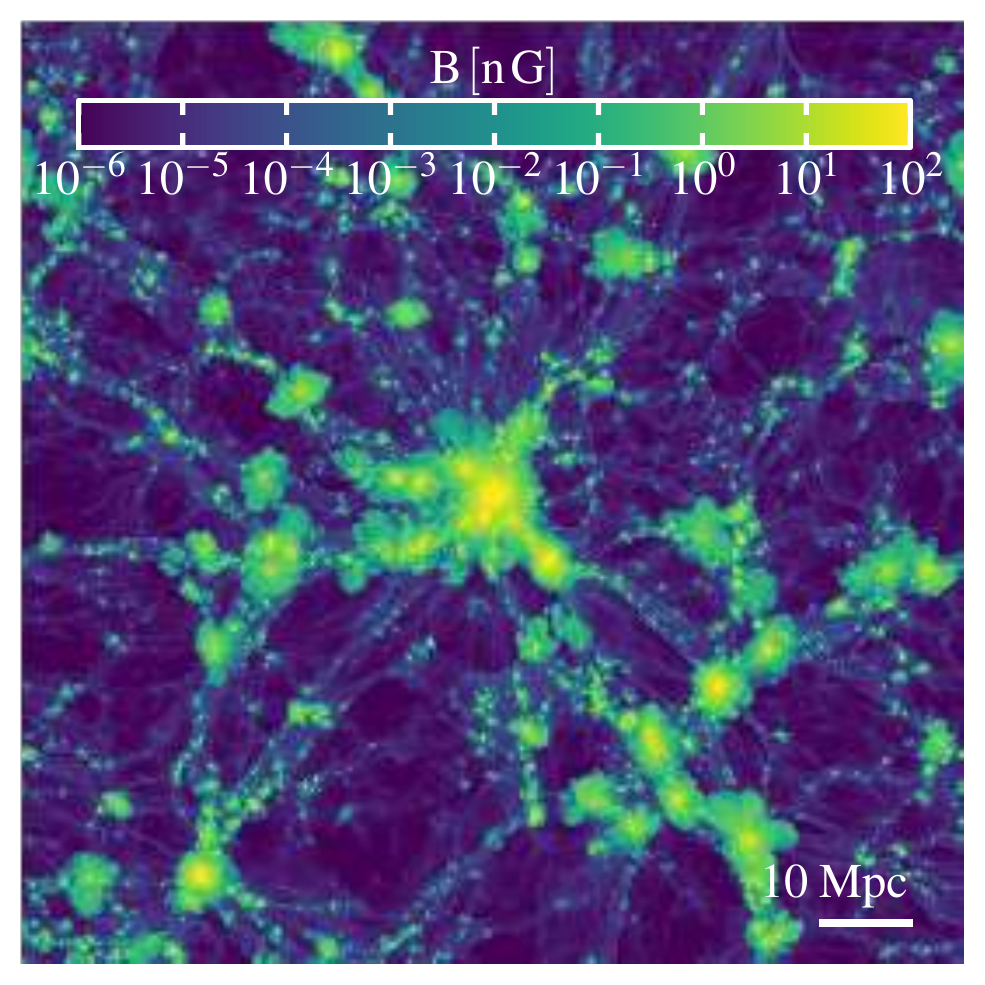}
\includegraphics[width=0.245\textwidth]{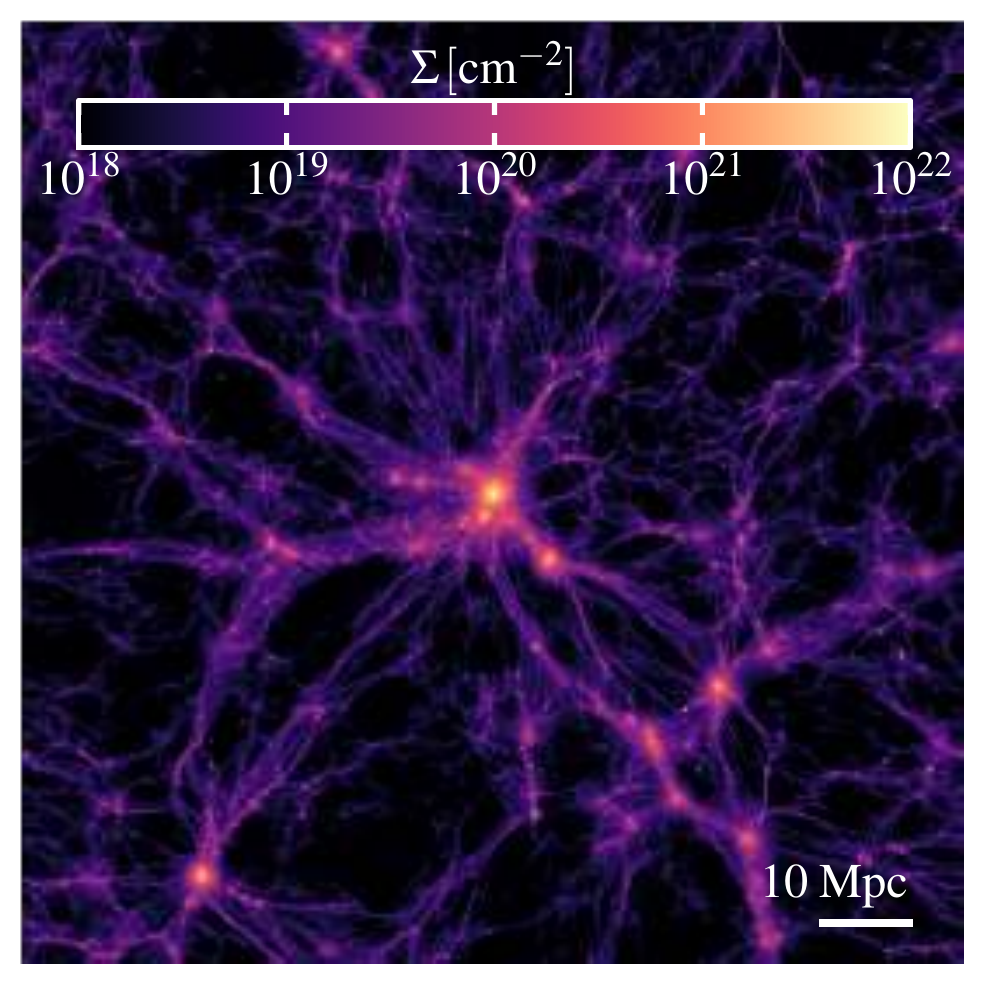}
\includegraphics[width=0.245\textwidth]{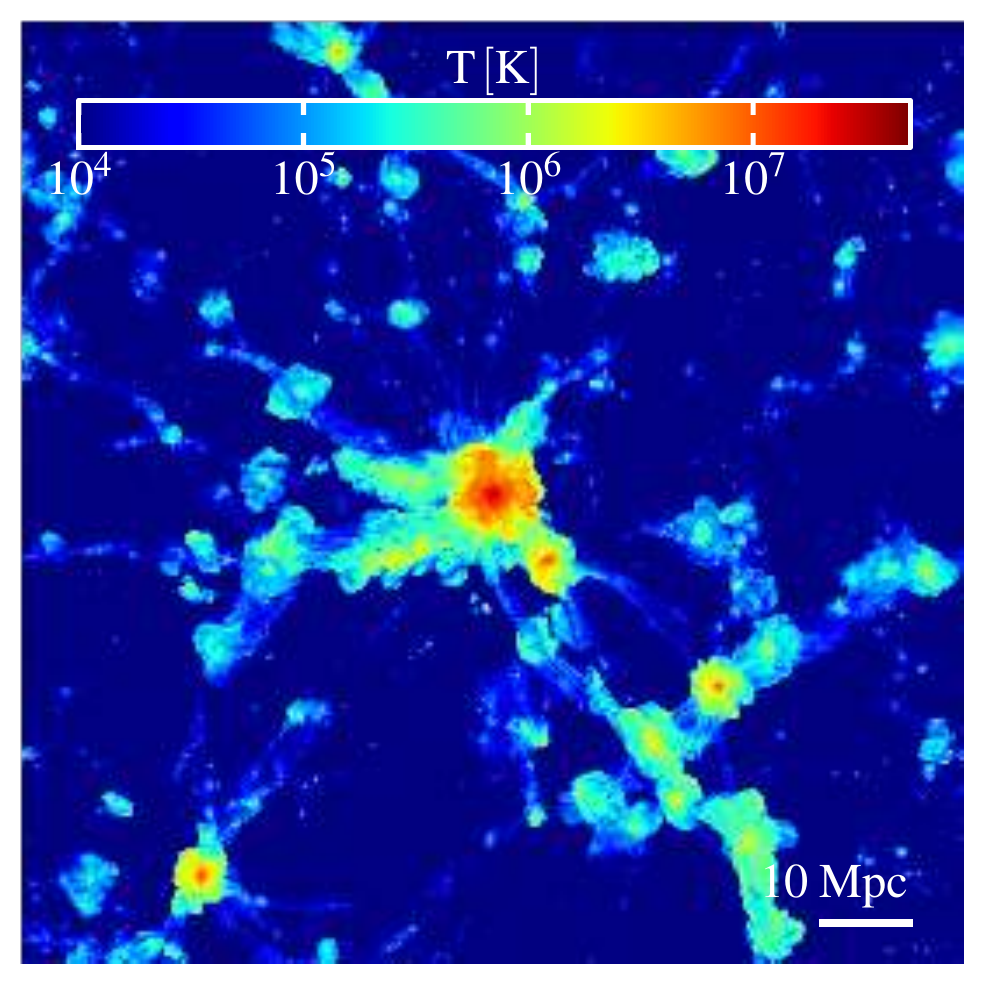}
\includegraphics[width=0.245\textwidth]{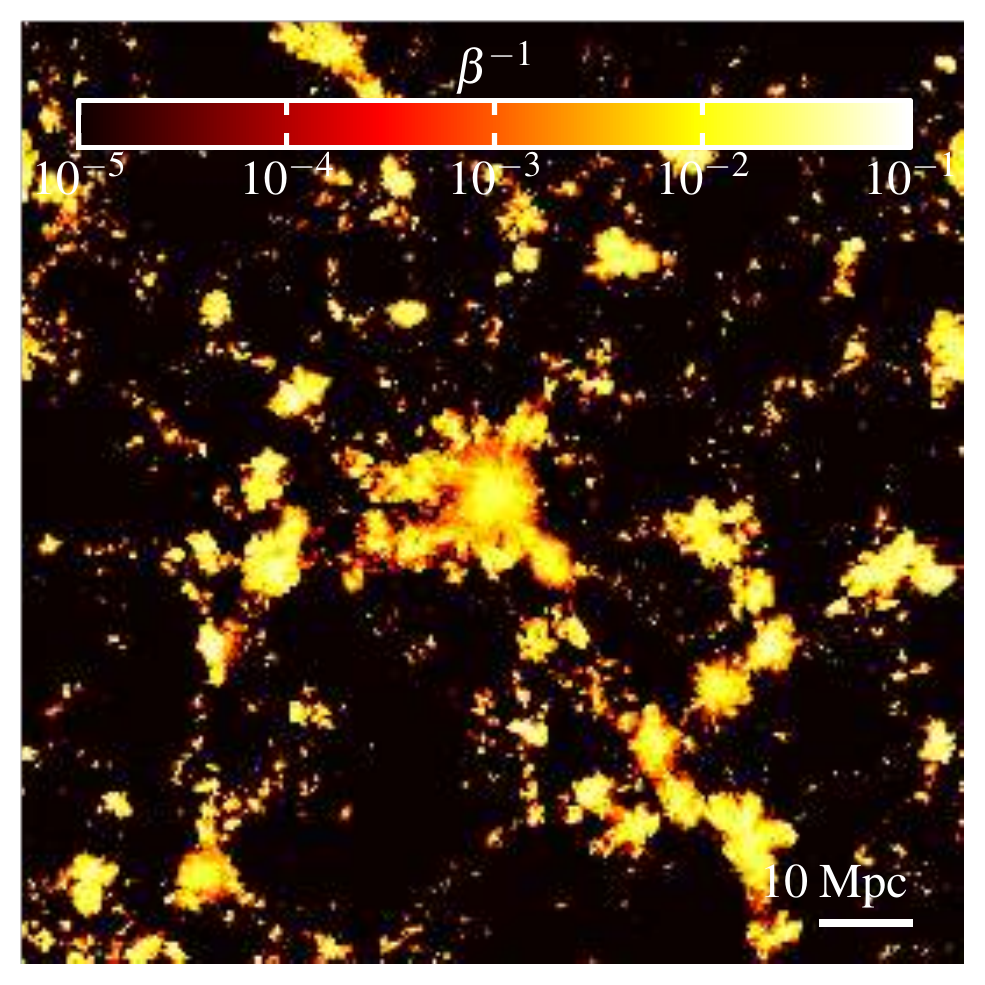} 
\includegraphics[width=0.245\textwidth]{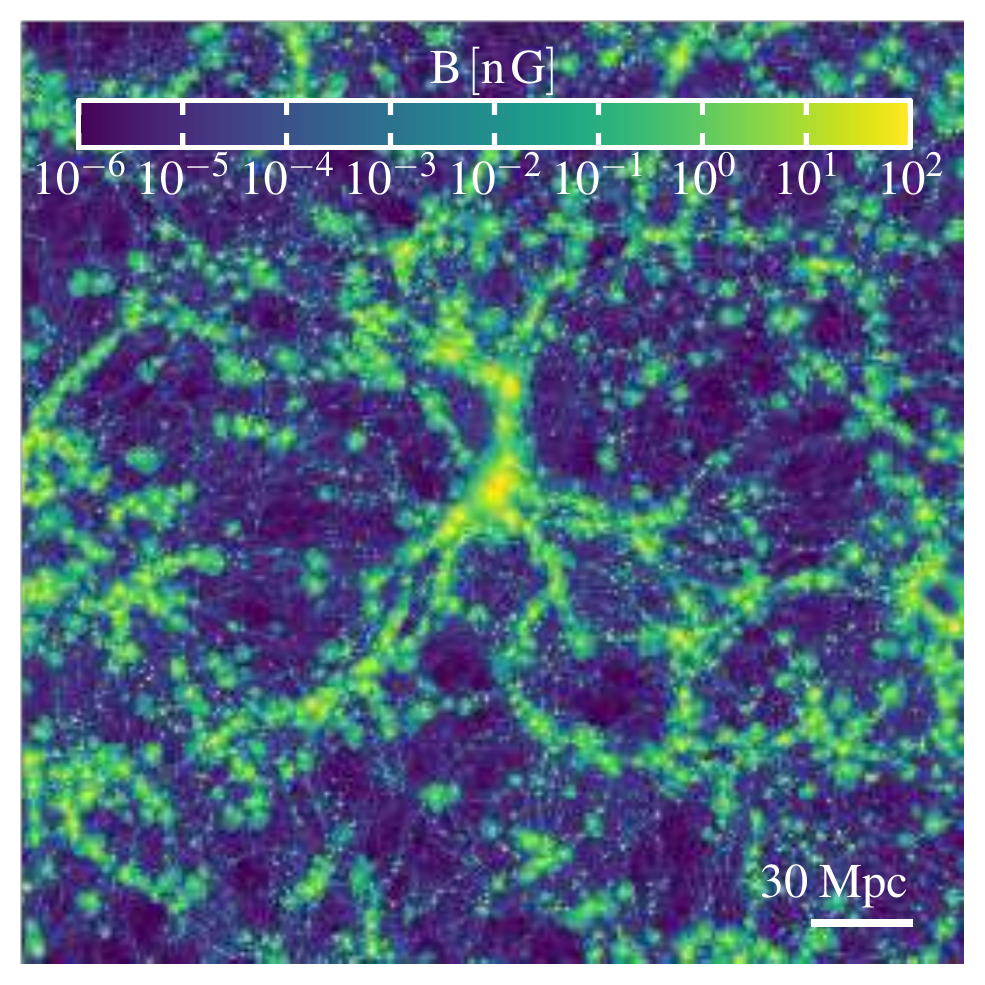}
\includegraphics[width=0.245\textwidth]{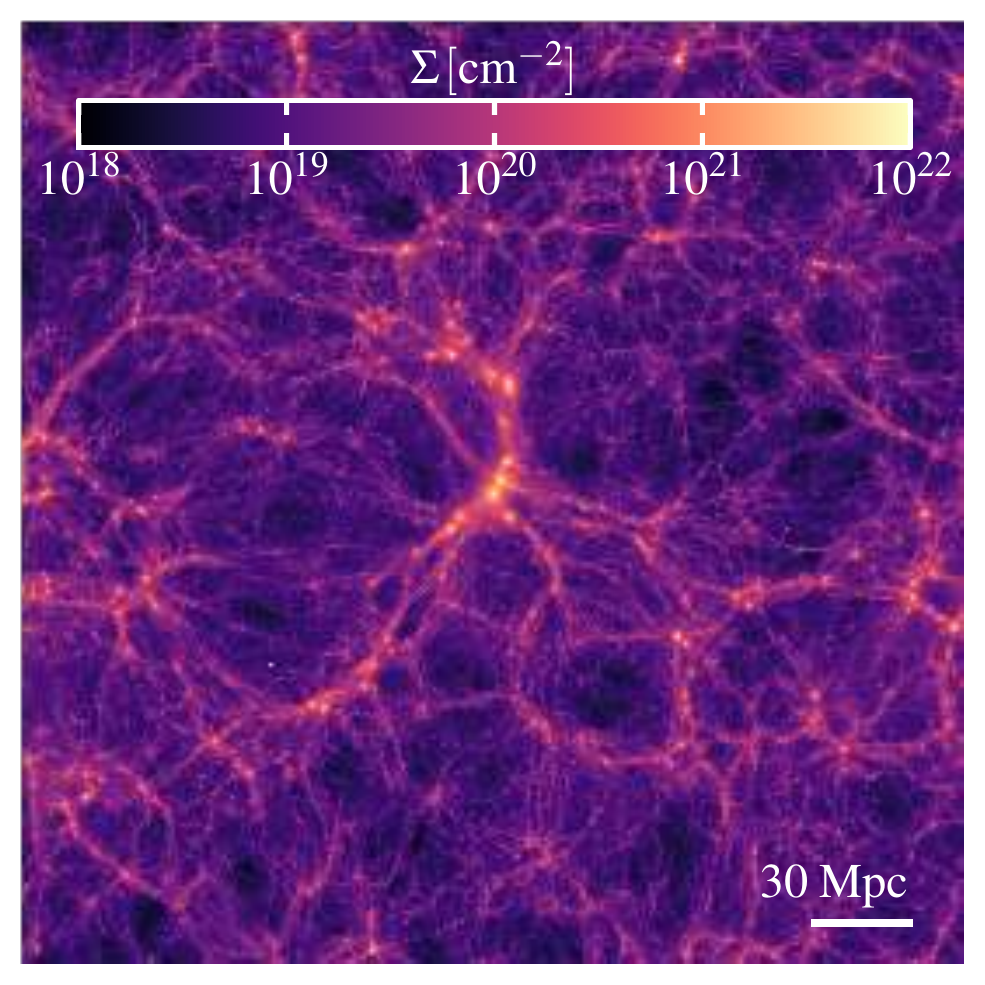}
\includegraphics[width=0.245\textwidth]{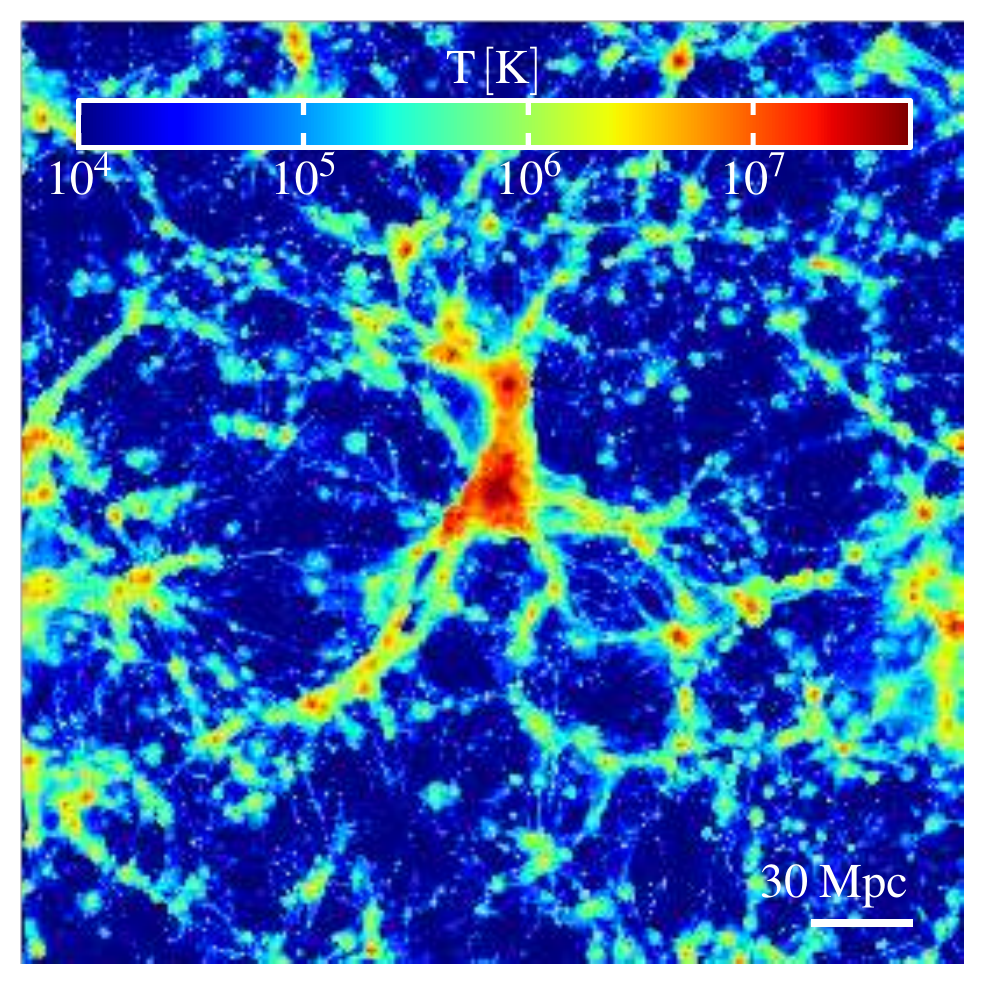}
\includegraphics[width=0.245\textwidth]{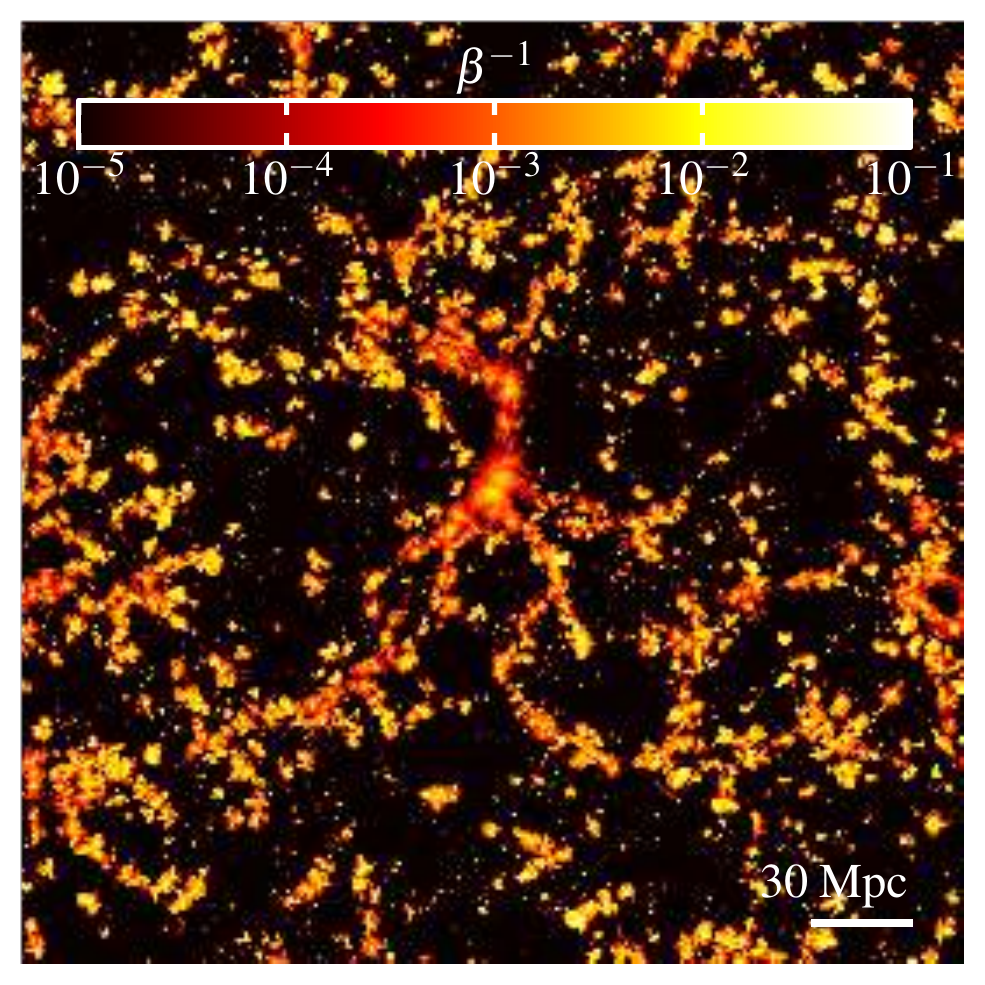}
\caption{Top row (from left-hand to right-hand side): Density-weighted projection of the 
magnetic field intensity, gas column density map, projected mass-weighted gas 
temperature and projected ratio between magnetic and thermal pressure at 
redshift zero for the TNG100 simulation. The centre of the projection region 
corresponds to the location of the most massive halo in the box and a slice of 
$\simeq 100\,{\rm Mpc}$ on a side and $\simeq 7.38\,{\rm Mpc}$ in depth is 
shown. Bottom row: Corresponding plots for TNG300 with a side length of $\simeq 
300\,{\rm Mpc}$ and a thickness of $\simeq 22.13\,{\rm Mpc}$.} 
\label{fig:Bprojection}
\end{figure*}

The paper is organized as follows. In Section \ref{sec:methods} we briefly
describe the IllustrisTNG simulation series and the numerical methodology. In
Section \ref{sec:statistics} we present the general properties of magnetic
fields in the IllustrisTNG series, both on large and on halo/galactic scales.
In Section \ref{sec:observations} we analyse the extended emission from radio
haloes in galaxy clusters and their observational scaling relations to get
further insights on the properties of magnetic fields in the simulations.
Finally, Section \ref{sec:conclusions} summarizes our findings. The modelling
of radio and X-ray emission for galaxy clusters that we have adopted in this
work is presented in \FMR{Appendixes \ref{sec:app}--\ref{sec:appc}}.

\section{The IllustrisTNG simulation series}\label{sec:methods}

The IllustrisTNG\footnote{\url{http://www.tng-project.org}} simulation series is 
the follow-up project of the Illustris simulation suite~\citep{Vogelsberger2014, 
Illustris, Genel2014, Nelson2015b, Sijacki2015}. IllustrisTNG is comprised of a 
set of cosmological MHD simulations that follow a $\Lambda$ cold dark matter 
cosmology with parameters $\Omega_{\rm m}$ = $\Omega_{\rm dm}$ + $\Omega_{\rm 
b}$ = 0.3089, $\Omega_{\rm b}$ = 0.0486, $\Omega_{\Lambda}$ = 0.6911, Hubble 
constant $H_0 = 100\,h\,\kms \Mpc^{-1}$ = 67.74 $\kms \Mpc^{-1}$, and power 
spectrum normalization and index $\sigma_{8} = 0.8159$ and $n_{\rm s} = 0.9667$ 
(\planckpappar).

IllustrisTNG consists of {\it three} primary simulations: TNG50, TNG100, and TNG300. 
The side length of {\bf TNG100} is $L \simeq 100\,\Mpc$ and its mass resolution 
is  $7.46\times 10^6\,{\rm M}_{\odot}$ and $1.39\times 10^6\,{\rm M}_{\odot}$ 
for the dark matter (DM) and baryonic components, respectively. The adopted 
values for the gravitational softening lengths are $\simeq 0.7$ kpc for DM and 
star particles (maximum physical value below $ z = 1$), while an adaptive 
comoving softening is used for gas cells with a minimum value of $\sim 0.18$ 
kpc. The larger box, {\bf TNG300}, has a side length of $L \simeq 300\,\Mpc$ and 
a mass resolution of $5.88\times 10^7\,{\rm M}_{\odot}$ and $1.1\times 
10^7\,{\rm M}_{\odot}$ for the DM and baryonic components, respectively. The 
softening lengths are a factor of two larger than in the TNG100 case. The 
smaller box, {\bf TNG50}, has a side length of $L \simeq 50\,\Mpc$ and a mass 
resolution of $4.43\times 10^5\,{\rm M}_{\odot}$ for DM and $8.48\times 
10^4\,{\rm M}_{\odot}$ for baryons. The softening lengths adopted are $\simeq 
0.3$ kpc for collisionless particles (maximum physical value below $z = 1$), and 
a minimum comoving softening of $\simeq 0.07$ kpc for gas cells. All the 
reported values refer to the highest resolution realizations of each box. In 
this work, we will present only results obtained from the TNG100 and TNG300 
series at the resolution quoted above since we are focusing on massive haloes.

For all the runs two additional simulations at lower resolution are 
also performed. Each resolution level differs from the other in terms of mass 
resolution and in the values of the softening lengths: in passing to a lower 
resolution level mass resolution is degraded by a factor of 8 and, accordingly, 
softening lengths are increased by a factor of 2. Finally, we have also run the 
associated DM-only simulations for all resolution levels. The primary 
numerical parameters of each realization can be found in Table~\ref{tab:sims}.

The simulations are carried out with the moving-mesh code \arepo\ \citep{Arepo}. 
The IllustrisTNG simulations employ a comprehensive module for galaxy formation 
physics, which is an updated version of the Illustris 
model~\citep{Vogelsberger2013}. The updated model is described in 
\citet{Weinberger2017} and \citet{Pillepich2017} to which we refer the reader 
for more details. The principal differences with respect to Illustris are: a new 
radio mode AGN feedback model~\citep[][]{Weinberger2017}, a revised SN wind 
model and refinements in the chemical evolution \citep[][]{Pillepich2017}, and 
the addition of ideal MHD \citep{Pakmor2011, Pakmor2013} \FMRR{with a \citet{Powell1999}
eight-wave divergence cleaning approach that yields results of comparable quality to the ones obtained 
with constrained transport schemes \citep{Evans1988} by keeping divergence errors under control 
\citep[see again][]{Pakmor2013}}. Furthermore, several 
algorithmic advances in the \arepo\ code have been employed in the new 
simulation suite, such as the use of a more flexible hierarchical time 
integration for gravitational interactions (Springel et al., in preparation) and 
improvements of the convergence rate of the underlying \FMRR{(magneto-)hydrodynamical solver 
\citep{Pakmor2016}}. In IllustrisTNG the code evolves the MHD equations starting 
from a homogeneous magnetic field of $10^{-14}$ (comoving) Gauss. \FMRR{In previous 
work \citep[e.g.][]{Marinacci2015, Marinacci2016, Pakmor2014, Pakmor2017} we have shown that 
the final outcome is insensitive to the actual value of the seed field over 
several orders of magnitude.} 

This work is part of a series of five papers introducing the IllustrisTNG 
project. Each one of these papers investigates different topics of the new 
simulations in order to illustrate their full scientific potential. In 
particular, in this paper we investigate the properties of the magnetic fields 
and diffuse radio emission in galaxy clusters. In \citet{Pillepich2017b}, we 
study the stellar content and distribution in massive haloes. In 
\citet{Nelson2017}, we show the colour distribution of the simulated galaxies 
and compare it with SDSS observational constraints. In \citet{Springel2017}, we 
examine the galaxy and matter clustering signal in the simulations and compare 
it with observations at low and high redshift. Finally, \citet{Naiman2017} 
explore different channels for metal enrichment in the simulations, focusing on 
chemical elements, such as europium, produced by r-processes in neutron 
star--neutron star mergers.

\section{Magnetic Field Statistics}\label{sec:statistics}
We start our analysis of the magnetic field properties in the IllustrisTNG
simulations by discussing statistics of the fields at large cosmological
and at halo/individual galaxy scales. 

\begin{figure*}
\centering
\includegraphics[width=0.49\textwidth]{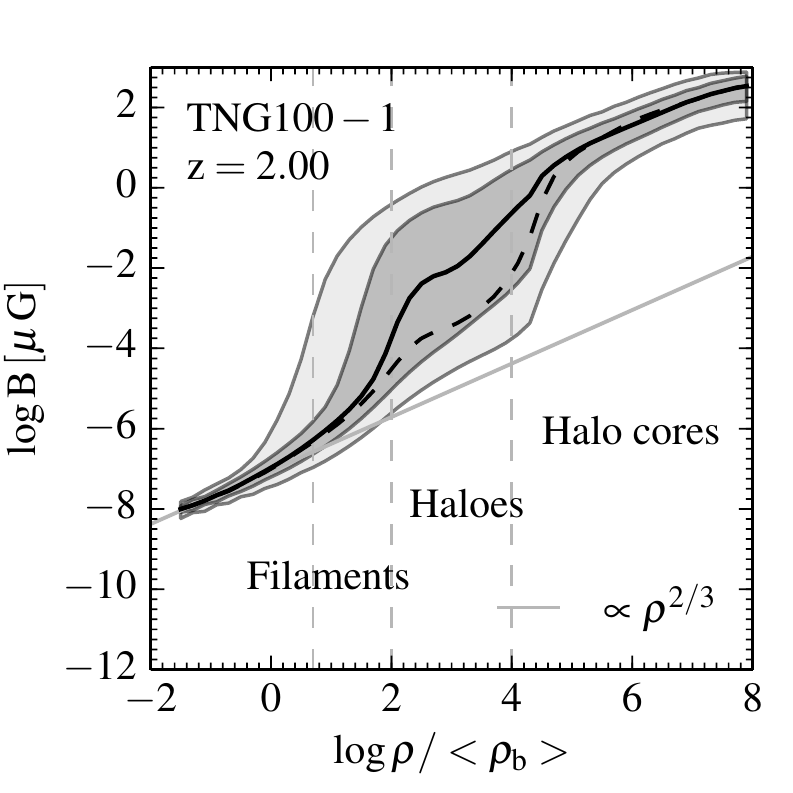}
\includegraphics[width=0.49\textwidth]{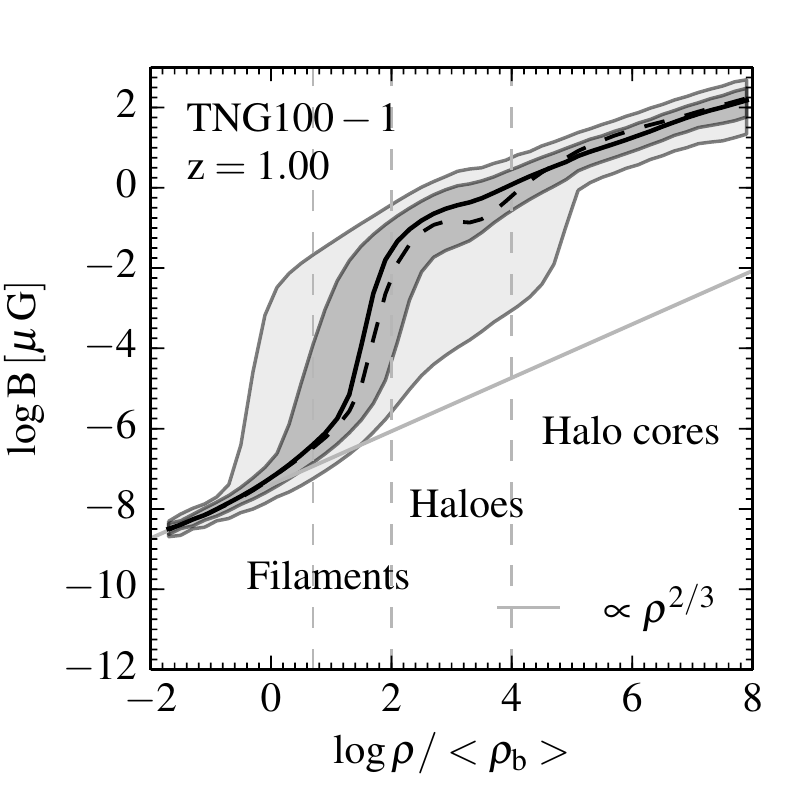}
\includegraphics[width=0.49\textwidth]{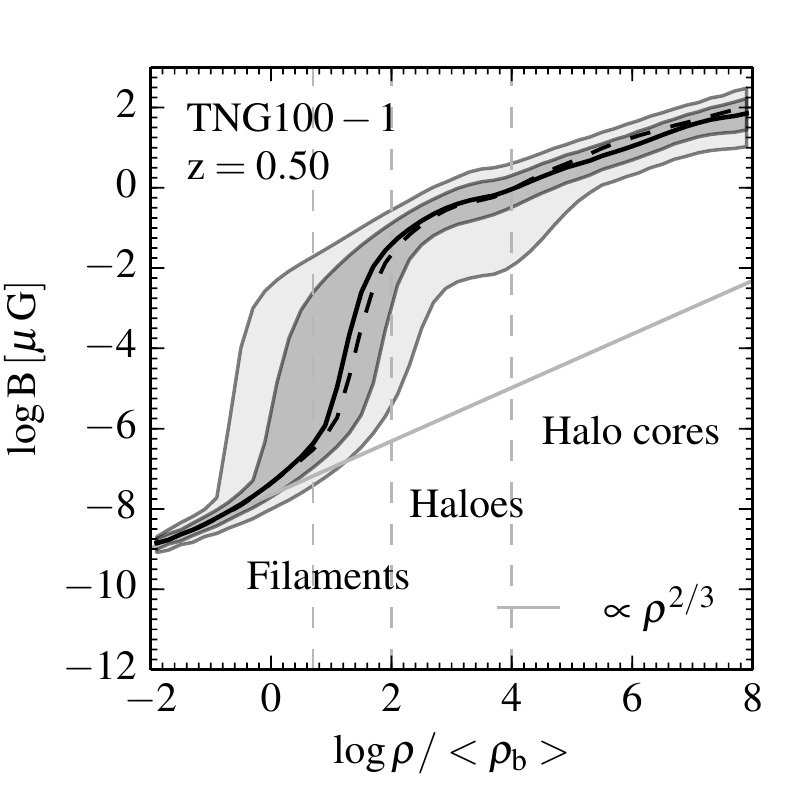}
\includegraphics[width=0.49\textwidth]{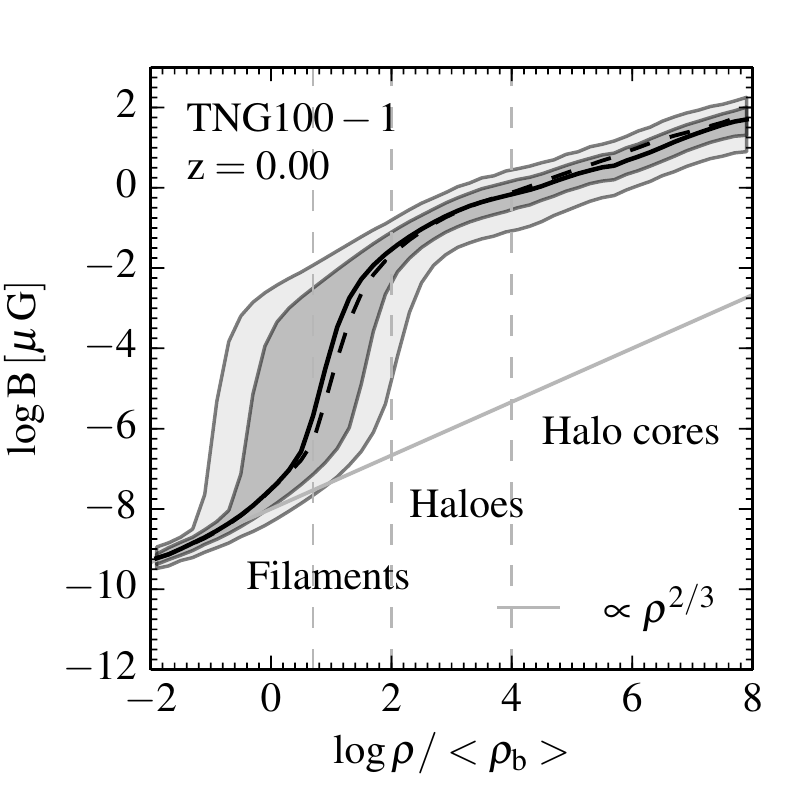} 
\caption{Redshift evolution of the magnetic field intensity versus baryon 
overdensity for the simulation TNG100. The black solid lines represent the 
median of the magnetic field distribution while the corresponding shaded areas 
indicate the $1\sigma$ (dark grey) and $2\sigma$ (light grey) spread around the 
median relation. We also show the density scaling of the magnetic field 
intensity ($\propto \rho^{2/3}$) expected in case of magnetic flux conservation 
(grey solid lines). At late times it is clearly visible how the amplification of 
the magnetic field is driven by the assembly of cosmic structures and enhanced 
by radiative cooling and feedback processes. In particular, at low overdensities 
[$\log(\rho/\rho_{\rm b}) \sim 0$] the median relationship follows closely the 
expected scaling for magnetic flux conservation but then steepens and forms a 
second branch at magnetic field intensities that are four to five orders of 
magnitude larger than this expectation at large [$\log (\rho/\rho_{\rm b}) \gsim 
1$] overdensities. The drop of the magnetic field values [$\propto (1+z)^{2}$] 
is again the effect of magnetic flux conservation and caused by the cosmological 
expansion. In all panels, the dashed black line represents the median trend 
obtained for the TNG300 box.}  
\label{fig:Bvsoverdensity}
\end{figure*}

\subsection{Large-scale statistics}\label{sec:largescale}

In Fig.~\ref{fig:Bprojection}, we present projections of a thick slice for the 
TNG100 (top row) and TNG300 (bottom row) runs, respectively. For each simulation 
four panels are displayed. From left to right they are: density-weighted 
projection of the magnetic field, gas column density map, mass-weighted gas 
temperature projection, and projected ratio between magnetic and thermal 
pressure ($\beta^{-1}$) of the gas. For both simulations, all the projections 
are centred on the most massive halo and show the full box size on the 
projection plane with a projection depth equal to $5\,h^{-1}{\rm Mpc}$ and 
$15\,h^{-1}{\rm Mpc}$ for TNG100 and TNG300, respectively.

The projections give a qualitative view of the large-scale distribution of the 
magnetic field in the simulations. We note that the main properties of the 
magnetic field on large scales (strength and spatial coincidence with haloes) 
are basically independent of the size of the simulated volume, even though the 
TNG300 run has a mass resolution of a factor of eight worse than the TNG100 box. 
Indeed, from Fig.~\ref{fig:Bvsoverdensity} we see that $||\boldsymbol{B}||$ as a 
function of the baryon overdensity differs by at most a factor of $0.25$ dex 
between TNG100 and TNG300 inside halo cores.

By comparing the projections of the magnetic field strength and gas density, it 
is apparent that the highest values of the field are found at the highest 
density peaks. This is in line with what has been reported in \citet[][and 
references therein]{Marinacci2015}, who found that magnetic fields are largely 
amplified from the initial seed value within haloes by the combined action of 
shear flows and turbulence induced by structure formation and stellar and AGN 
feedback processes. Differently from \citet{Marinacci2015}, the distribution of 
the magnetic fields tracks more closely the density distribution of gas. This is 
due to the difference in the implementation of the AGN feedback model 
\citep[see][]{Weinberger2017}, which is less bursty in the IllustrisTNG 
simulations compared to the original Illustris model used in 
\citet{Marinacci2015}. This is also apparent by comparing the temperature 
projections. In IllustrisTNG the high temperature regions are only found within 
massive haloes, while due to the radio mode feedback implementation in Illustris 
\citep[see][]{Sijacki2007, Vogelsberger2013} those regions extend much further 
out. The ratio between magnetic and gas thermal pressure shows that the highest 
values are reached around density peaks within haloes while they drop 
dramatically in filaments and voids. The largest $\beta^{-1}$ values are 
achieved in low-mass haloes, with the central value of $\beta^{-1}$ is declining 
with increasing virial mass for both boxes (see also discussion of 
Fig.~\ref{fig:BetaandBstackedfullphys}). This is due to the larger gas 
temperatures in massive haloes that tend to decrease the maximum value of 
$\beta^{-1}$. In TNG300 the coarser resolution also plays a role because the B 
field amplification is somewhat reduced. Generally, magnetic fields are 
dynamically unimportant on large scales.

In Fig.~\ref{fig:Bvsoverdensity}, inspired by \citet{Dolag2005}, we show more 
quantitatively the redshift evolution of the magnetic field intensity as a 
function of the baryon overdensity for TNG100. Vertical dashed lines roughly 
separate regions such as filaments, haloes and halo cores at  selected values of 
baryon overdensities. The solid grey line indicates the expected scaling under 
the assumption of magnetic flux conservation. The solid black line shows the 
median trend of magnetic field intensity, while the shaded areas indicate the 
$1\sigma$ (dark grey) and the $2\sigma$ (light grey) dispersion around the 
median value. 

\begin{figure*}
\centering
\includegraphics[width=0.29\textwidth]{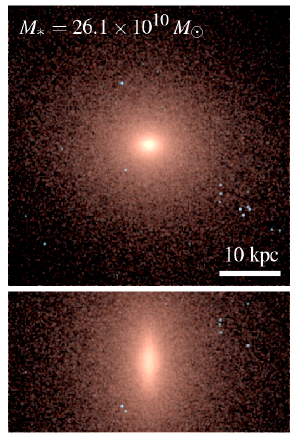}
\includegraphics[width=0.29\textwidth]{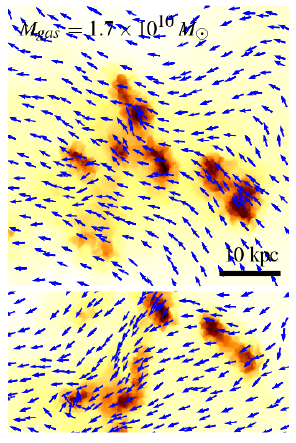}
\includegraphics[width=0.29\textwidth]{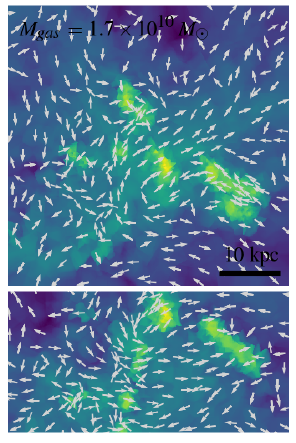}
\includegraphics[width=0.29\textwidth]{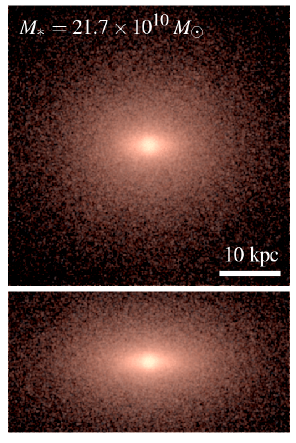}
\includegraphics[width=0.29\textwidth]{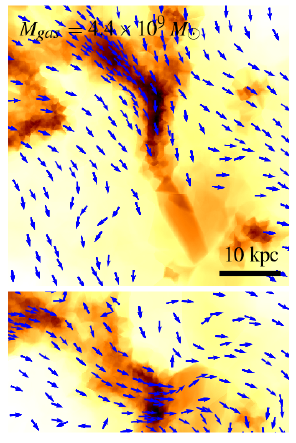}
\includegraphics[width=0.29\textwidth]{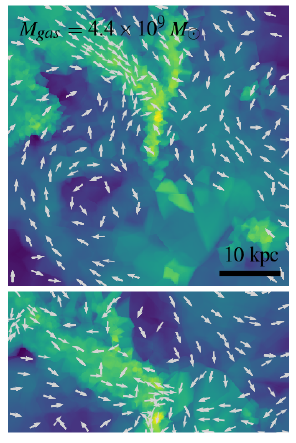}
\includegraphics[width=0.29\textwidth]{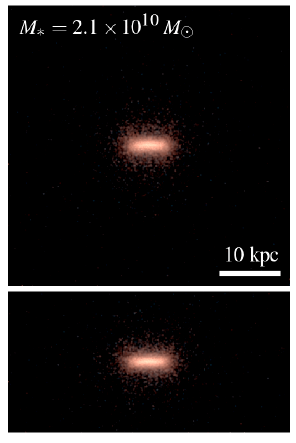}
\includegraphics[width=0.29\textwidth]{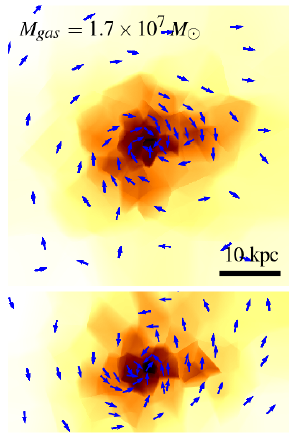}
\includegraphics[width=0.29\textwidth]{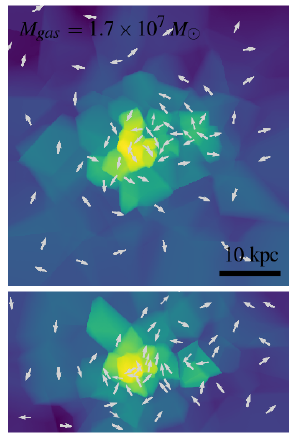}
\caption{Stellar (left-hand panels), gas density (middle), and volume-weighted 
magnetic field (right-hand panels) projections of three early-type galaxies in the TNG100 
simulation. The arrows show the direction of the velocity and magnetic fields
in the gas density and magnetic field panels, respectively.} 
\label{fig:Bmorphologies}
\end{figure*}

\begin{figure*}
\centering
\includegraphics[width=0.29\textwidth]{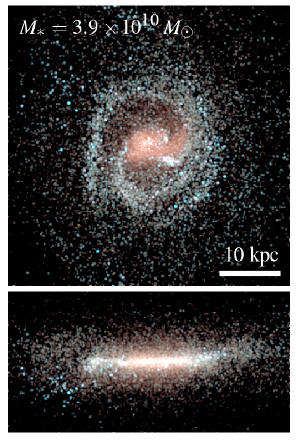}
\includegraphics[width=0.29\textwidth]{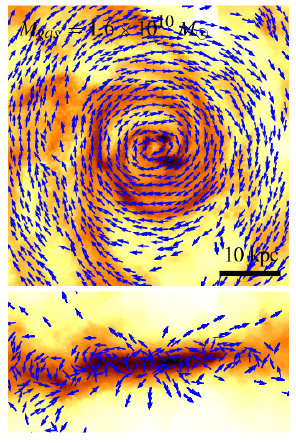}
\includegraphics[width=0.29\textwidth]{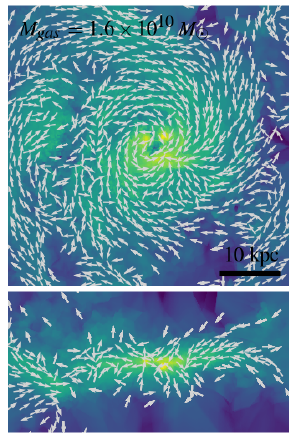}
\includegraphics[width=0.29\textwidth]{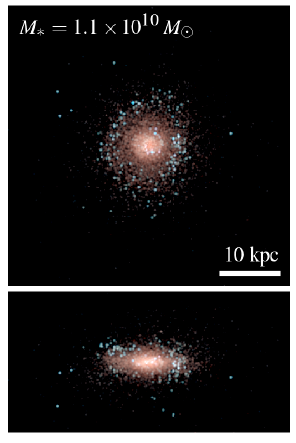}
\includegraphics[width=0.29\textwidth]{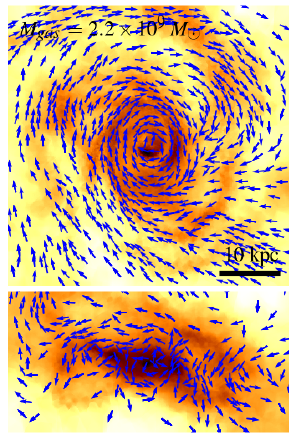}
\includegraphics[width=0.29\textwidth]{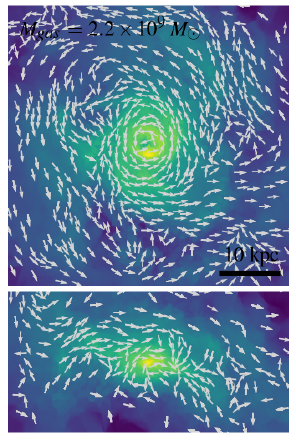}
\includegraphics[width=0.29\textwidth]{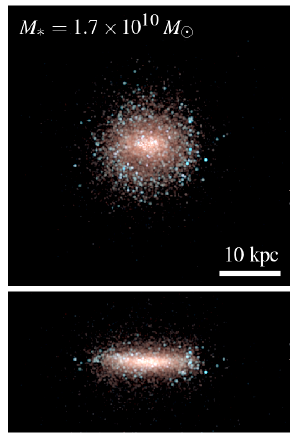}
\includegraphics[width=0.29\textwidth]{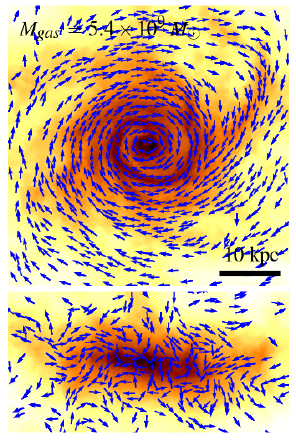}
\includegraphics[width=0.29\textwidth]{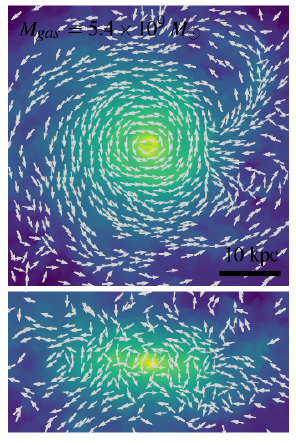}
\caption{As in Fig.~\ref{fig:Bmorphologies} but for three late-type galaxies 
in the TNG100 simulation.} 
\label{fig:Bmorphologies2}
\end{figure*}

At all redshifts, it can be seen that the amplification of magnetic fields from 
their seed value is driven by the assembly of structures. At low baryon 
overdensities, the magnetic field strength follows the relation expected from 
magnetic flux conservation arguments, i.e. the amplification is primarily driven 
by the compression of the gas. This trend changes dramatically for overdensities 
above $\simeq 10$ where a sudden increase of the field strength sets in, up to 
about four or five orders of magnitude above the flux conservation predictions. This 
is indicative that additional processes -- e.g. shear flows and turbulence 
\citep[see also][]{Dolag1999, Dolag2002} -- are at play to amplify magnetic 
fields. The field strength after the sudden increase at overdensities $\sim 10$ 
keeps increasing with baryon overdensity. Trends are similar at all redshift, 
with the $z = 2$ panel showing the least abrupt change in magnetic field 
intensities from low to high overdensities. However, even at $z=2$ most of the 
amplification at large overdensities has already occurred. We also find the same 
trends (shown by the dashed black lines in Fig.~\ref{fig:Bvsoverdensity}) for 
the TNG300 box. Given the lower resolution of this run, the magnetic field 
amplification is less pronounced at the transition region between low-density 
gas and gas within haloes (especially at $z=2$). However, at low and large 
overdensities the predictions of the TNG100 and TNG300 simulations agree.
\FMRR{To conclude, given the similarity in the shape of the median magnetic field strength-gas overdensity relation 
with the one found by \citet{Marinacci2015} in their full-physics simulations, 
we would like to point out that the full spectrum 
of baryon physics, and in particular the presence of feedback loops, is needed to amplify 
the initial seed field to values of $\sim 10\,\muG$ observationally found inside galaxies
(see also Section \ref{sec:galscale})}.

\subsection{Halo and galaxy statistics}\label{sec:galscale}

We investigate the connection between galaxy morphology and magnetic field 
topology in Figs.~\ref{fig:Bmorphologies} and \ref{fig:Bmorphologies2}, where we 
select three examples of early-type galaxies (Fig.~\ref{fig:Bmorphologies}) and 
late-type galaxies (Fig.~\ref{fig:Bmorphologies2}) from the TNG100 simulation. 
For each galaxy class, the three columns show from left to right: stellar 
density, gas density and volume-weighted magnetic field intensity. In the two 
latter columns, superimposed arrows show the direction of the velocity and 
magnetic fields, respectively. 

Early-type galaxies show predominantly a spheroidal distribution of stars and a 
highly irregular gas distribution, which is due to the interaction of the AGN 
kinetic feedback with the surrounding gas \FM{\citep[see also][]{Gaspari2012, 
Weinberger2017}}. An exception is represented by the least massive galaxy of the 
sample, in which a thick and old stellar disc and a more regular spheroidal gas 
distribution are present. The magnetic field intensity traces the gas density 
distribution very closely and its topology is highly irregular. Magnetic field 
lines tend to prevalently orient along the gas filaments visible in the 
projections, which, in the second case, also correlate with the predominant 
direction of the velocity field presented in the middle panel.

\begin{figure*}
\centering
\includegraphics[width=0.93\textwidth]{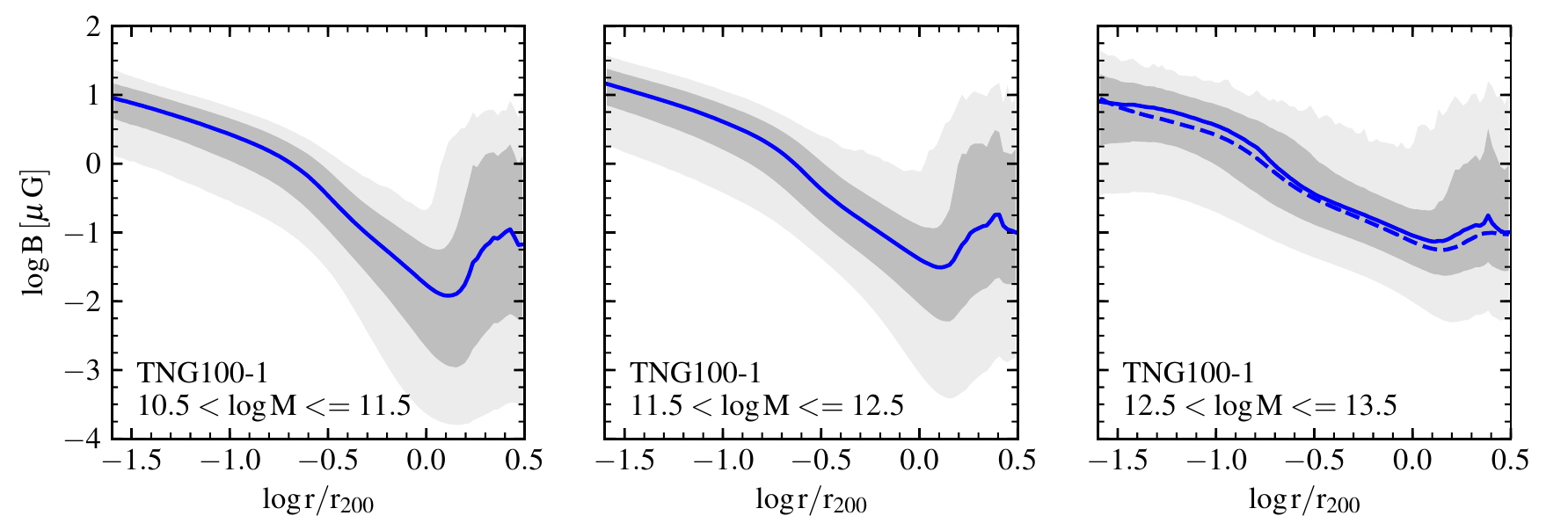}
\includegraphics[width=0.93\textwidth]{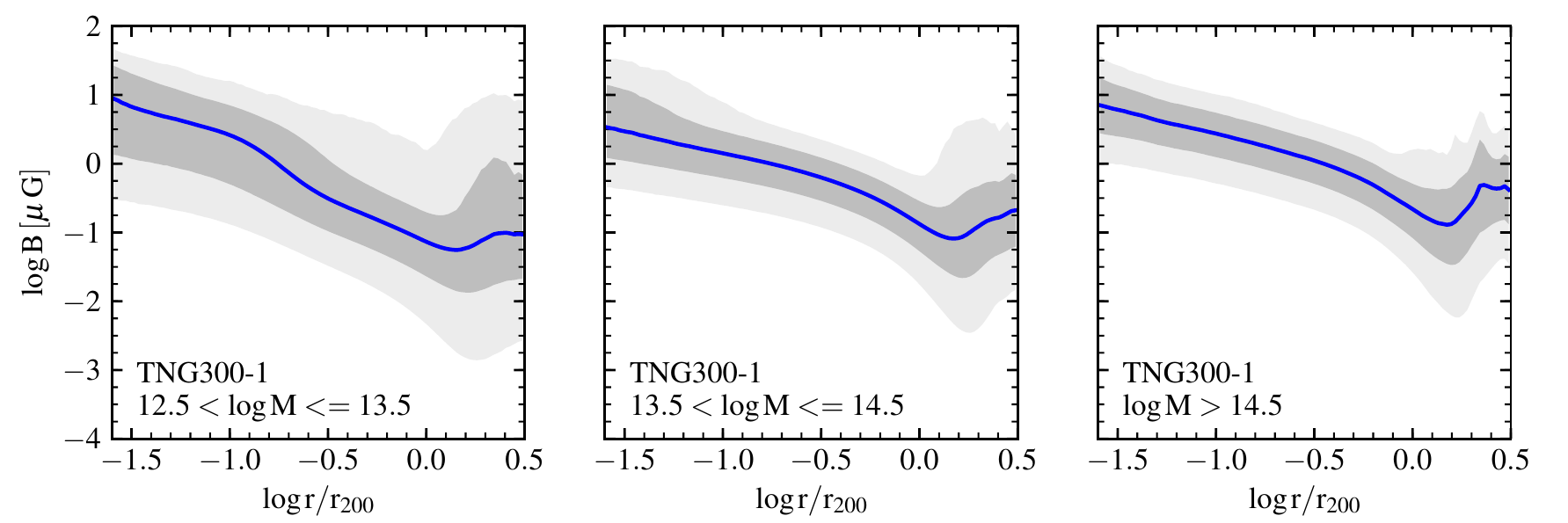}
\includegraphics[width=0.93\textwidth]{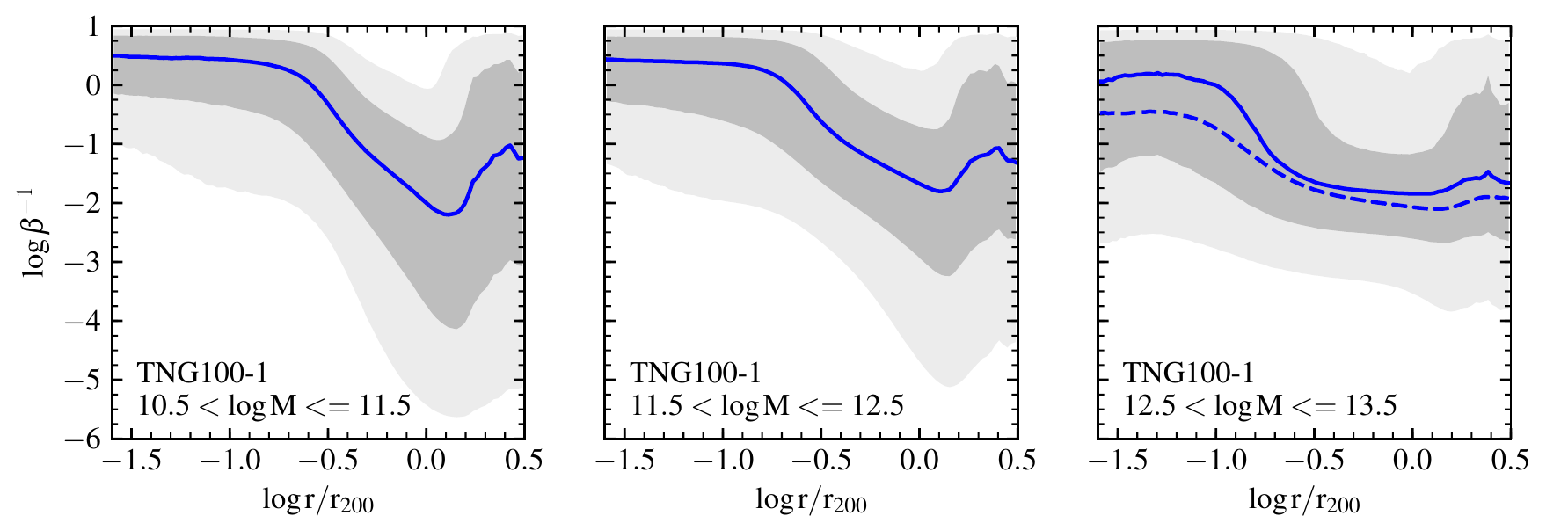}
\includegraphics[width=0.93\textwidth]{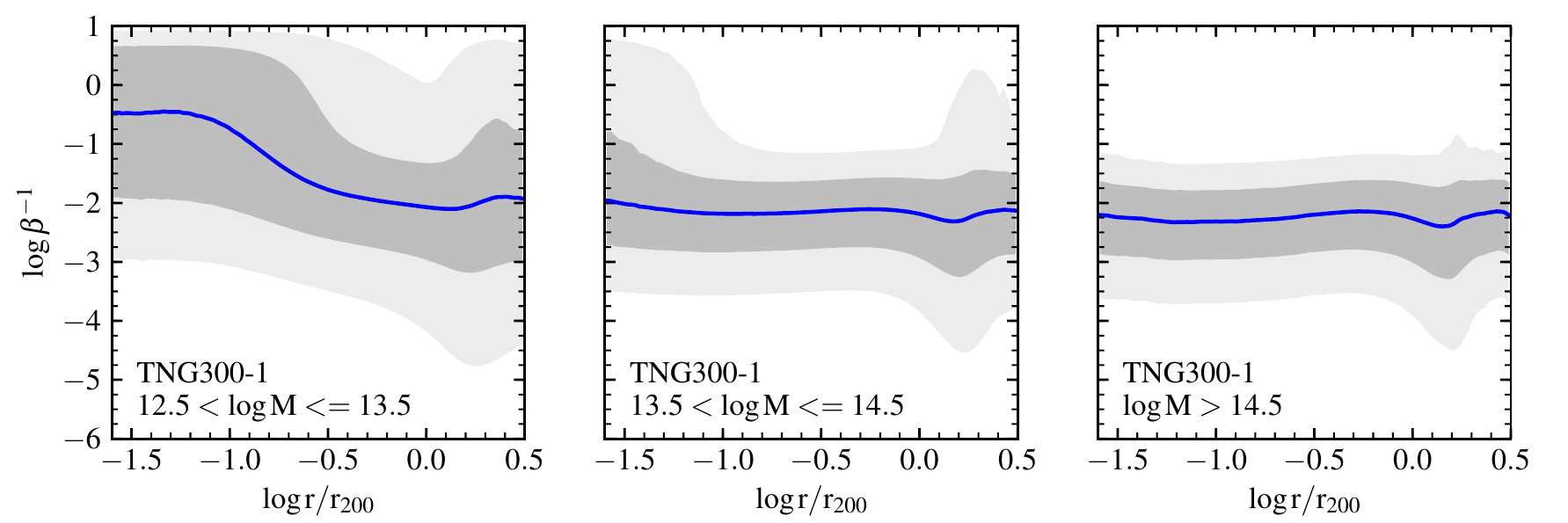}
\caption{Top six panels: Radial magnetic field intensity profiles at redshift 
zero. The figure presents the stacked profiles of the FOF groups in the virial 
mass bins indicated in each panel. Less massive haloes have been taken from the 
TNG100 run (top row), while, at the more massive end, haloes from the TNG300 run 
(bottom row) have been used. The solid blue line represents the (mass-weighted) 
median trend while the dark and light grey regions show the $1\sigma$ and $2\sigma$ 
spread around the median value, respectively. Bottom six panels: Radial profiles 
of the ratio between magnetic and thermal gas pressures at redshift zero. To 
facilitate the comparison in the overlapping mass bins, we report the median 
relation for the TNG300 run in the TNG100 results as a blue dashed line.} 
\label{fig:BetaandBstackedfullphys}
\end{figure*}

For late-type galaxies stars are organized in a comparatively thin disc with 
respect to the early-type galaxies. The disc still shows residual star 
formation, as the blue stellar colours indicate. The gas is also organized in a 
similar structure, which is rotationally supported and extends far beyond the 
stellar disc of the galaxy, as the density projection panels clearly show. For 
the first galaxy presented here the bipolar configuration of the galactic wind 
generated by stellar feedback is also visible in the edge-on gas projection. The 
magnetic field intensity follows closely the gas distribution. However, the most 
striking difference compared to early-type galaxies is that the topology of the 
field is much more regular and the field is predominantly oriented with the 
gaseous disc. In fact, within discs the magnetic field is mostly toroidal and 
anti-aligned with the direction of the velocity field. We would like to note 
that the magnetic field orientation could switch back and forth between aligned 
and anti-aligned configurations \citep[see also][]{Pakmor2014}. Moreover, the 
differential rotation of the gaseous disc \FM{can provide} the source for field 
amplification via a galactic dynamo \FM{\citep[see also][]{Pakmor2017}}\FMRR{, although
it is unclear if this can be directly modelled at the resolution achieved in IllustrisTNG}. The 
final magnetic field intensities ($\approx 1-10\,\muG$) are consistent with 
observational findings for late-type galaxies \FMRR{\citep{Basu2013, Beck1996, 
Beck2009, Beck2015}}.

\begin{figure*}
\centering
\includegraphics[width=0.33\textwidth]{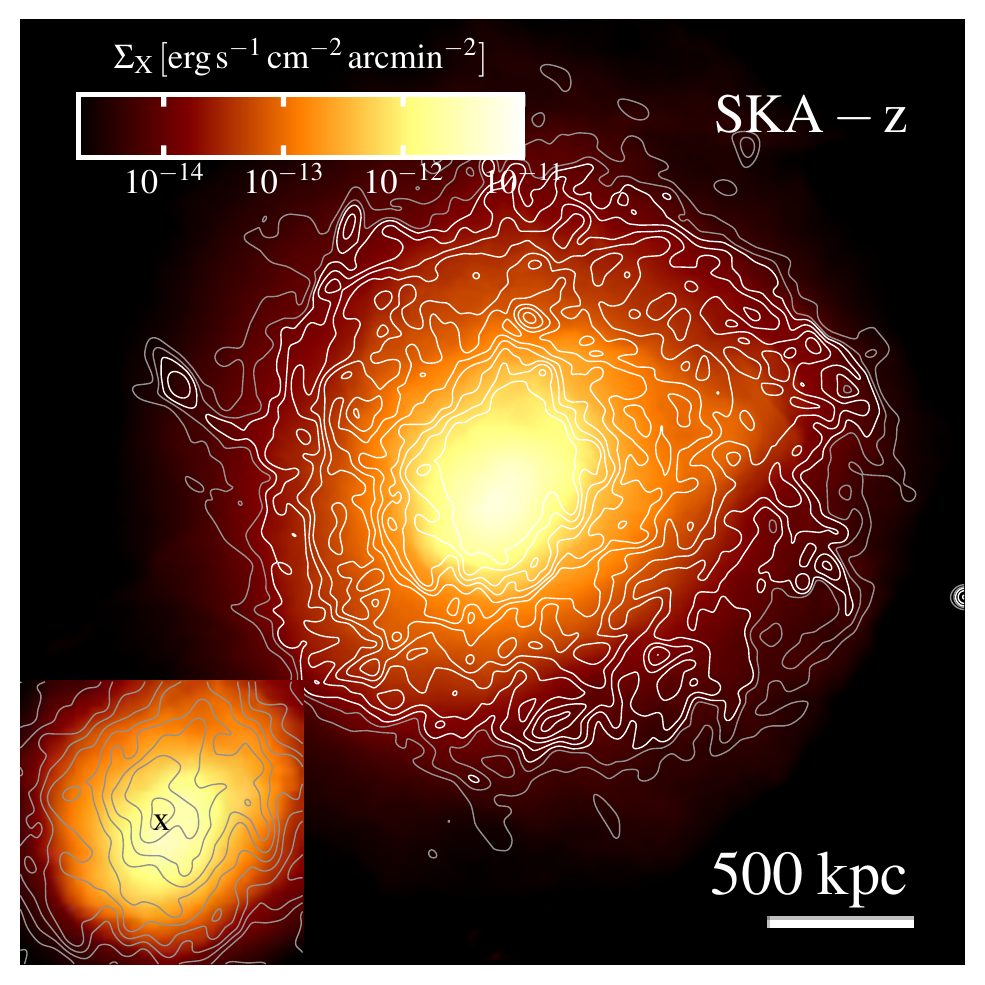}
\includegraphics[width=0.33\textwidth]{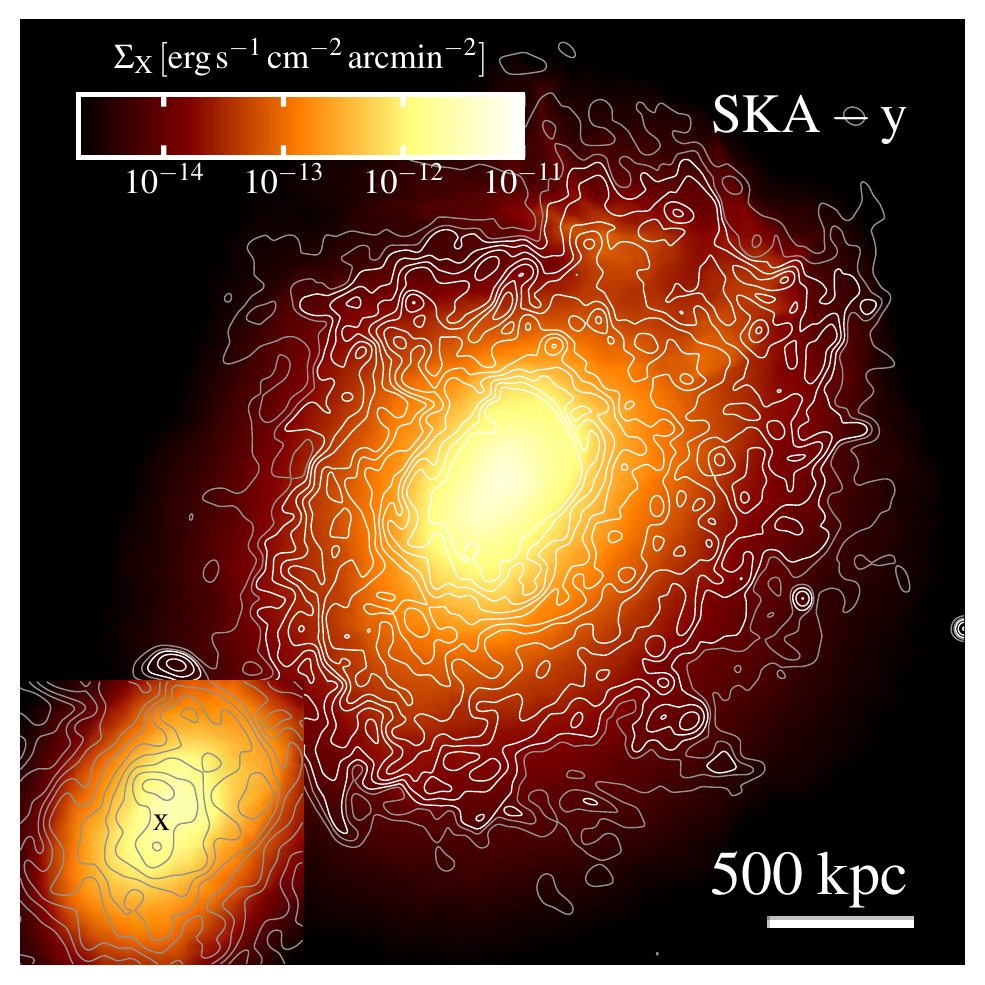}
\includegraphics[width=0.33\textwidth]{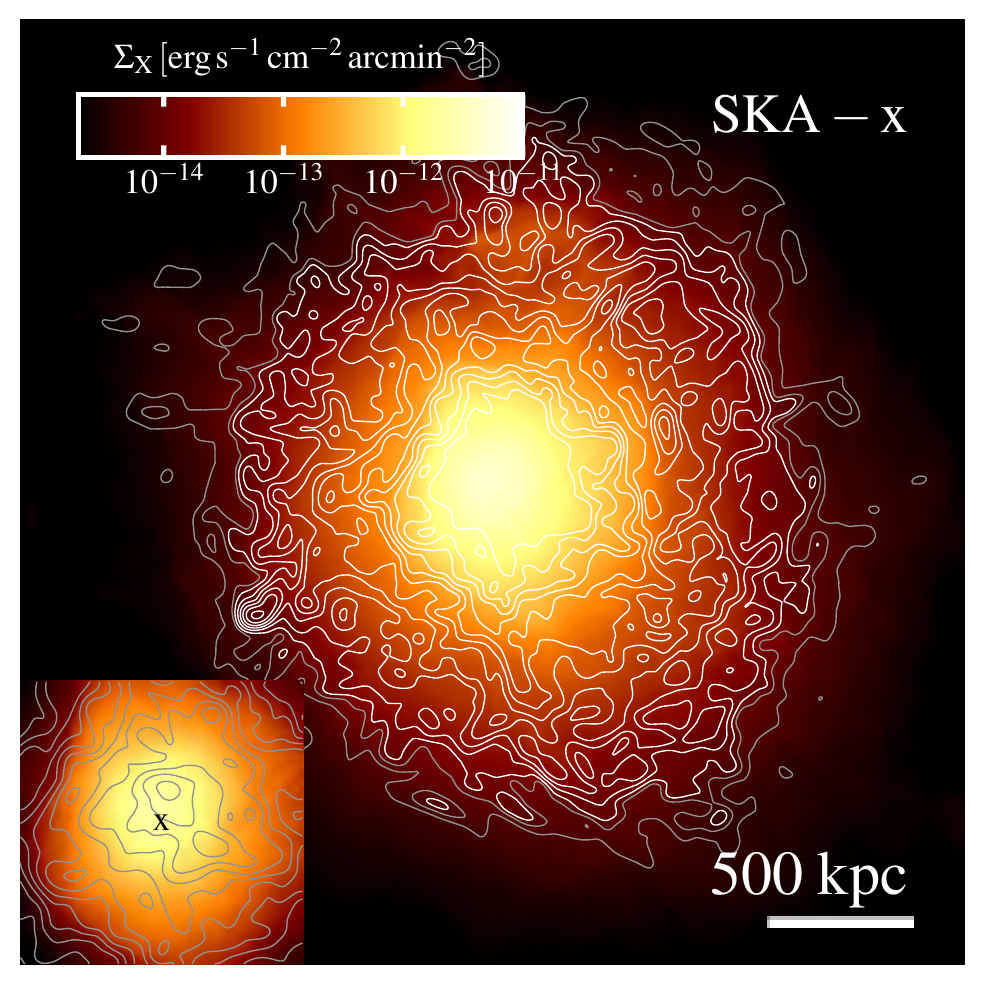}
\caption{X-ray map (colour) overlayed with synchrotron emission contours 
\FMR{(model 1 in Section~\ref{sec:model1})} for the 
most massive halo of the TNG300 simulation taken from the $z = 0$ snapshot. The 
panels are 3.5 Mpc on a side and in projection depth and show projections in the 
$xy$, $xz$, and $yz$ planes from left to right, respectively. Note how the X-ray 
and radio morphologies match one another, indicative of the fact that the gas 
and large-scale magnetic field morphologies are similar. Note also the offset 
between the centres of X-ray and radio emissions. X-ray maps are displayed to a 
minimum surface brightness value of $1.95\times10^{-15} \erg\,{\rm 
s^{-1}}\,\cm^{-2}\,{\rm arcmin^{-2}}$, assuming a telescope effective area of 
$200\,\cm^{2}$ \citep{Anderson2010}. Radio maps are smoothed on a scale of 
10 arcsec (FWHM) with a Gaussian kernel. Contour levels are placed at $0.06\, 
{\rm mJy\,beam^{-1}}$ spaced by a $\sqrt{3}$ factor from one another (white, 
representing detections), while grey contours are at 0.02 and 0.04 ${\rm 
mJy\,beam^{-1}}$. These values are representative of the typical beam size and 
sensitivity (0.02 ${\rm mJy\,beam^{-1}}$ is the $1\sigma$ noise level) of future 
SKA observations of radio haloes in galaxy clusters \citep[see][]{Vazza2015a}. 
The assumed redshift used to generate the X-ray and radio maps is 0.2. More 
details about map creation can be found in Appendix~\ref{sec:app} and 
Table~\ref{tab:mapsproperties}.} 
\label{fig:XrayRadioport}
\end{figure*}

In Fig.~\ref{fig:BetaandBstackedfullphys}, we show stacked profiles of the 
mass-weighted magnetic field intensities (top six panels) and the ratio between 
magnetic and thermal pressure (bottom six panels). Solid lines in each panel 
show the median trend and dark and light shaded areas the associated $1\sigma$ and 
$2\sigma$ spreads. Regardless of the halo mass bin\footnote{Unless otherwise 
stated, halo masses (i.e. virial masses) are expressed in terms of 200 times the 
critical density for closure $\rho_{\rm crit} = 3 H_0^2/ (8\pi G)$ throughout 
the paper.}, the magnetic field intensity is a declining function of radius and 
is declining more slowly in more massive haloes. The intensity of the field 
tends to decline faster once a radial distance $r \sim 0.3\,r_{\rm 200}$ is 
reached. Past the virial radius, magnetic field intensities tend to increase 
again, which is due to the contribution of substructures. Magnetic field 
intensities in the centres are $\simeq 1-10\,\muG$ with a drop of about three orders 
of magnitude in the external regions. These values have to be compared with a 
magnetic field intensity of $\simeq 10^{-3}\,\muG$ (i.e. the value of the seed 
field scaled by adiabatic compression at the highest baryon overdensities), 
which demonstrates the efficient amplification of the field within haloes due to 
structure formation. The maximum value reached by the magnetic field intensity 
is consistent across all mass bins. The TNG300 simulation shows in general good 
agreement, in both the central B field value and the slope of the relation, for 
the overlapping virial mass bins despite its coarser resolution when compared to 
TNG100 (see the blue dashed line for a direct comparison). If anything, the 
magnetic field profile in the overlapping mass bin declines slightly faster with 
radius, which is likely due to a less efficient amplification caused by numerical 
diffusion on the coarser grid scale \citep[e.g.][]{Cho2009, Jones2011, 
Vazza2014}. 

\begin{figure*} 
\centering
\includegraphics[width=0.495\textwidth]{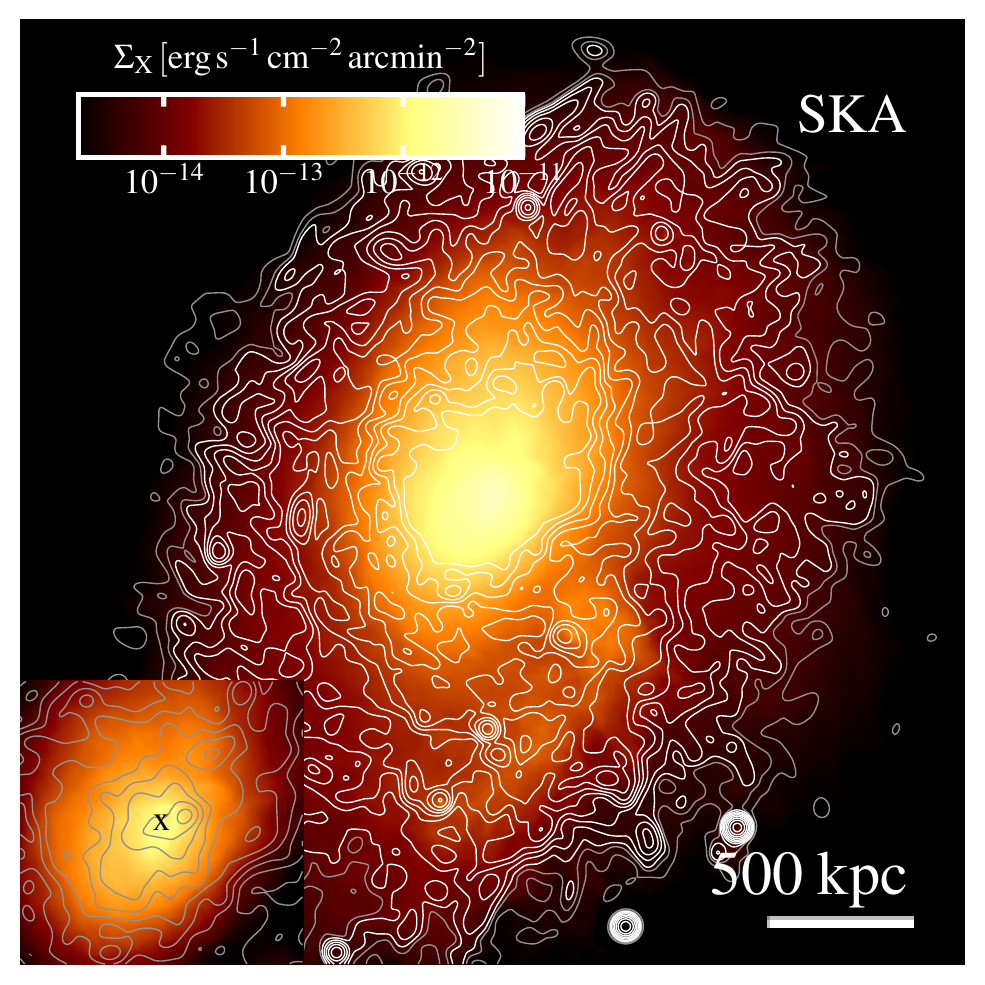}
\includegraphics[width=0.495\textwidth]{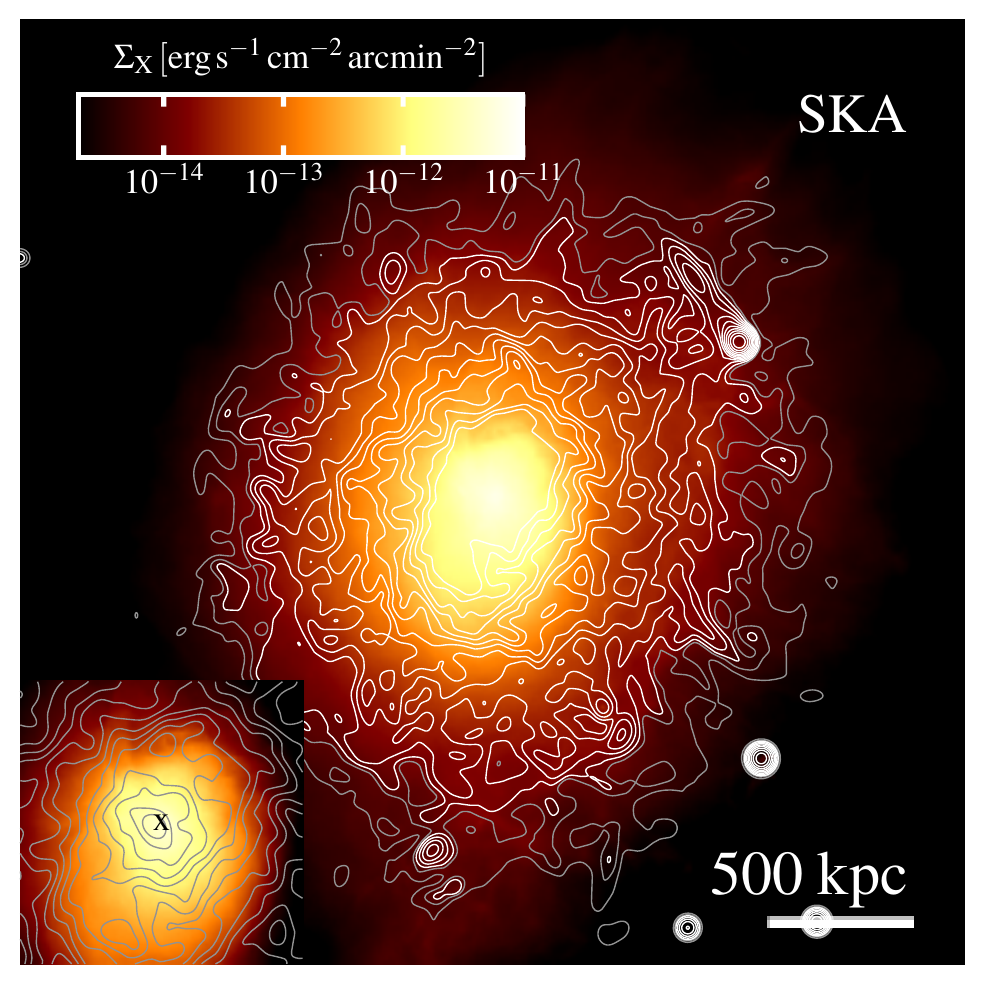}
\includegraphics[width=0.495\textwidth]{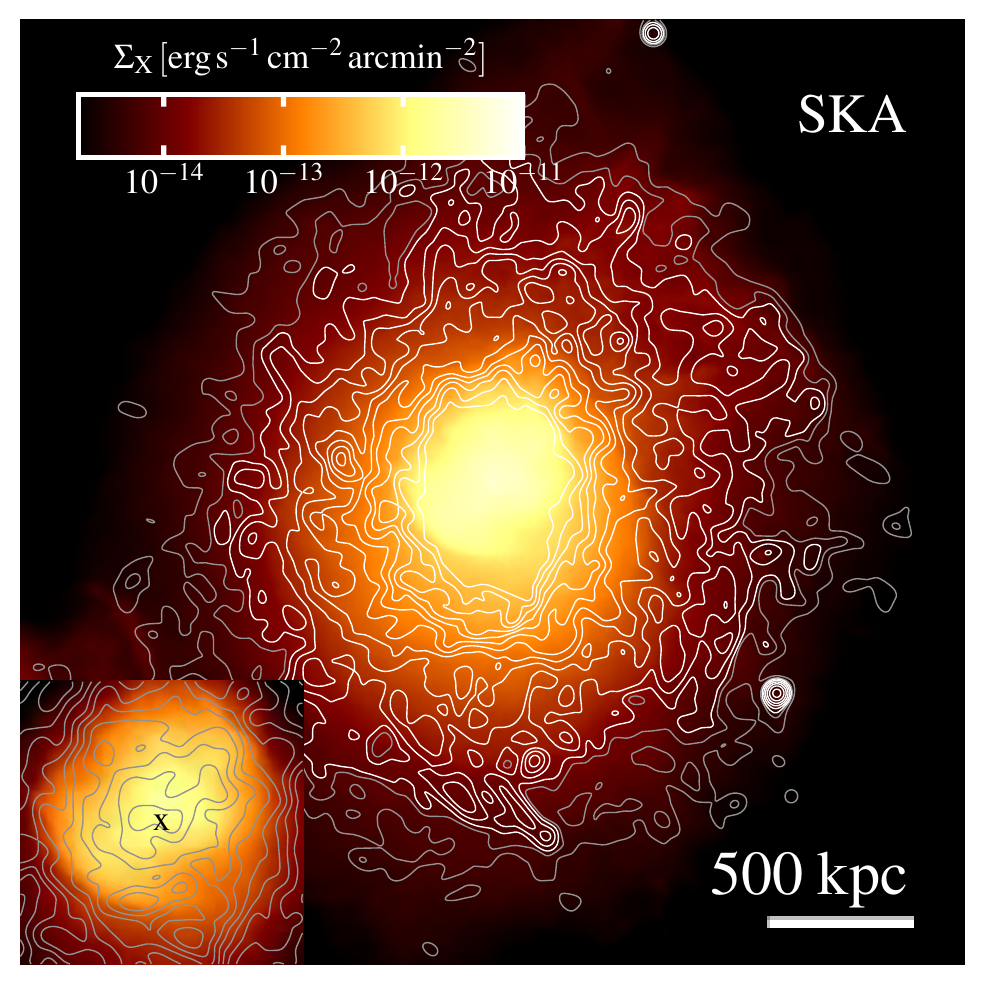}
\includegraphics[width=0.495\textwidth]{fig8d}
\caption{As in Fig.~\ref{fig:XrayRadioport}, but showing the evolution of the 
most massive halo. Note at earlier times the presence of compact regions of 
radio emission without any significant X-ray counterpart originating from 
substructures. Maps are generated for the $z = 0.3$, $z = 0.2$, $z=0.1$, and $z = 
0$ (from top left to bottom right) outputs, respectively.} 
\label{fig:XrayRadioev}
\end{figure*}

The ratio between magnetic and thermal pressure ($\beta^{-1}$) gives an 
indication of the importance of magnetic fields on the dynamics of the gas. From 
the bottom six panels in Fig.~\ref{fig:BetaandBstackedfullphys}, it can be seen 
that in the external regions of the haloes, for distances $\sim 0.3\,r_{\rm 
200}$, magnetic fields are sub-dominant and the $\beta^{-1}$ ratio is at most 
0.1. Moving to the inner regions, the value of the field gets largely amplified 
\FM{by the combined action of radiative cooling (leading to high baryon 
overdensities) together with the increased level of shear and turbulent gas 
motions triggered by stellar and AGN feedback \citep[see also][]{Marinacci2015},}
\FMR{while the thermal content of the gas declines due to the efficient 
cooling occurring at the halo centres. It is currently unclear whether 
with the present resolution it is possible to model the emergence of a 
small-scale dynamo amplification \citep[see e.g.][]{Rieder2017},
although measurements of the turbulence spectra \citep{Pakmor2017} 
in higher resolution cosmological simulations performed with \arepo\ are consistent with 
this scenario. Strong magnetic field 
amplification and effective gas cooling at the halo centres both} contribute to increase 
the $\beta^{-1}$ up to a value of $\sim 3$, as it is visible for the panels 
analysing the TNG100 run. Note also That in this case the contribution of the 
magnetic field to the dynamics of the gas is in general sub-dominant, since the 
magnetic pressure is typically smaller (of the order of $\sim 30$ per cent) than 
the kinetic energy density content of the gas \citep[see also][for a calculation 
at higher resolution focusing on the Milky Way mass scale]{Pakmor2017}. For the 
panels showing the results of the TNG300 simulation, this increase of the 
$\beta^{-1}$ ratio is also present, but the maximum value never exceeds a few 
percent or tens of percent. We ascribe this difference between TNG100 and TNG300 
to two effects: the lower numerical resolution of the latter simulation, which 
yields lower or more rapidly declining values for B fields (note that the most 
massive bin of the TNG100 and the least massive bin of the TNG300 runs overlap; 
the two simulations can be directly compared here with the help of the blue 
dashed line); and the mass of the haloes analysed in the TNG300 runs that host 
high-temperature gas, which is kept hot by AGN feedback processes, in the halo 
regions.

\section{Observational signatures in massive haloes}\label{sec:observations}

In this section we explore several mock observations and scaling relations that 
depend on magnetic fields within massive haloes. We focus on the massive end of 
the TNG300 box by adopting a lower cut in virial mass of $10^{14}\,{\rm 
M_\odot}$, which is below the value explored by current radio observations, but 
it will be probed by future SKA surveys \citep{Cassano2015}. This leaves us with 
a sample of 280 haloes in total. \FMR{A description of how the X-ray and radio 
emission are modelled can be found in Appendices~\ref{sec:app} and ~\ref{sec:ff}. 
In particular, for radio \FMRR{halo} emission we have compared two different models in our analysis.
They differ in the way the distribution of relativistic particles (and in particular electrons) is parametrized
and we refer the reader to Sections~\ref{sec:model1} and \ref{sec:Sarazin} for their
complete characterization.}

\FMRR{We would like to caution that the models for the relativistic electron 
distribution that we adopt in this analysis are rather simplistic. However, the 
scope of this work is not to find nor suggest a solution to all the outstanding 
issues related to  the complex phenomenon of radio halo emission in galaxy 
clusters. Instead, with the present analysis we aim at exploiting the unique opportunity 
that the IllustrisTNG suite offers in terms of number of galaxy clusters 
simulated with a cutting-edge galaxy formation physics model that includes 
self-consistently the evolution of magnetic fields. To do so it is natural to 
first use simple models for the distribution of relativistic particles to explore
them and assess their weak points under the 
assumption that the magnetic field properties predicted by the simulations can 
be trusted. Trying more complex and advanced models for relativistic particles 
and comparing them in detail with the observations would be desirable, but it is 
outside the scope of the present investigation. The optimal solution would be to 
self-consistently include cosmic rays in the calculations 
\citep[e.g.][]{Pakmor2016, Pfrommer2017}, but in large cosmological volumes, as 
the ones probed by IllustrisTNG, this is a  very challenging numerical task that 
is still very far from a satisfactory solution.}

\subsection{Radio and X-ray mock observations}

Figure~\ref{fig:XrayRadioport} shows a composite X-ray image overlayed with 
radio emission contours (see Appendix \ref{sec:app} for details on how X-ray and 
radio \FMR{-- model 1 in Section~\ref{sec:model1} --} emissions are computed) for the 
most massive halo of the TNG300 simulation ($M_{200} = 1.53 \times 10^{15}{\rm 
M_\odot}$). This galaxy cluster would observationally be classified as having 
extended radio halo emission.  The three panels show different projections 
(along the $z$, $y$, and $x$ axis). The inset on the bottom left-hand corner of each 
panel is $800\,\kpc$ across and shows a zoom of the central region.  The inset 
is centred on the potential minimum of the halo, and the `x' indicates the 
centre of the X-ray emission, which coincides with the centre of the projection. 
The colour map showing the X-ray brightness displays only values above the 
detection limit for typical {\it Chandra} parameters. Radio contours are placed 
at $1\sigma$ and $2\sigma$ rms level (grey, corresponding to non-detection) and at 
levels $>3\sigma$ spaced by factors of $\sqrt{3}$ (white, corresponding to 
detection) and are computed for an SKA observation at 120 MHz, with the 
associated resolution and sensitivity (see Table~\ref{tab:mapsproperties}). Only 
gas cells that are not eligible for star formation in our galaxy formation 
physics model (i.e. gas with $n \lsim 0.1\,{\rm cm^{-3}}$) and that are cooling 
are considered for computing radio and X-ray emission. We apply this selection 
criterion in all the figures presenting radio and/or X-ray results.

The maps indicate that X-ray and radio emissions are spatially coincident. The 
morphology of both emission maps is regular and roughly spherical, although in 
the $y$ projection both radio and X-ray emissions are elongated in the lower 
left to upper right direction. Radio contours follow quite closely the shape of 
the X-ray emission, showing the close link between the physics generating the 
two processes. In particular, X-rays probe the thermodynamic state of the hot 
gas, while radio emission probes the strength and amplification of the magnetic 
fields. Both the thermodynamics of the gas and the field strength are set by the 
assembly of the cluster and feedback processes due to AGN and stellar feedback. 
So a connection between these two emission mechanisms is expected. 

However, by looking more closely at the central regions it can be seen that 
there is an offset between the radio and X-ray signals. This shift depends on 
the orientation chosen for the projection: for example, the two maxima of the 
emission are almost coincident in the left-hand panel, while the offset becomes more 
noticeable in the other two orientations. In the middle panel, the centre of the 
X-ray emission is in between two local maxima of the radio map. An offset 
between the maxima of X-ray and radio emissions is also often found in radio 
halo observations \citep{Govoni2012}. Radio haloes can be quite asymmetric and 
the offsets are more pronounced for more asymmetric distributions and smaller 
radio halo sizes \citep{Feretti2010}. A possible explanation for this offset can 
be the presence of large spatial variations in the intracluster magnetic field 
or a non-uniform distribution of relativistic particles. Note that our modelling 
 does not include the latter effect, which implies that we might be 
underestimating the offsets in this analysis.

\begin{figure*}
\centering
\includegraphics[width=0.247\textwidth]{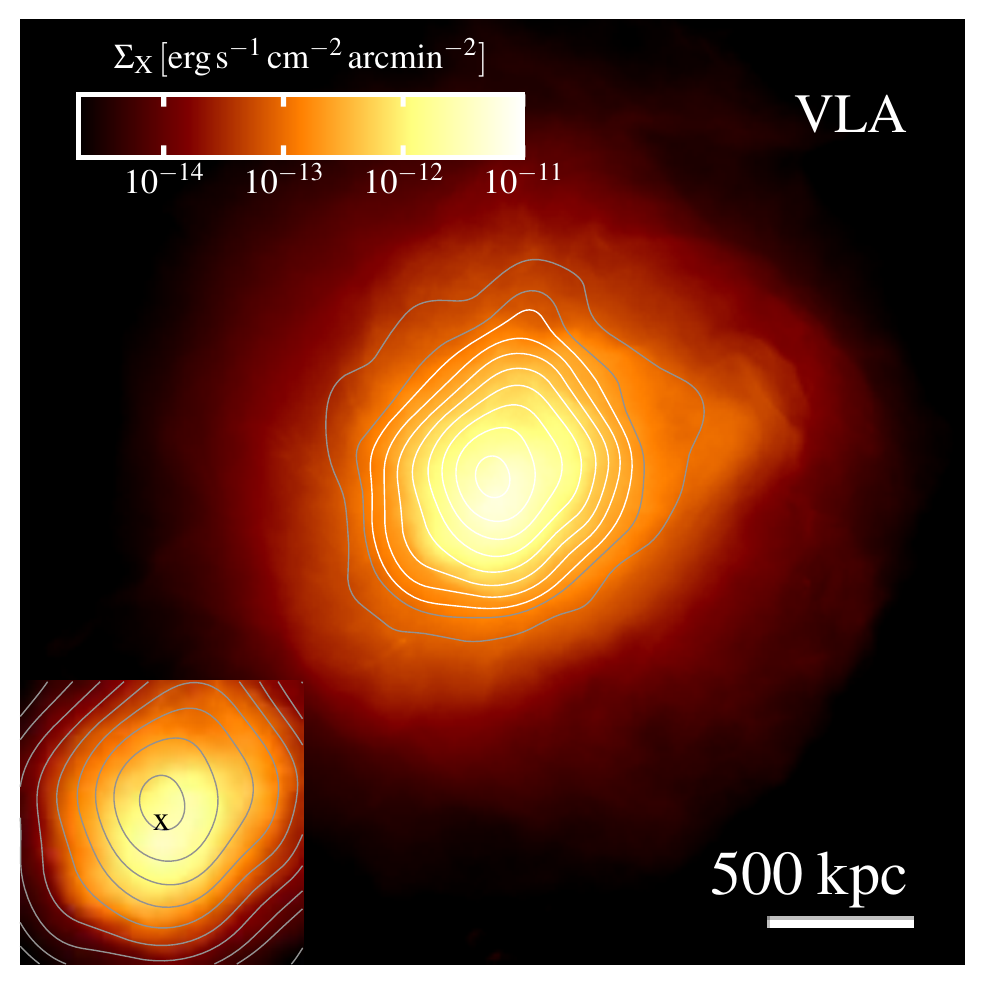}
\includegraphics[width=0.247\textwidth]{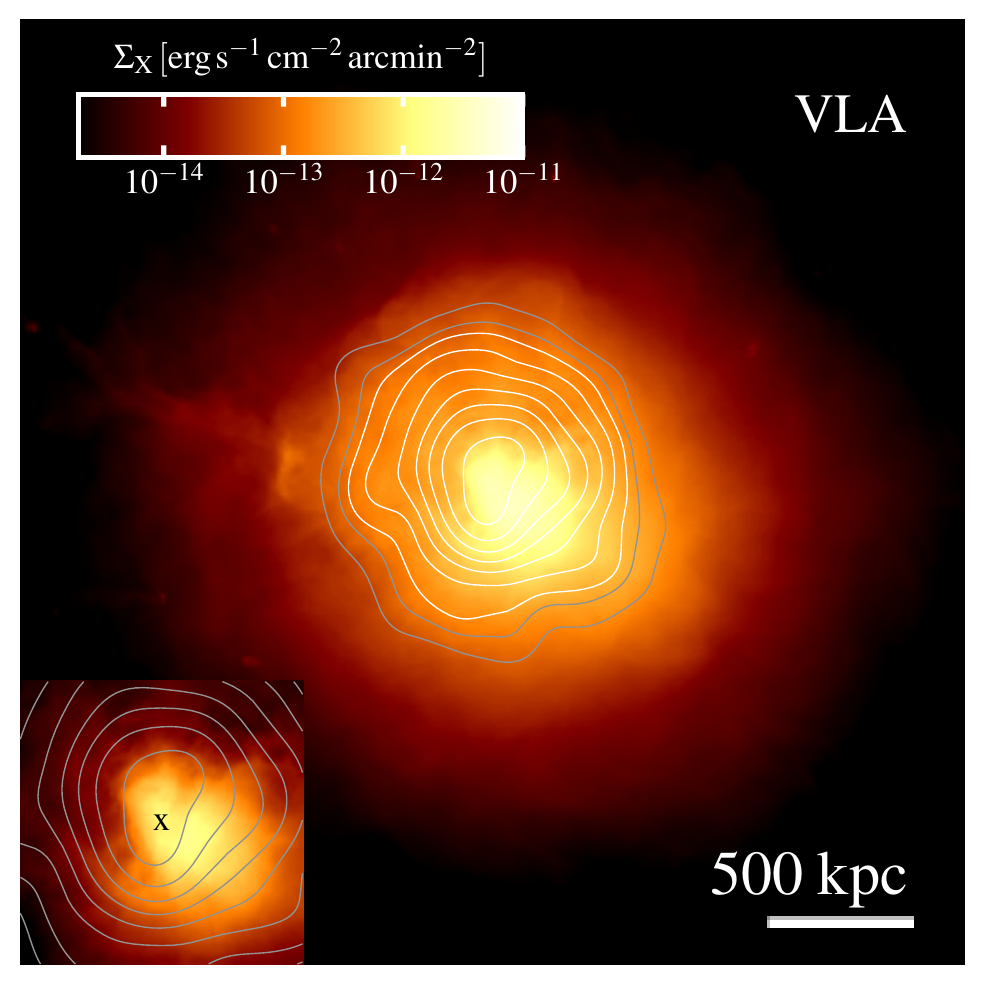}
\includegraphics[width=0.247\textwidth]{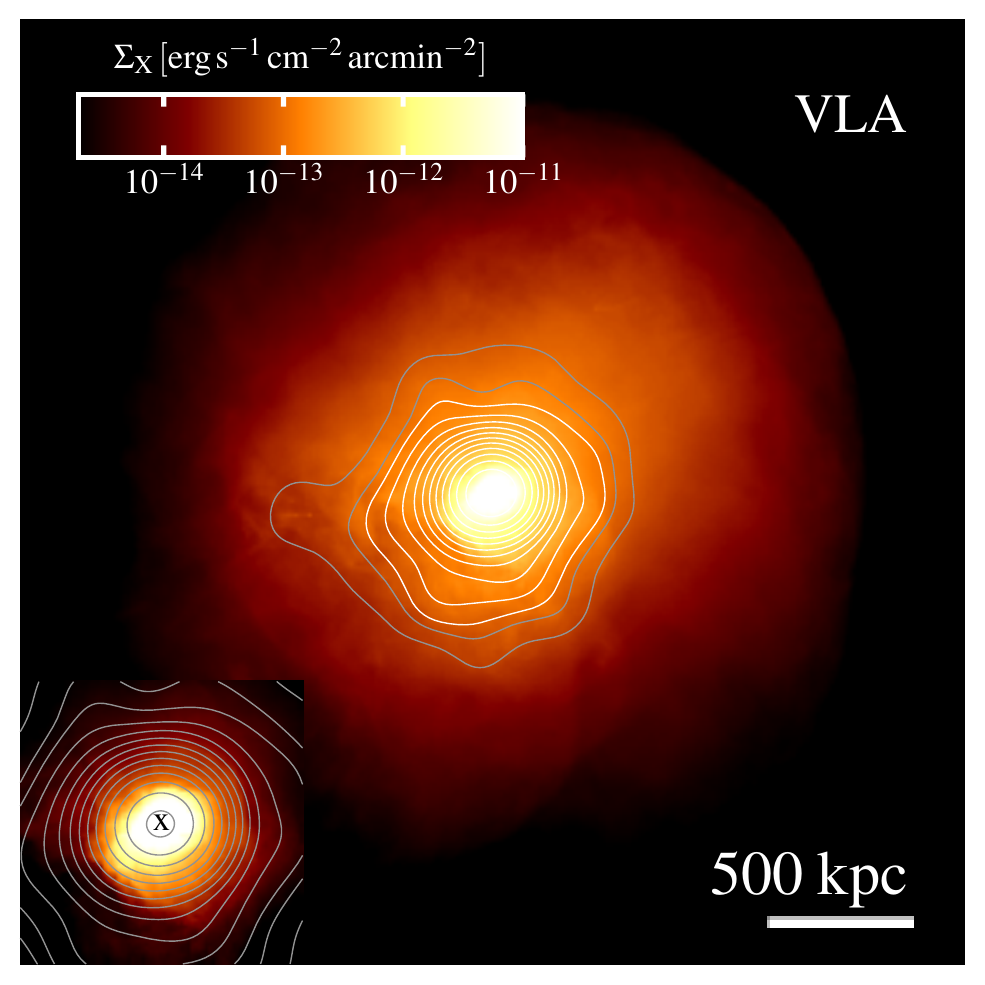}
\includegraphics[width=0.247\textwidth]{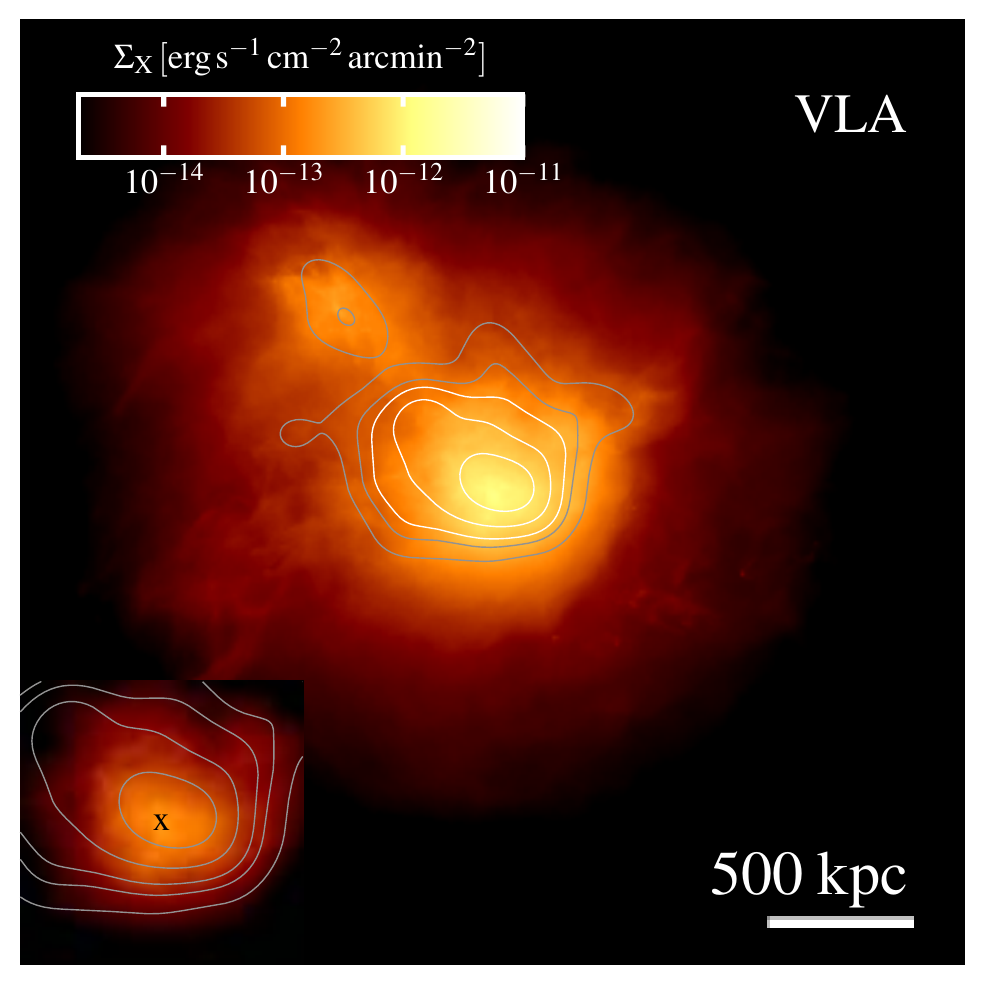}
\includegraphics[width=0.247\textwidth]{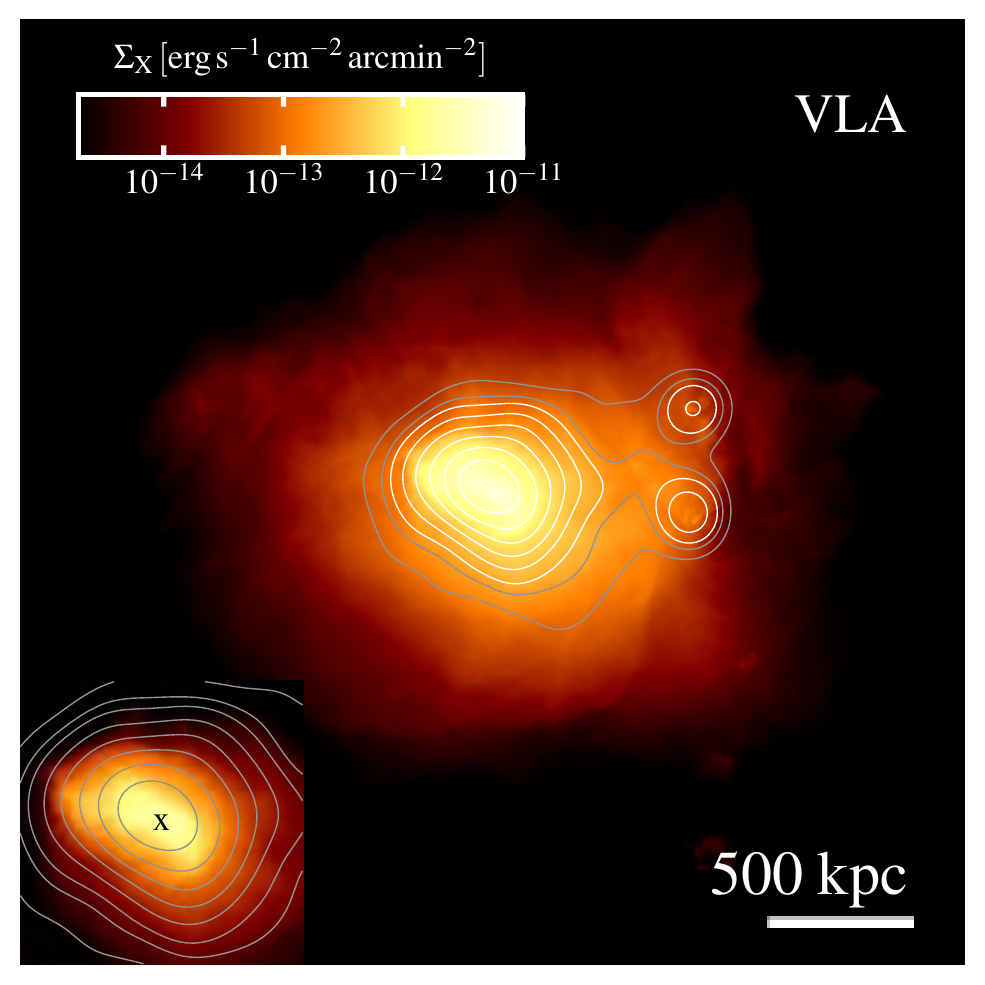}
\includegraphics[width=0.247\textwidth]{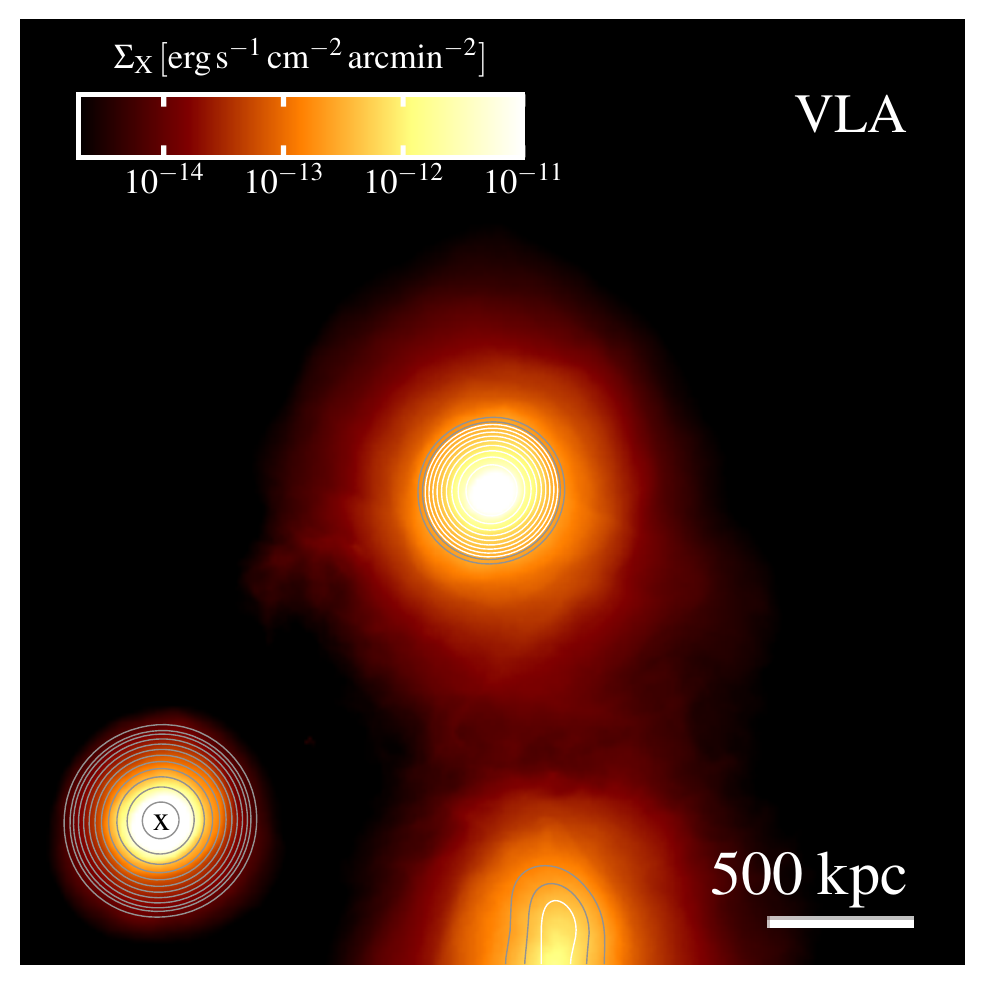}
\includegraphics[width=0.247\textwidth]{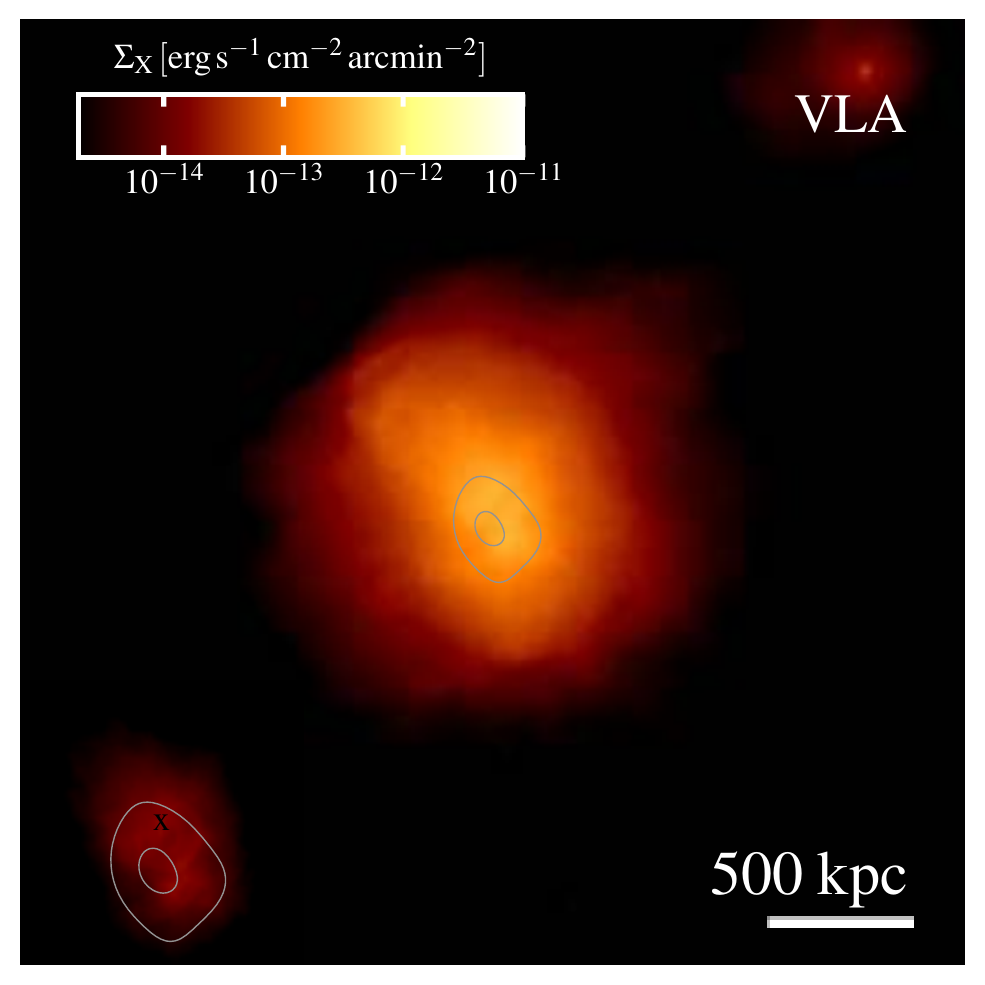}
\includegraphics[width=0.247\textwidth]{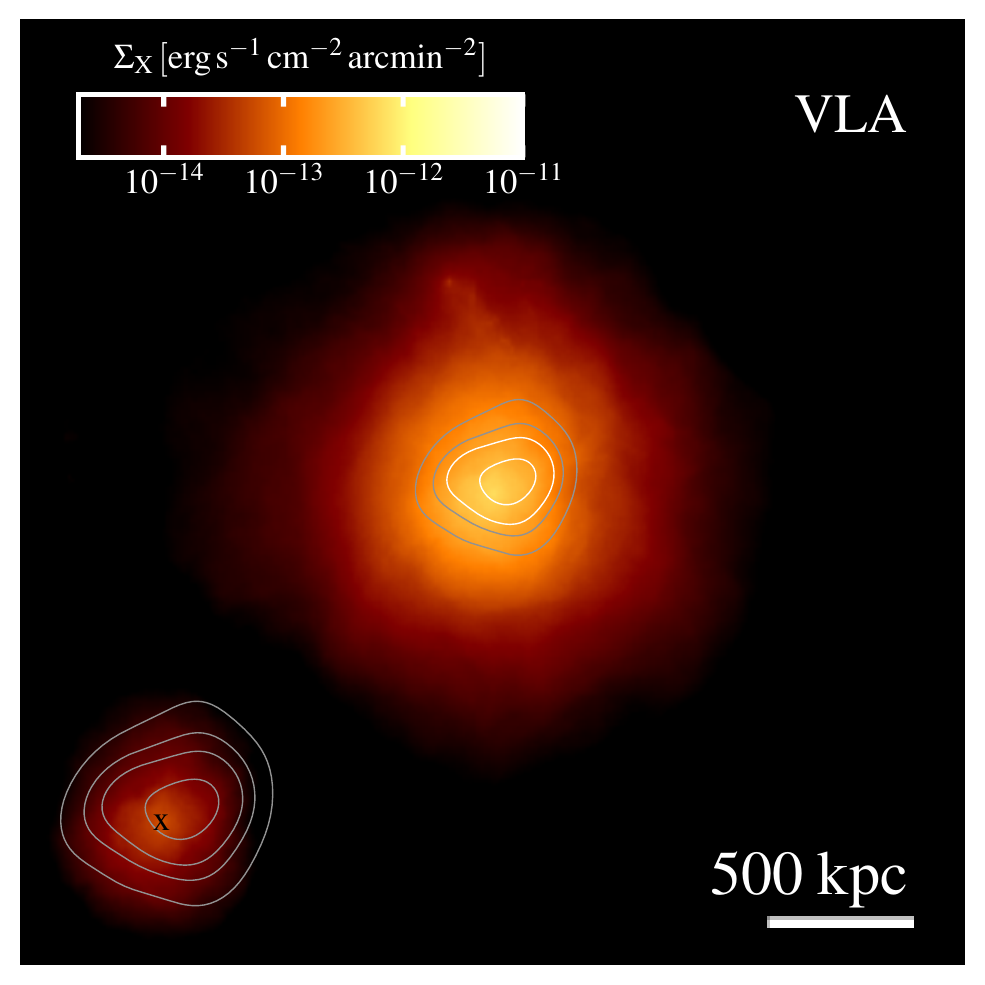}
\includegraphics[width=0.247\textwidth]{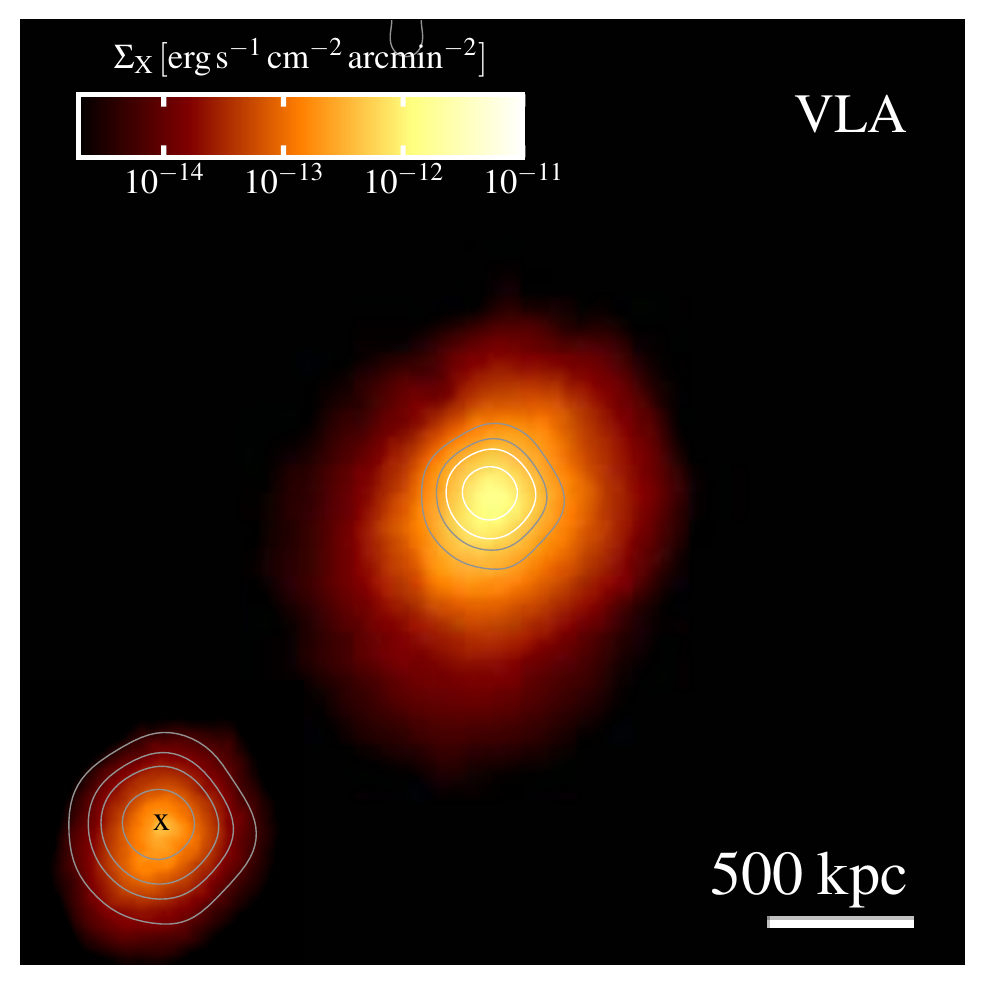}
\includegraphics[width=0.247\textwidth]{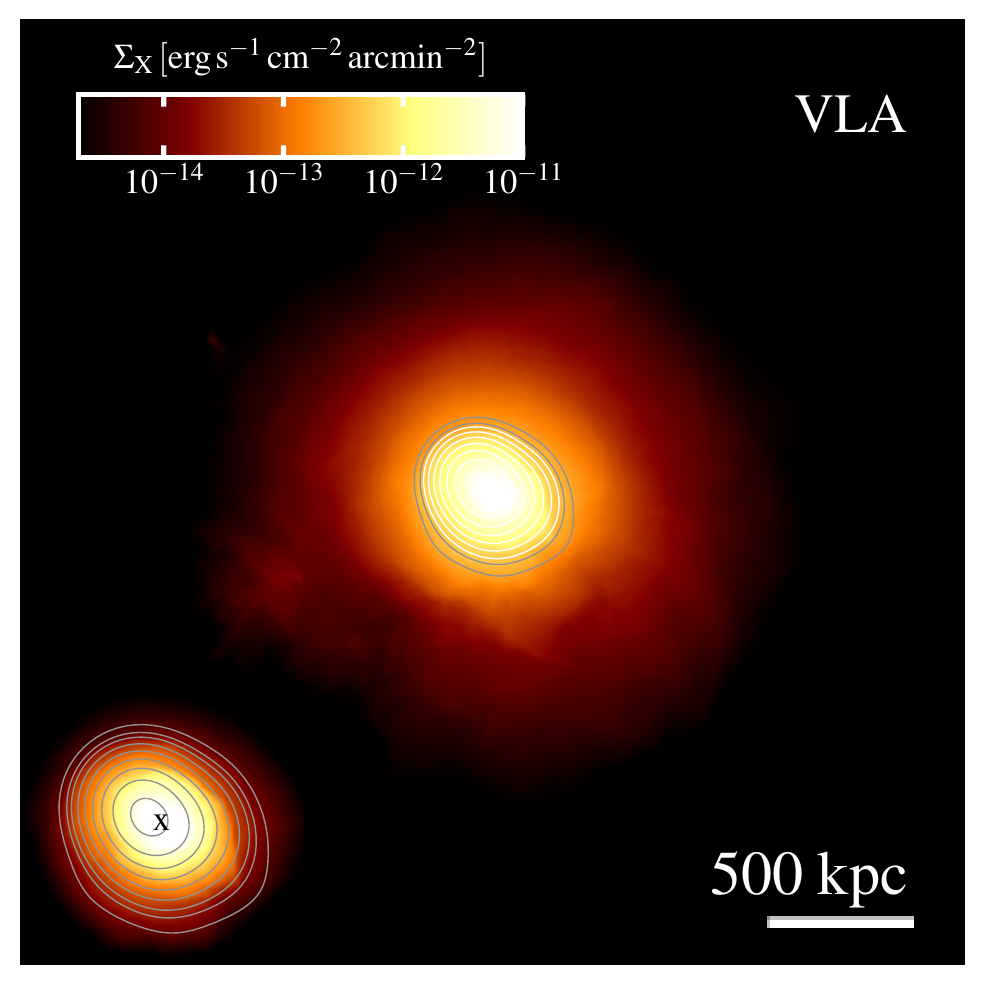}
\includegraphics[width=0.247\textwidth]{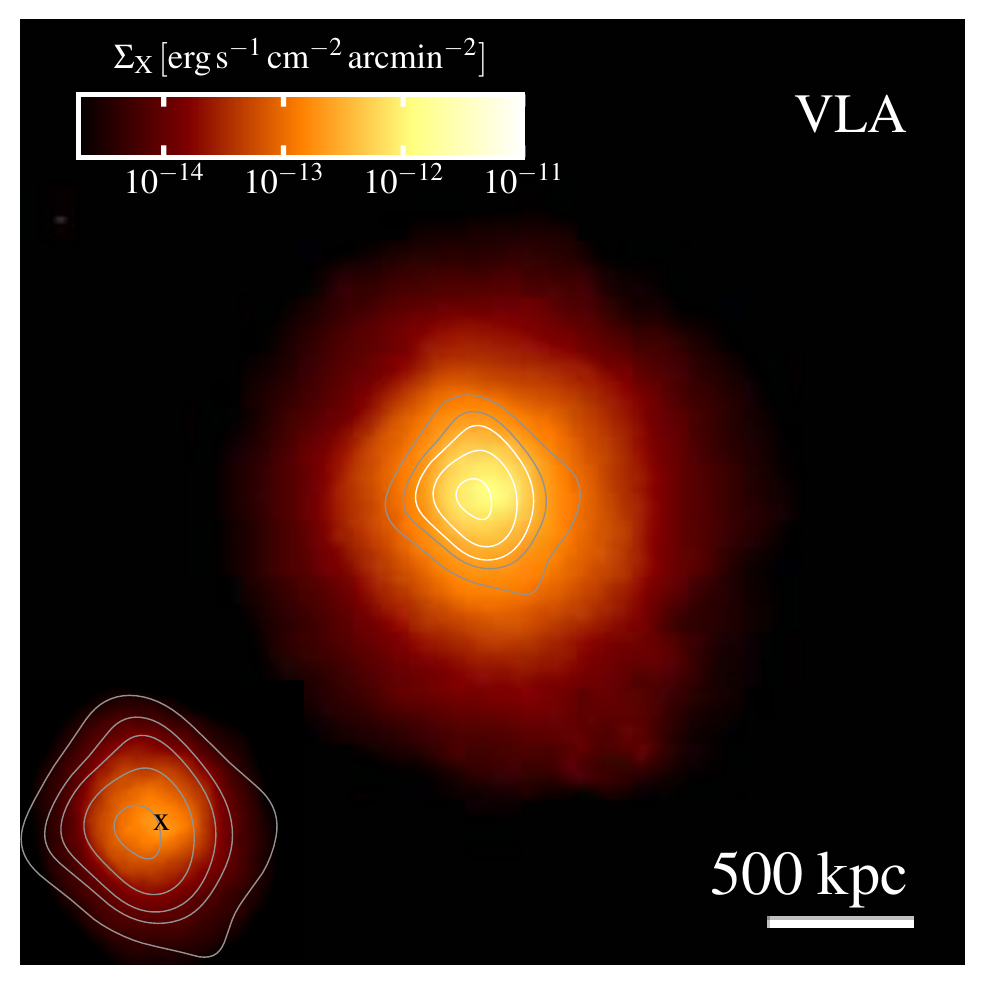}
\includegraphics[width=0.247\textwidth]{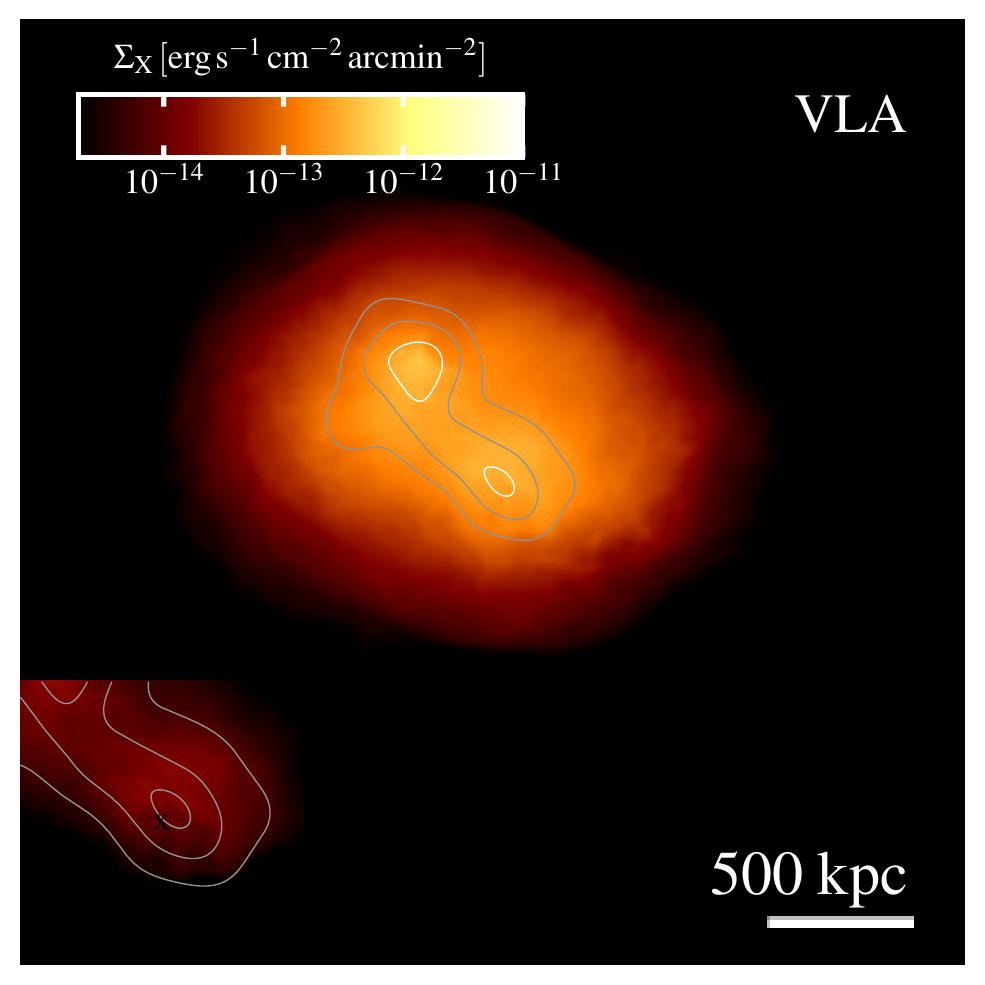}
\includegraphics[width=0.247\textwidth]{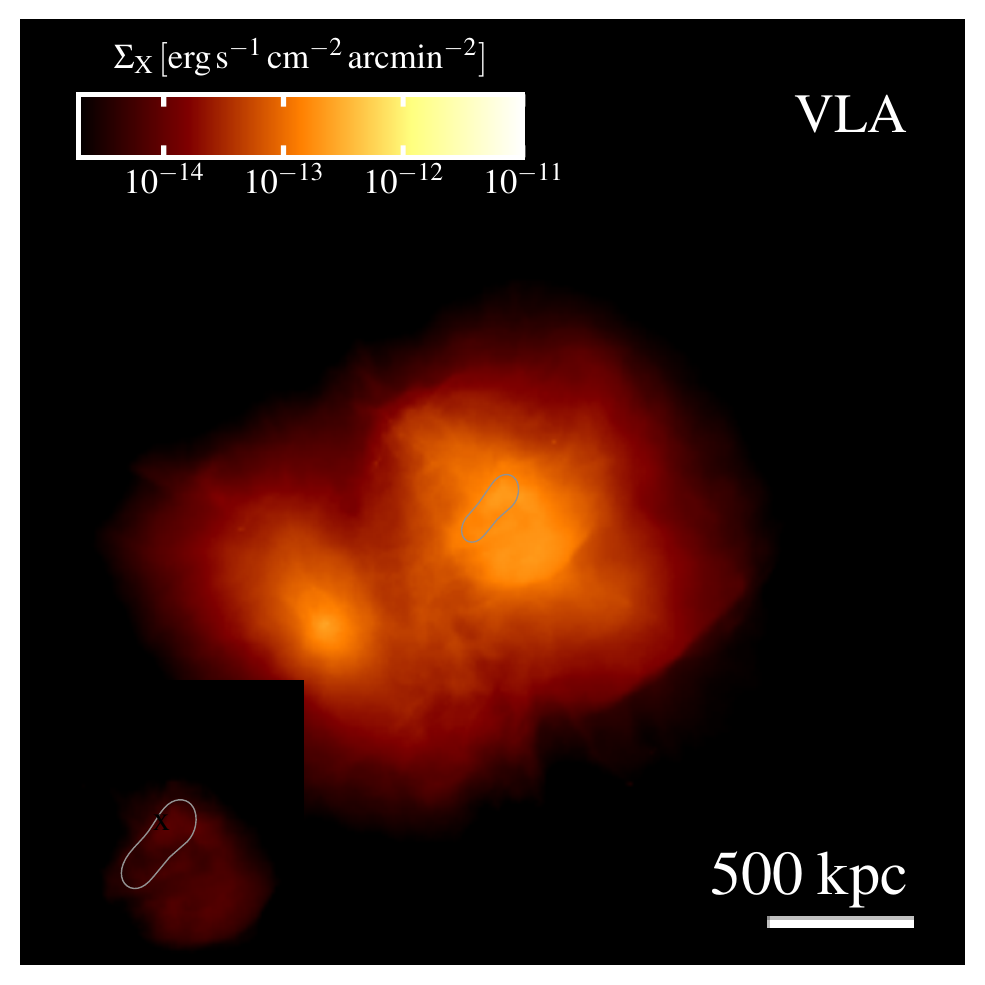}
\includegraphics[width=0.247\textwidth]{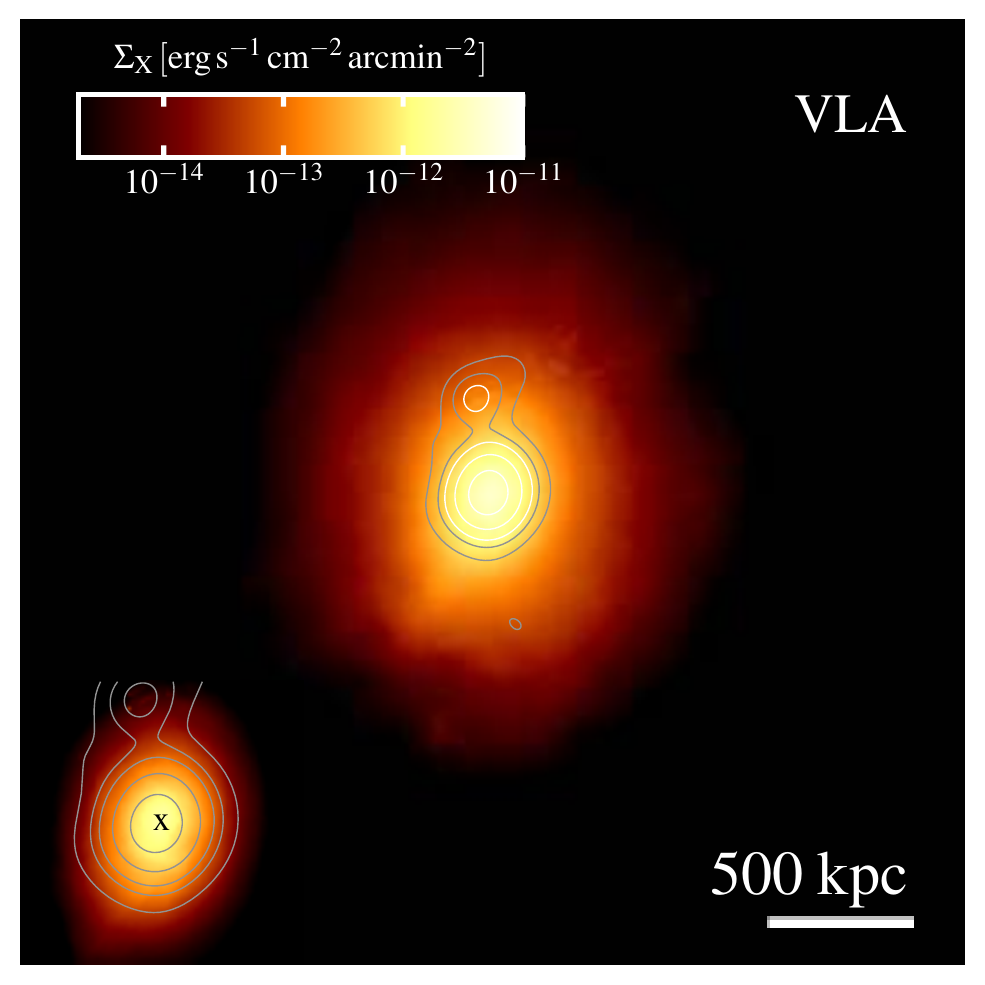}
\includegraphics[width=0.247\textwidth]{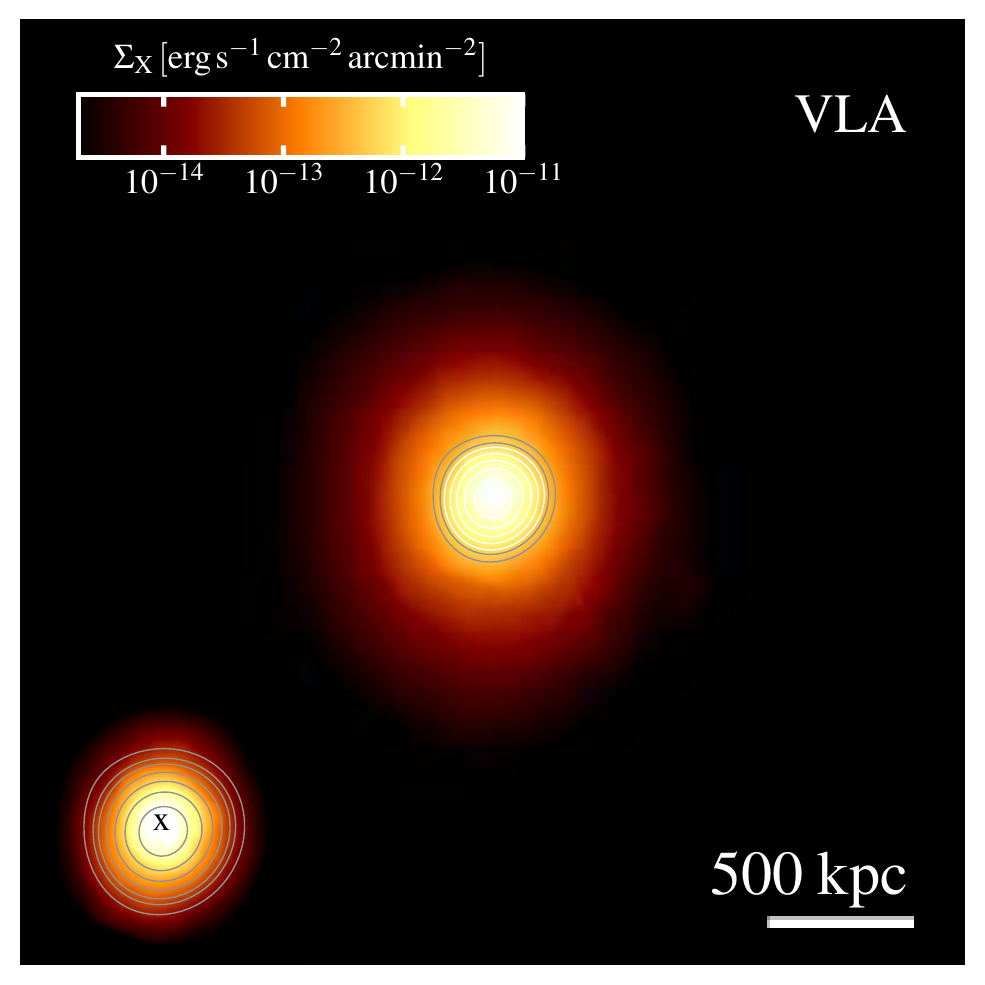}
\includegraphics[width=0.247\textwidth]{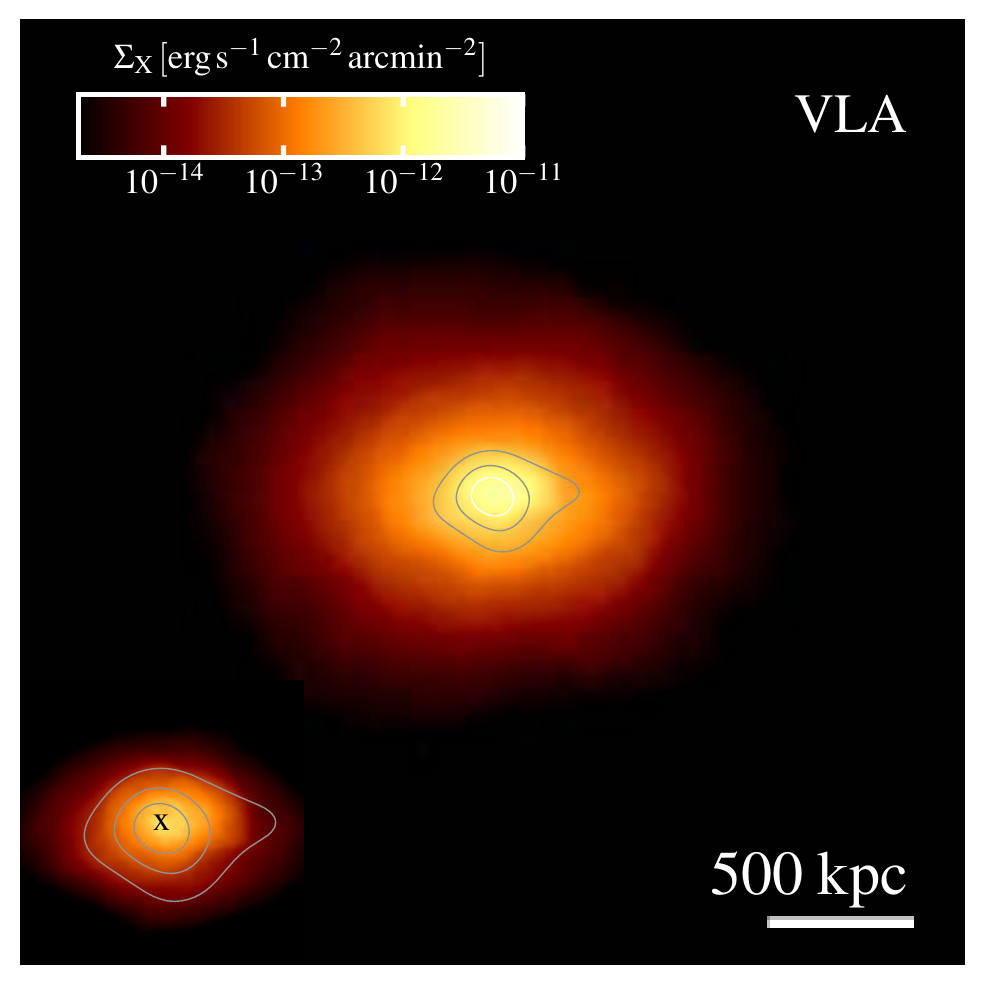}
\includegraphics[width=0.247\textwidth]{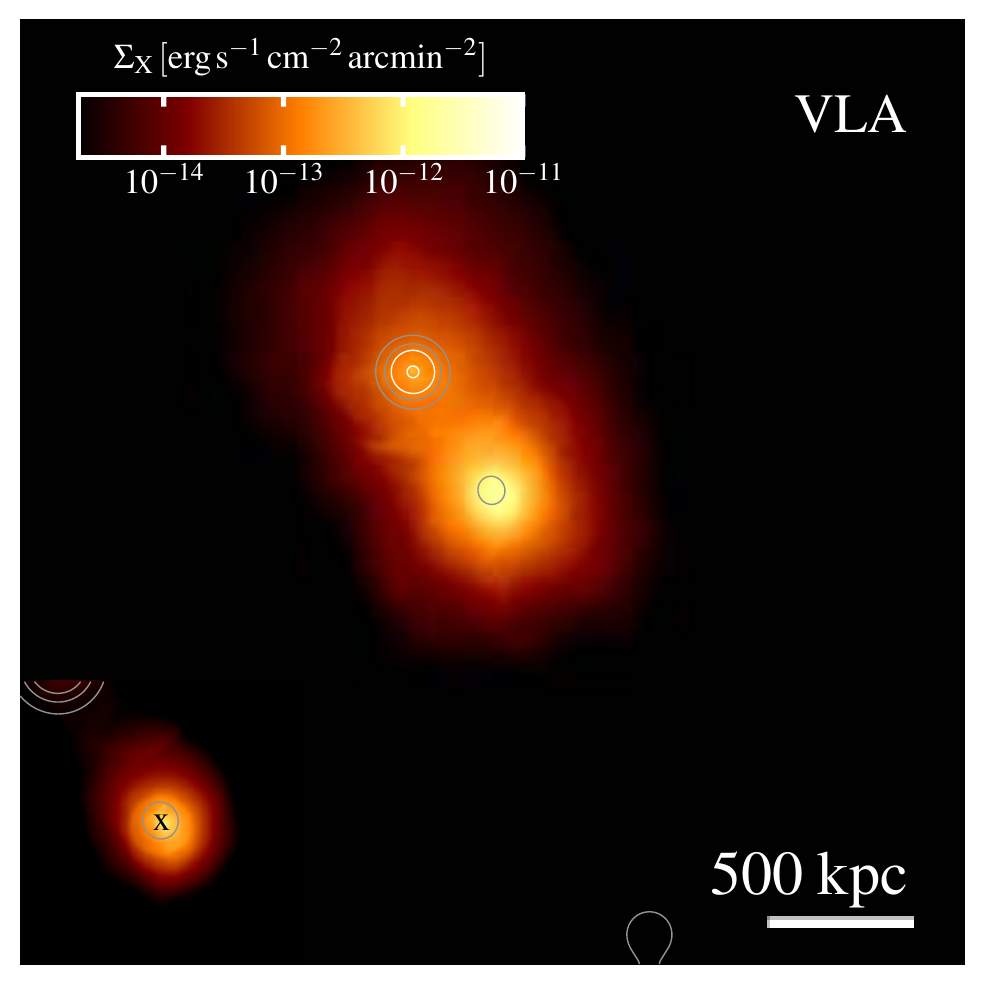}
\includegraphics[width=0.247\textwidth]{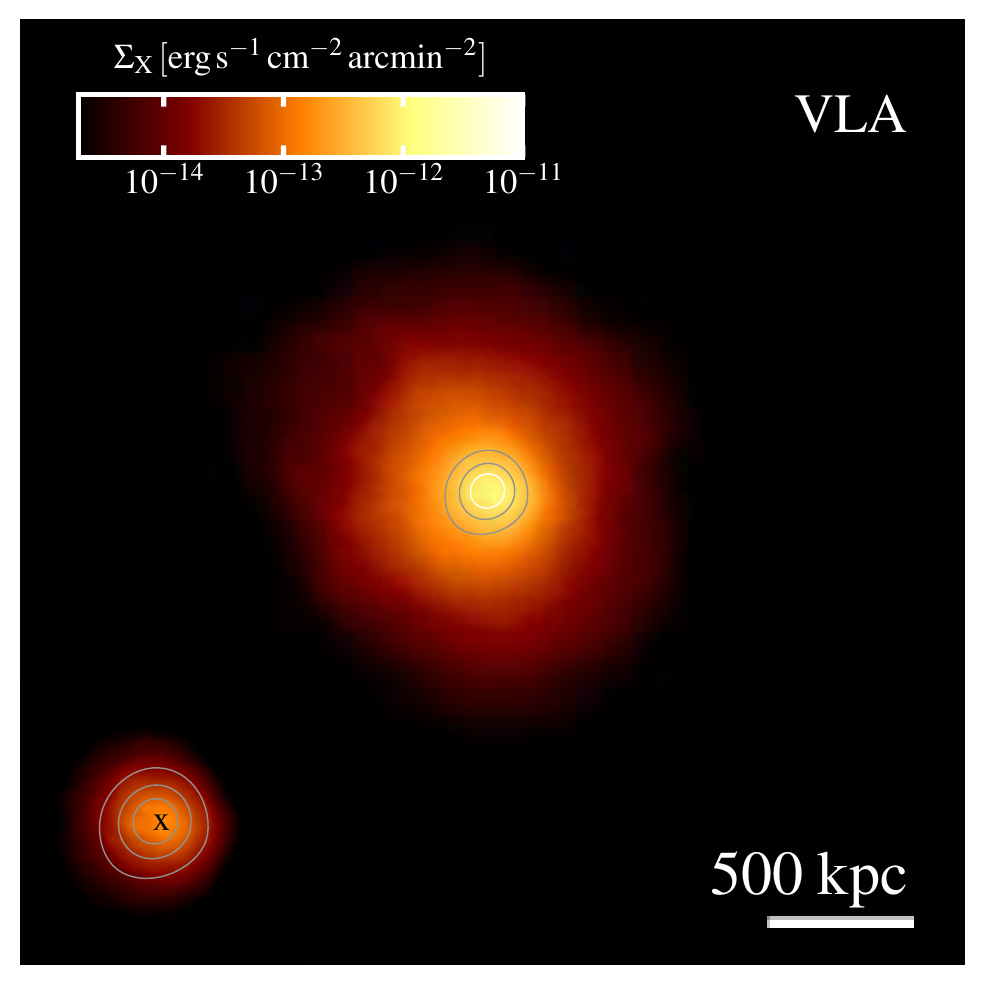}
\includegraphics[width=0.247\textwidth]{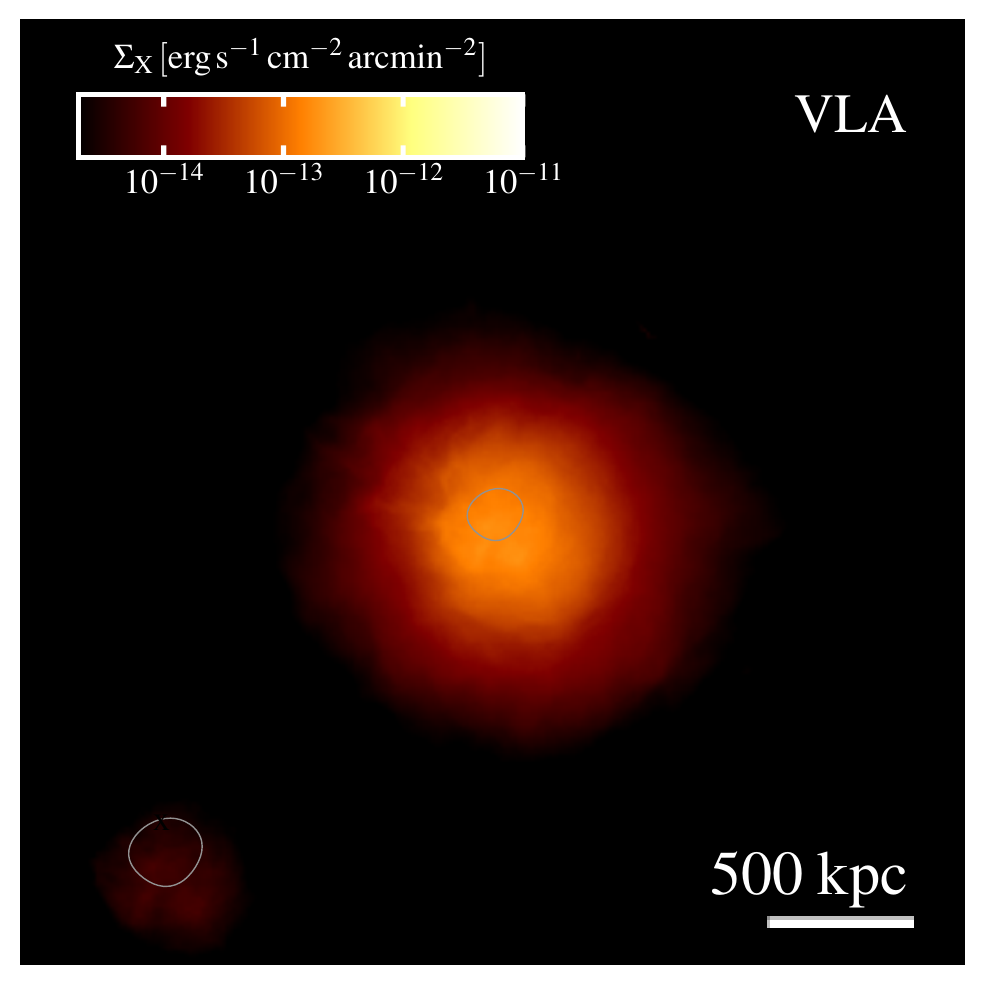}
\includegraphics[width=0.247\textwidth]{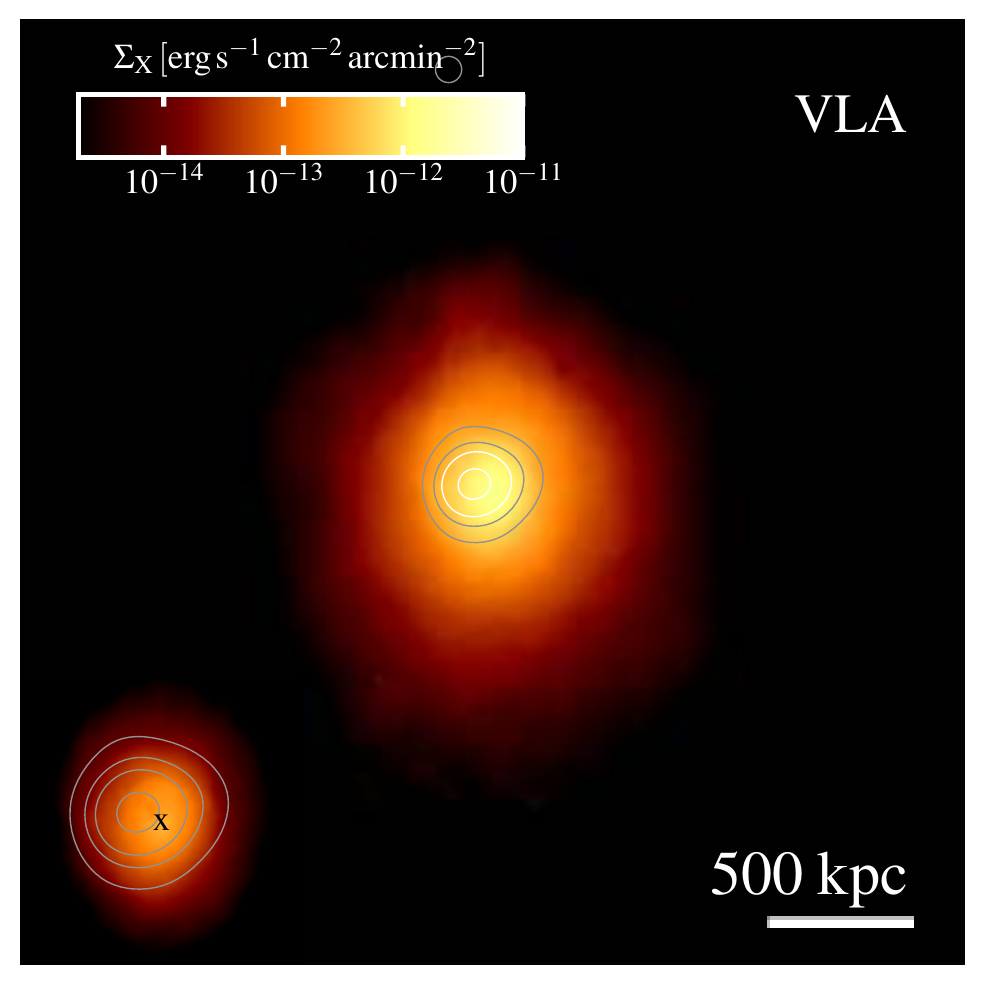}
\caption{As in Fig.~\ref{fig:XrayRadioport}, but for the 20 most massive haloes 
of the simulation TNG300. In this case, however, radio maps have been obtained 
by creating mock VLA observations of radio haloes \citep[see 
e.g.][]{Govoni2001,Giovannini2009,Vacca2011}, with the parameters listed in 
Table~\ref{tab:mapsproperties}. Note that several objects in the sample would 
not be detected as having extended emission from radio haloes.} 
\label{fig:XrayRadio}
\end{figure*}

\begin{figure*}
\centering
\includegraphics[width=0.495\textwidth]{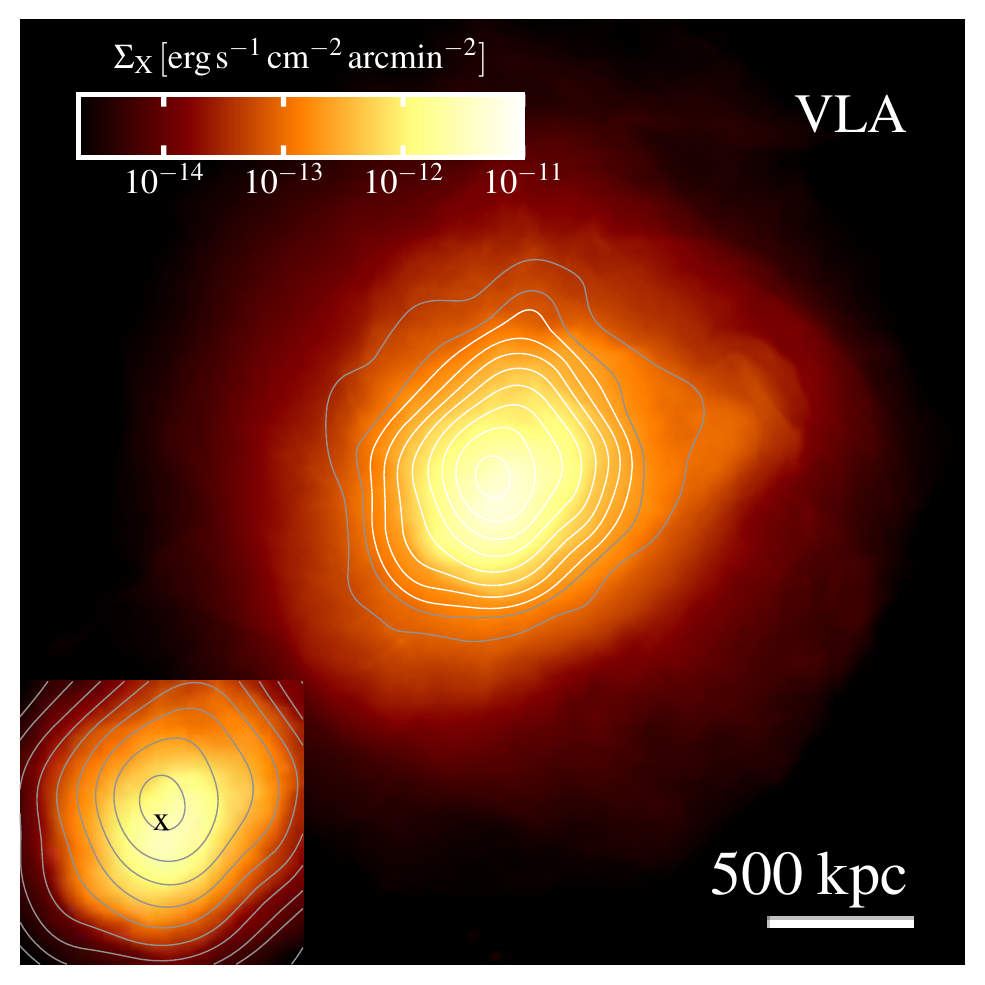}
\includegraphics[width=0.495\textwidth]{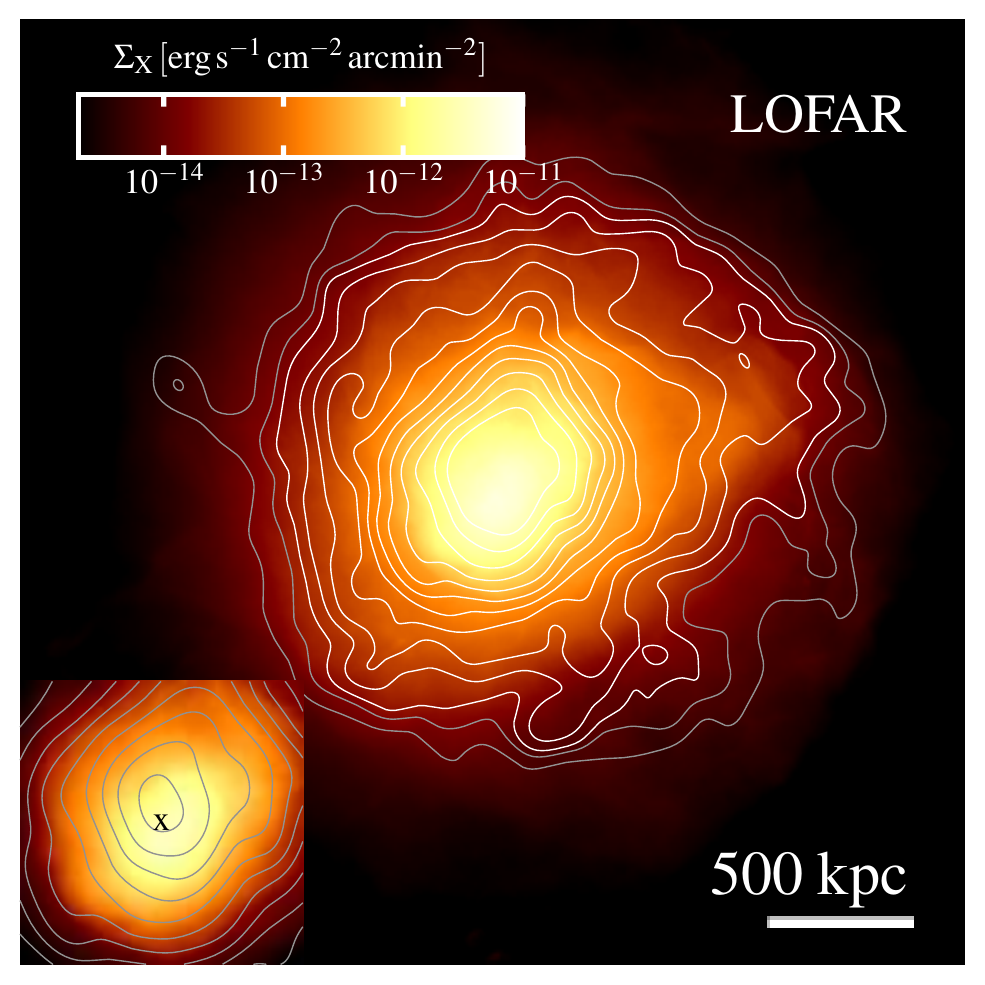}
\includegraphics[width=0.495\textwidth]{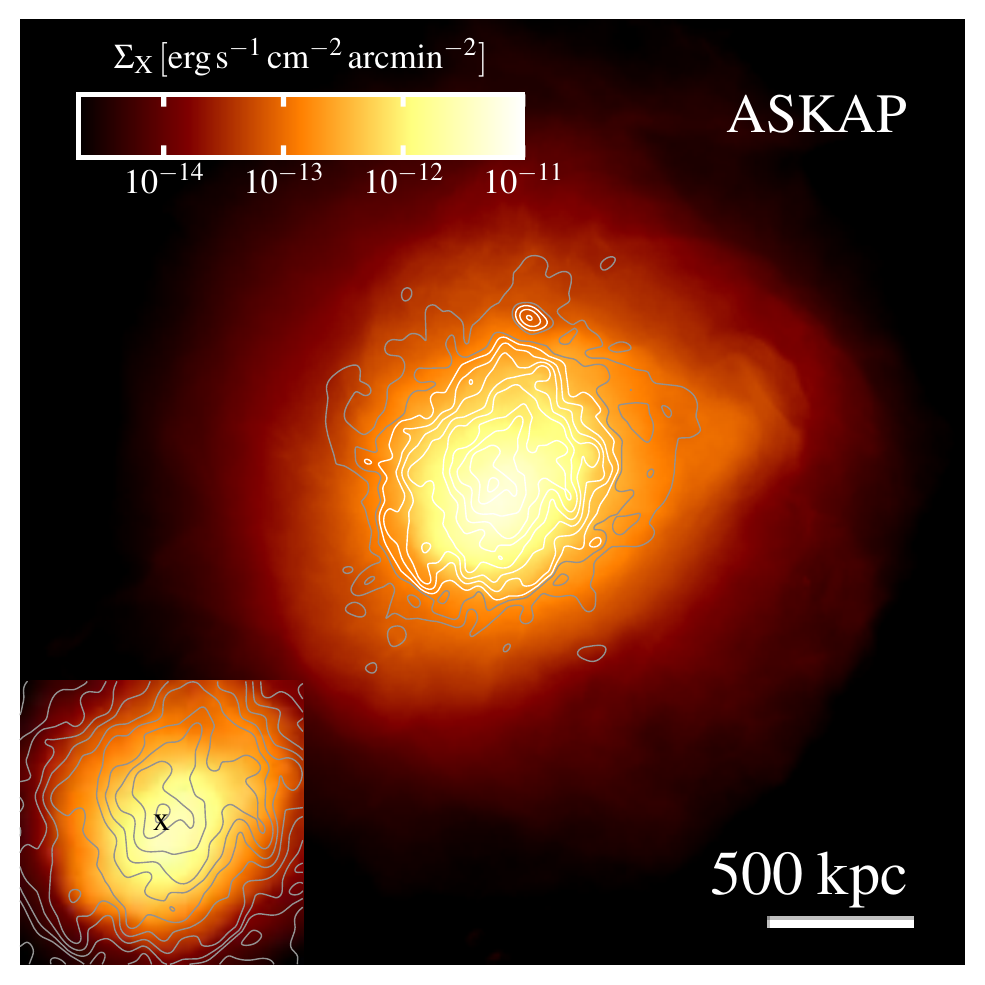}
\includegraphics[width=0.495\textwidth]{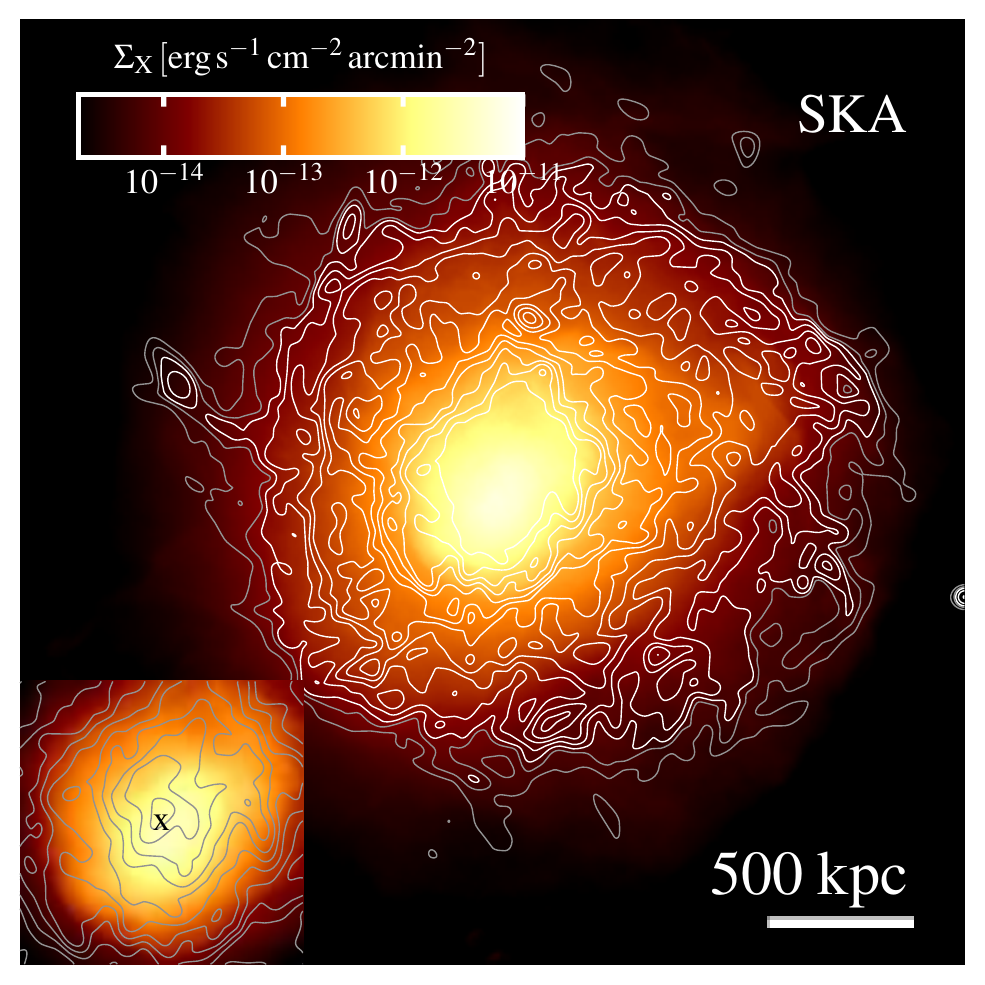}
\caption{ X-ray map (colour) overlayed with synchrotron emission contours 
\FMR{(computed by assuming model 1 -- see Section~\ref{sec:model1})} for 
the most massive halo of the TNG300 simulation. Each panel is 3.5 Mpc on a side 
and in projection depth and represents the observation of a different radio 
telescope.  From top left to bottom right these are VLA, LOFAR, ASKAP, and SKA. 
Contour levels are placed in terms of the noise level at $1\sigma$ and $2\sigma$ 
(grey) and $3\sigma$ spaced by a $\sqrt{3}$ factor from one another (white). 
More details on the radio telescope configurations can be found in 
Table~\ref{tab:mapsproperties}. All maps have been computed for a fiducial 
redshift of 0.2.} 
\label{fig:XrayRadiotel}
\end{figure*}

In Fig.~\ref{fig:XrayRadioev}, we present the evolution of the X-ray and radio 
\FMR{(model 1 in Section~\ref{sec:model1})}
significant X-ray counterpart, originating from substructures, is also present. 
Consistent with the observations the X-ray/radio offset is more pronounced where 
the radio (and the X-ray) emission is more asymmetric. The offset between the 
X-ray and radio can be interpreted as the presence of magnetic field variations 
in the ICM \FM{likely caused by shear and turbulent gas motions -- triggered by 
radiative cooling instabilities and stellar and AGN feedback, which may initiate 
a dynamo process -- that contribute to the amplification of the magnetic field 
and are not necessarily coincident with the gas density peak where the 
bremsstrahlung mechanism is more efficient.}

\begin{figure*}
\centering
\includegraphics[width=0.495\textwidth]{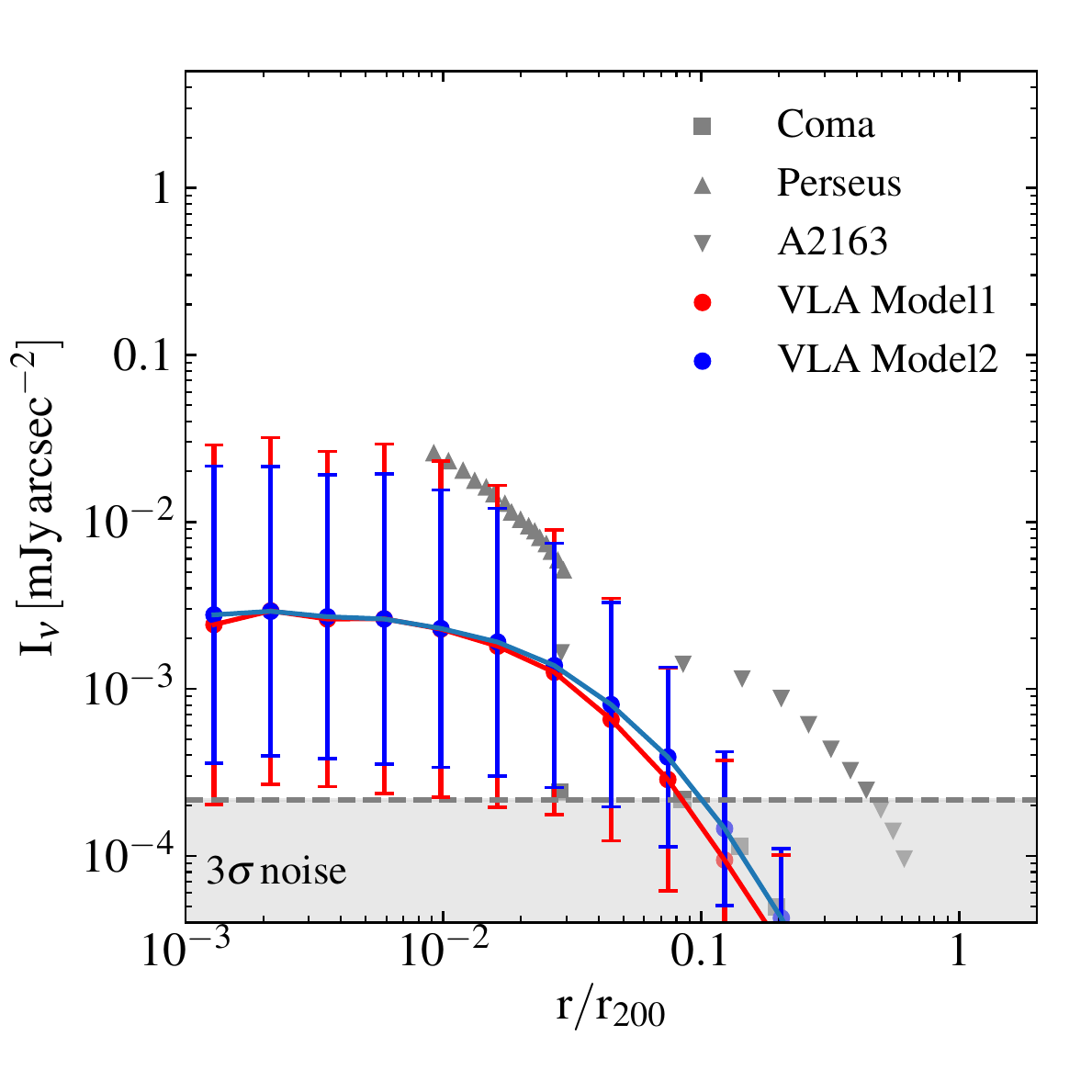}
\includegraphics[width=0.495\textwidth]{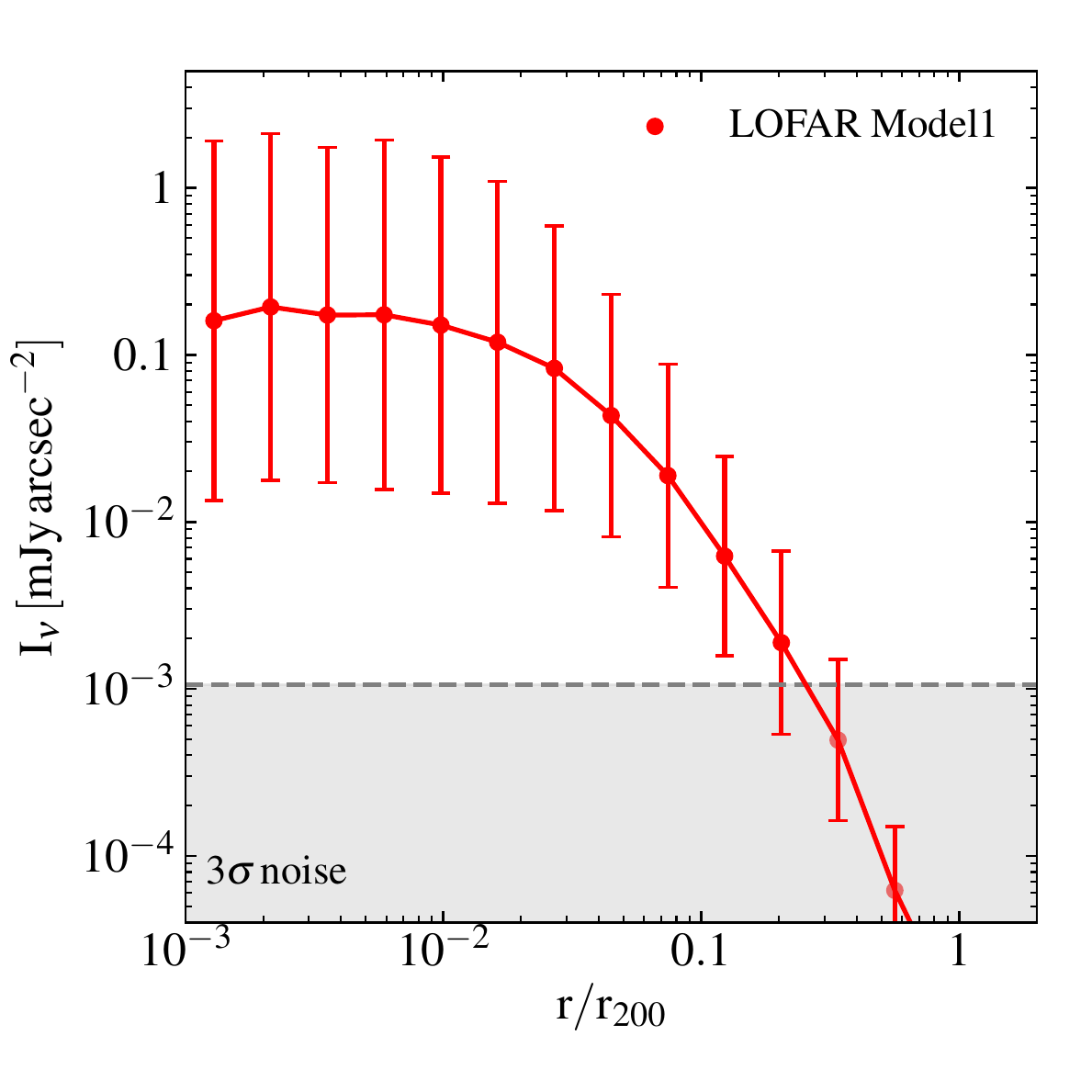}
\includegraphics[width=0.495\textwidth]{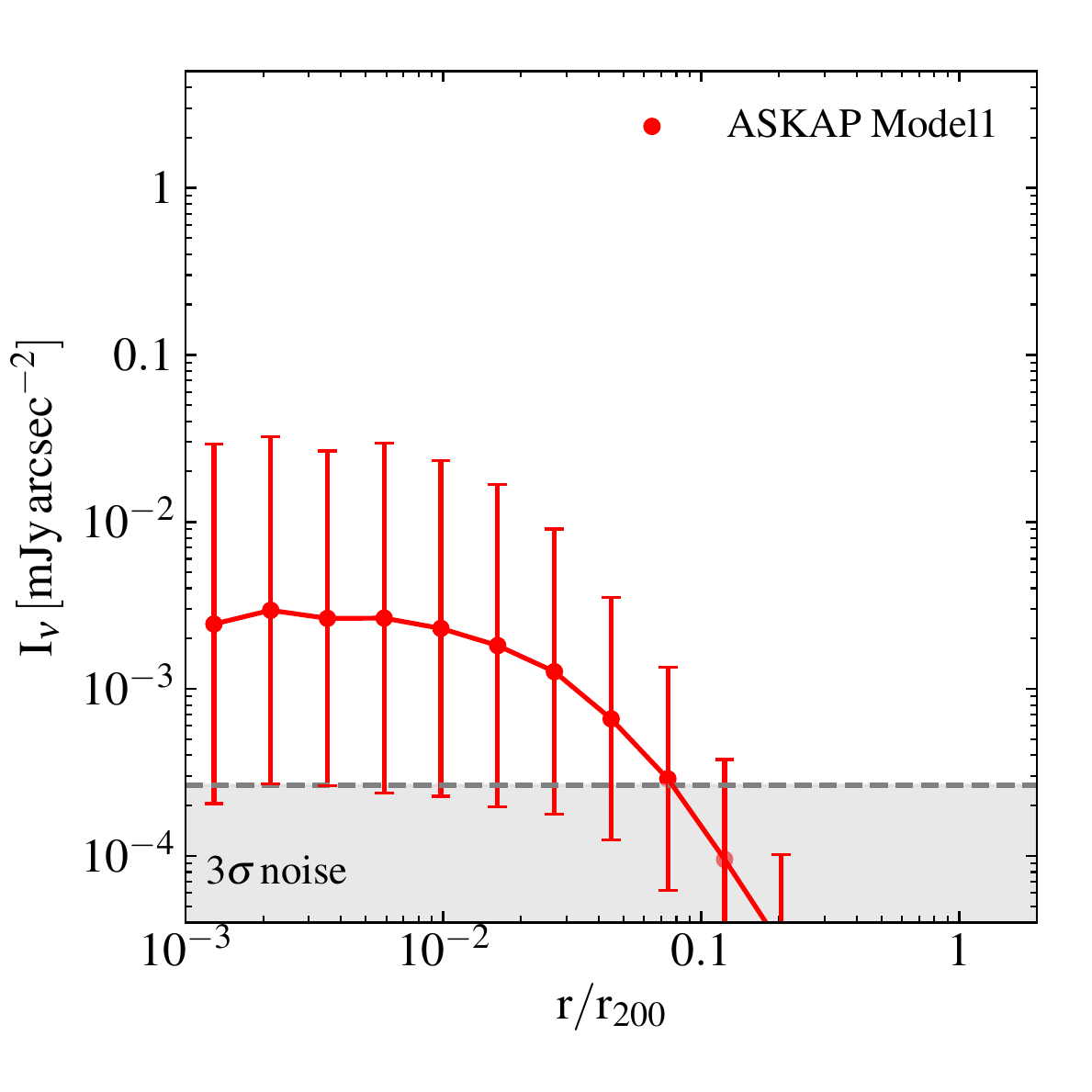}
\includegraphics[width=0.495\textwidth]{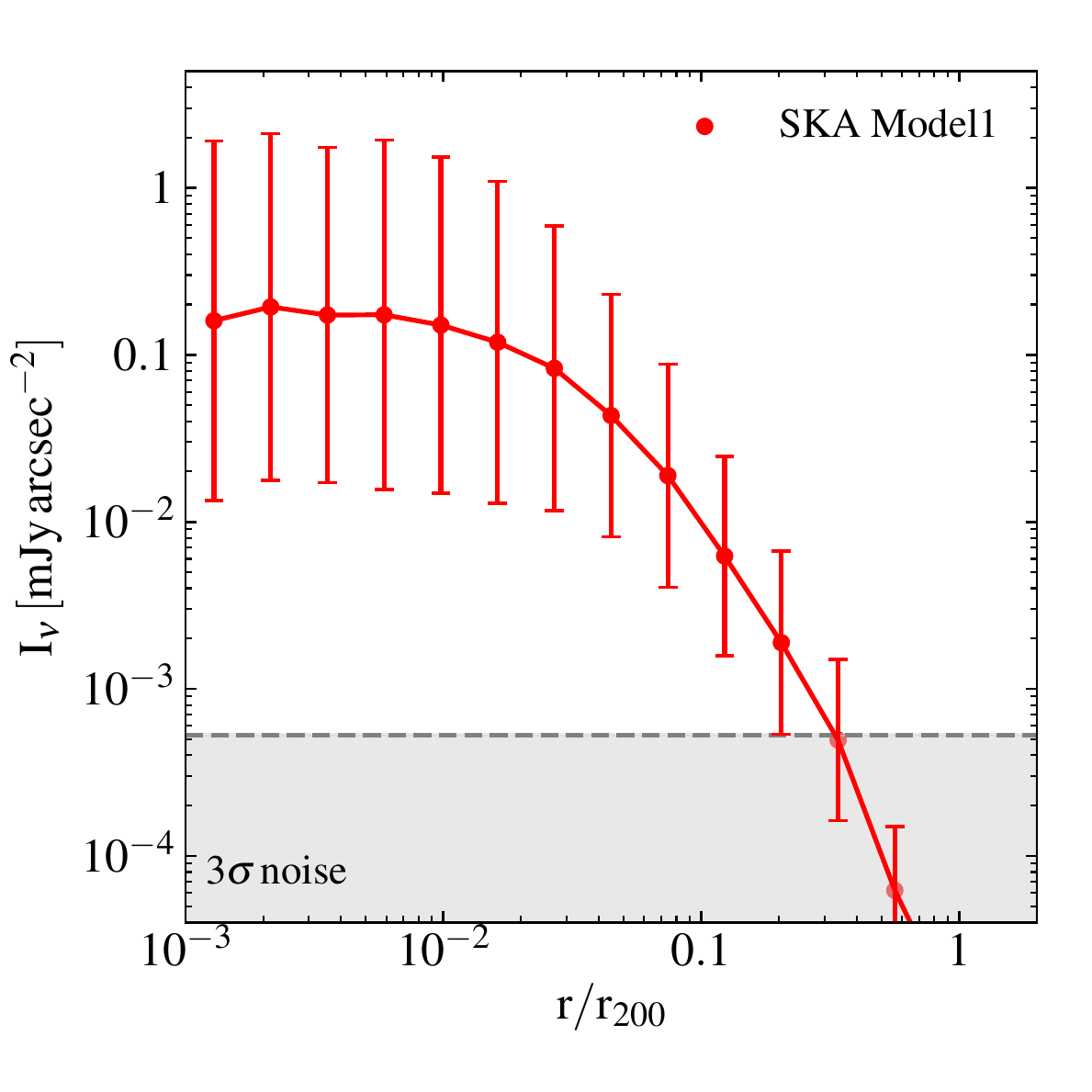}
\caption{Average radio surface brightness profiles for the radio haloes of the 
20 most massive haloes in the simulation TNG300 at redshift zero. Synchrotron 
emission has been averaged in annuli centred on the minimum potential of each 
halo and results have been binned in radial bins normalized to the virial radius 
of each object. Error bars indicate the $1\sigma$ dispersion around the mean 
value. The synchrotron emissivity has been converted into surface brightness 
units by assuming a redshift of 0.2 (and determining the luminosity and angular 
diameter distance given the cosmology) for each source.  The shaded area 
represents the $3\sigma$ noise level of typical observations of radio haloes in 
galaxy clusters with the telescopes indicated in the legend (see also 
Table~\ref{tab:mapsproperties}). Emission in the shaded region would not be 
detected. VLA mock profiles are compared to the actual observed profiles of Coma 
\citep{Deiss1997}, Perseus \citep{Pedlar1990}, and A2163 \citep{Murgia2009}, as 
indicated in the legend. \FMR{Red and blue colours in the top left-hand panel show the predictions 
of radio model 1 and 2, respectively.}}
\label{fig:radioprof}
\end{figure*}

In Fig.~\ref{fig:XrayRadio}, we show radio and X-ray maps of the 20 most massive 
haloes ($M_{200} = 3.79\times 10^{14}-1.53\times 10^{15}\,{\rm M_{\odot}}$) in 
TNG300. These radio maps were generated \FMR{by assuming model 1 (see Sec.~\ref{sec:model1}) for the distribution 
of relativistic electrons and with parameters typical of VLA 
observations. Thus,} their resolution is lower and their sensitivity smaller 
than the SKA maps shown before (see Table~\ref{tab:mapsproperties}). Also, they 
probe higher frequency synchrotron radiation (at $1.4\,{\rm GHz}$), and this, 
given the spectral form of the radiation that we have assumed ($\propto 
\nu^{-1.7}$), implies a lower radiation flux at a fixed distance. We note that 
only these 20 massive haloes are presented because, with only a handful of 
exceptions, less massive objects have non-detectable extended radio emission 
with the typical VLA parameters that we have adopted to generate the radio maps. 
Indeed, only the five most massive galaxy clusters, out of 280 with virial mass 
above $10^{14}\, {\rm M}_{\odot}$, would be detected as having extended radio 
haloes according to the criterion expressed in equation~(\ref{eq:extemission}) 
below.

\begin{figure}
\centering
\includegraphics[width=0.46\textwidth]{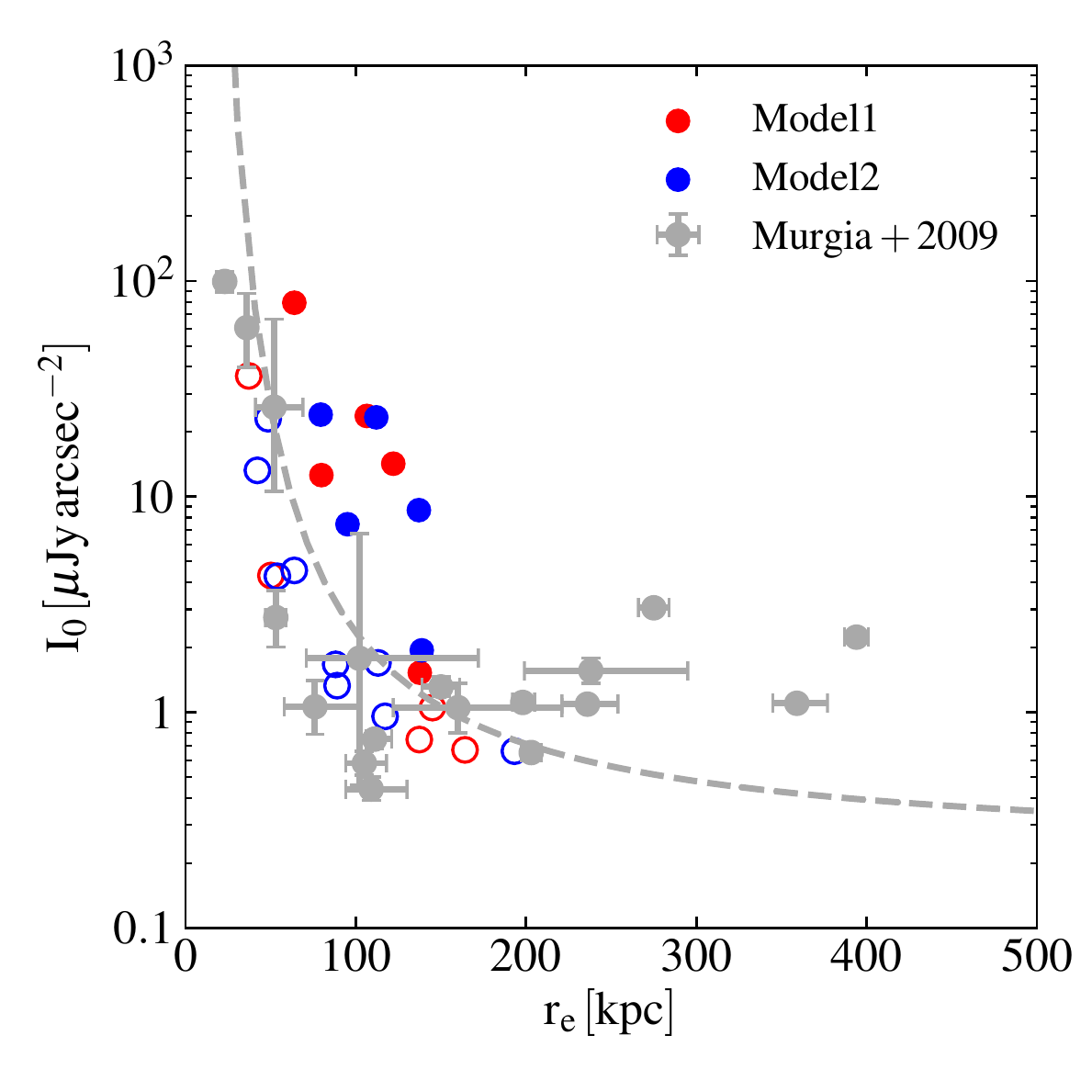}
\includegraphics[width=0.46\textwidth]{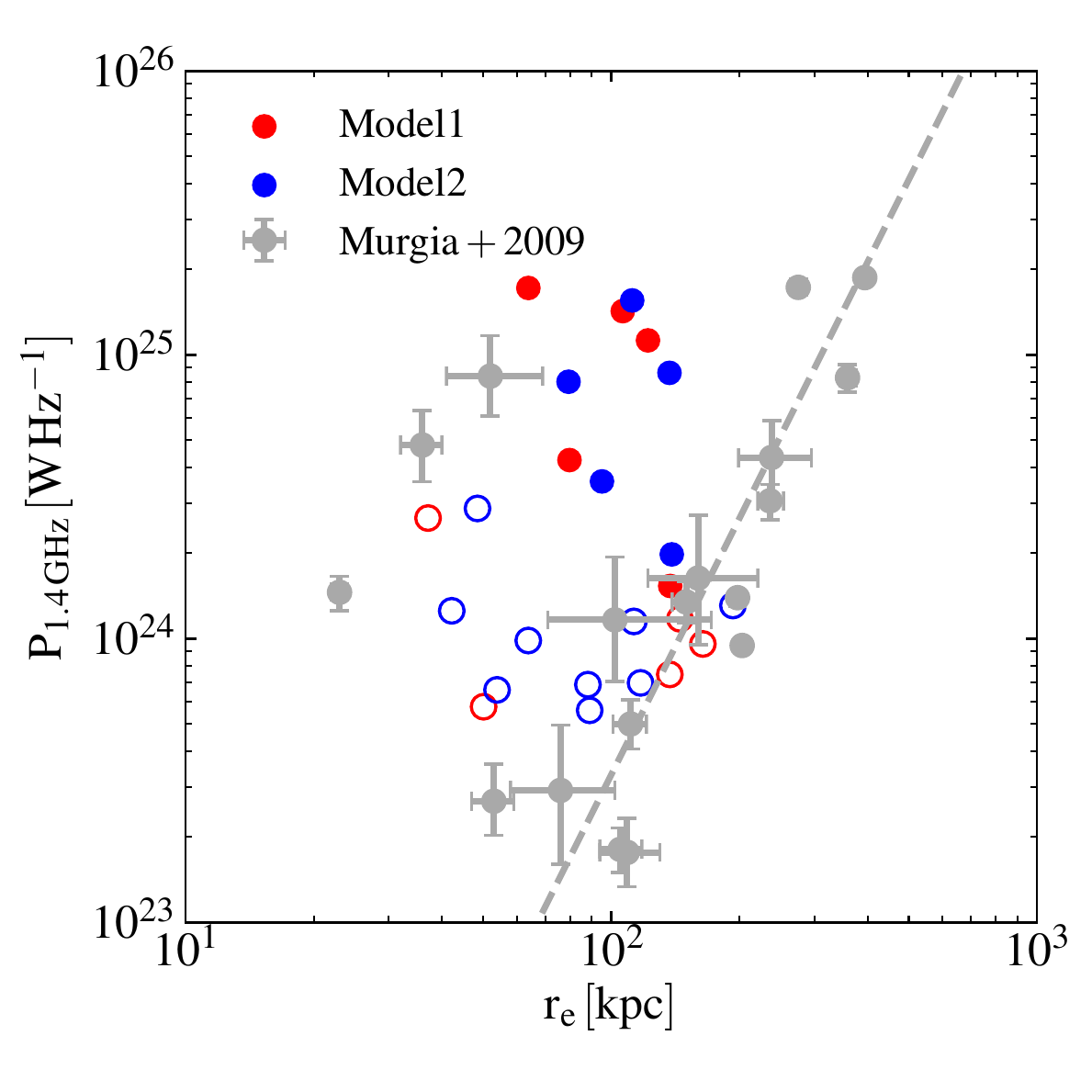}
\caption{Top panel: Exponential fit parameters to the $1.4$ GHz radio halo 
emission of the objects more massive than $10^{14}\,{\rm M_\odot}$ in the TNG300 
run. Only objects with at least three detectable points in their mock VLA radio 
surface brightness have been fitted with a profile of the form $I(r) = I_0 
\exp(-r/r_e)$. The dashed line shows the threshold for detectability of these 
sources as extended haloes at a fiducial redshift of 0.2 (see text for details). 
The simulated objects are divided in detected (filled circles) and non-detected 
(empty circles) radio haloes according to this criterion. \FMR{Red and blue colours 
refer to synchrotron model 1 and model 2, respectively (see Appendix~\ref{sec:app} for details)}. Observational 
constraints have been taken from \citet{Murgia2009}. Bottom panel: Relation 
between the radio power at 1.4 GHz and $r_e$. The dashed line represents the 
relation $\log\,P_{\rm 1.4\,GHz} = 23.52 + 3\log(r_e/100\,\kpc)$, derived for 
observed \FMR{giant radio haloes by assuming an average synchrotron emissivity 
of $10^{-42} {\rm erg\,s^{-1}\,cm^{-2}\,Hz^{-1}}$ \citep{Murgia2009}.}} 
\label{fig:Radiofits}
\end{figure}

For the most massive haloes, the majority of the analysed objects presents a 
radio emission whose morphology matches the one of the X-ray emitting gas. The 
extent of the radio emitting region, however, varies on a halo by halo basis. We 
analyse this aspect in more detail in Section~\ref{sec:scaling}. Visually, it is 
also possible to distinguish two classes of radio haloes: (i) more extended 
($\gsim 1\,\Mpc$ across) haloes with relatively low central surface brightness 
and (ii) more compact haloes ($\lsim 1\,\Mpc$ across) with a much steeper radial 
profile and large central brightness. This division is reminiscent of the 
observational classification of giant radio haloes and mini haloes \citep[][see 
also Fig.~\ref{fig:Radiofits}]{Bravi2016, Cassano2008, Feretti1996, 
Giacintucci2017}. In several cases, the radio emission has a very low surface 
brightness, which would be extremely difficult to detect observationally, or it 
is absent altogether. 

To compare the performances of different, current, and forthcoming, radio 
instruments, we present in Fig.~\ref{fig:XrayRadiotel} X-ray and radio maps of 
the most massive halo in the TNG300 simulation for different instruments. Radio 
maps have been computed \FMR{by assuming model 1 (see Section~\ref{sec:model1})} and 
for four different instruments, as indicated in the top 
right-hand corner of each panel, with sensitivity, resolution, and observing frequency 
given in Table~\ref{tab:mapsproperties}. The panels are arranged in such a way 
that columns display instruments observing at the same frequency, but with 
increasing resolution moving from top to bottom, and rows instruments with 
similar spatial resolution. 

By looking at the columns it is readily apparent that the increasing resolution 
level brings out finer details in the radio maps. For the left-hand column, at an 
observing frequency of $1.4\,{\rm GHz}$, VLA and ASKAP have similar 
sensitivities so the extent of the emitting radio regions remains approximately 
the same. For the right-hand column, at a much lower frequency of $120\,{\rm MHz}$, 
the superior sensitivity of SKA allows to map radio emission at larger 
distances. Comparing across rows, i.e. changing the observation frequency, 
results in an increase in size of the radio emitting region. Since the telescope 
sensitivity is similar, the large emitting region is explained by the fact that 
the emitted radio flux is larger at lower frequency, because of the spectral 
shape assumed in the map creation. Due to the similar beam size of the 
instruments, the level of fine details in the radio maps across rows is 
comparable. 

In Fig.~\ref{fig:radioprof} we present average radio surface brightness profiles 
for the 20 most massive haloes of the simulation TNG300 at redshift zero. 
Profiles are computed in annuli centred on the potential minimum of each halo 
with radial bins normalized to the virial radius of each halo. Error bars 
indicate the  $1\sigma$ dispersion around the mean value. Radio brightness has 
been computed for the fiducial redshift $z=0.2$. The shaded area indicates 
values of the radio surface brightness below the 3$\sigma$ rms noise level of 
each instrument. Simulated emission in this region is considered as 
non-detected. 
\FMR{Red and blue colours in the top left-hand panel present the predictions of radio models 1 and 2,
respectively (see Appendix~\ref{sec:app} for details).}

Though there are comparatively large deviations from the mean trend, as shown by 
the substantial dispersion around the mean values \FMR{(with model 2 showing generally a smaller
scatter, see also the discussion in Section~\ref{sec:scaling})}, the brightness of the 
synchrotron radiation generally declines with increasing radius. The shape of 
the decline is roughly exponential. A fit of the mean profiles of the form 
$A\exp(-br/r_{\rm 200})$ demonstrates that the cut-off radius of the profiles 
($1/b \simeq 0.03-0.05$) is approximately the same for all instruments. The 
central surface brightness depends on the observation frequency. Given the 
spectral shape of the synchrotron radiation, it is larger for observations at 
$120\,{\rm MHz}$ (LOFAR and SKA) by about a factor of 60 compared to instruments 
observing at $1.4\,{\rm GHz}$ (VLA and ASKAP). In the mean profile generated 
with mock VLA observations, we also report real observed profiles in galaxy 
clusters \citep{Deiss1997, Murgia2009, Pedlar1990}, as indicated in the legend. 
The values of the simulated profiles are roughly consistent with the observed 
ones. \FM{In particular, the average profile provides a good match to the Coma 
data, while being marginally consistent with the observed radio profile of 
Perseus} within the relatively large scatter of the simulated \FM{data}. \FM{The 
discrepancy between the simulated and observed profiles is severe only for the 
case of A2163, which features a more extended emission than any of the simulated 
clusters, signalling a general lack of very extended radio haloes in TNG300 (see 
also Fig.~\ref{fig:Radiofits})}. \FM{Overall}, it can be seen in the 
observations that radio haloes showing a rather high brightness are more compact 
than our mean profile. Also, observed objects with lower brightness tend to be 
more extended than our average brightness profile. \FMR{No significant differences are
found between the findings of radio models 1 and 2.}

In Fig.~\ref{fig:Radiofits}, top panel, we present the results of exponential 
fits of the form $I_0\exp(-r/r_{\rm e})$ to the surface brightness profiles of 
the 280 haloes more massive than $10^{14}\,{\rm M_\odot}$ in the simulation 
TNG300. The profiles to be fit have been generated for the fiducial redshift 
$z=0.2$ and with VLA observing frequency and resolution \FMR{for synchrotron model 1 (red)
and mode 2 (blue), respectively}. They extend radially up 
to a maximum radius of 2 Mpc and they are binned with a spacing given by half 
the radio FWHM beam size \citep{Murgia2009}. In the plot, we report the values 
for the best-fitting parameters, together with results obtained from actual 
observations of radio haloes taken from \citet{Murgia2009}. A given halo is 
present in the plot only if at least three points of the computed radio surface 
brightness profile are above the instrument sensitivity (calculated as 
$3\sigma$, where sigma is the rms noise level). In this way its brightness 
profile can be meaningfully fitted with an exponential function. This gives a 
total of 10 haloes out of the original sample of 280. The grey dashed line 
separates objects detected as radio haloes (towards large central surface 
brightness values and characteristic radii) from those that are not detected 
(lower central brightness values and characteristic radii), but for which it is 
still possible to perform the surface brightness fit. The line is computed by 
considering as a detection radio halo profiles for which $r_{\rm max} > 
2\,b_{\rm FWHM}$. $r_{\rm max}$ is defined as \begin{equation} 3\sigma = I_0 
\exp\left(-\frac{r_{\rm max}}{r_{\rm e}}\right), \end{equation} where $\sigma$ 
is the instrumental rms noise, while $2\,b_{\rm FWHM}$ is twice the beam size of 
the telescope (in this case VLA). This implies the following \textit{minimum} 
central surface brightness for detection \citep{Murgia2009} 
\begin{equation}
 I_0 > 3\sigma\exp\left(\frac{2\,b_{\rm FWHM}}{r_{\rm e}}\right).
 \label{eq:extemission}
\end{equation}
Simulated objects detected as extended radio sources according to this criterion 
are indicated with filled circles, while non-detections are represented by empty 
circles. 

\begin{figure}
\centering
\includegraphics[width=0.49\textwidth]{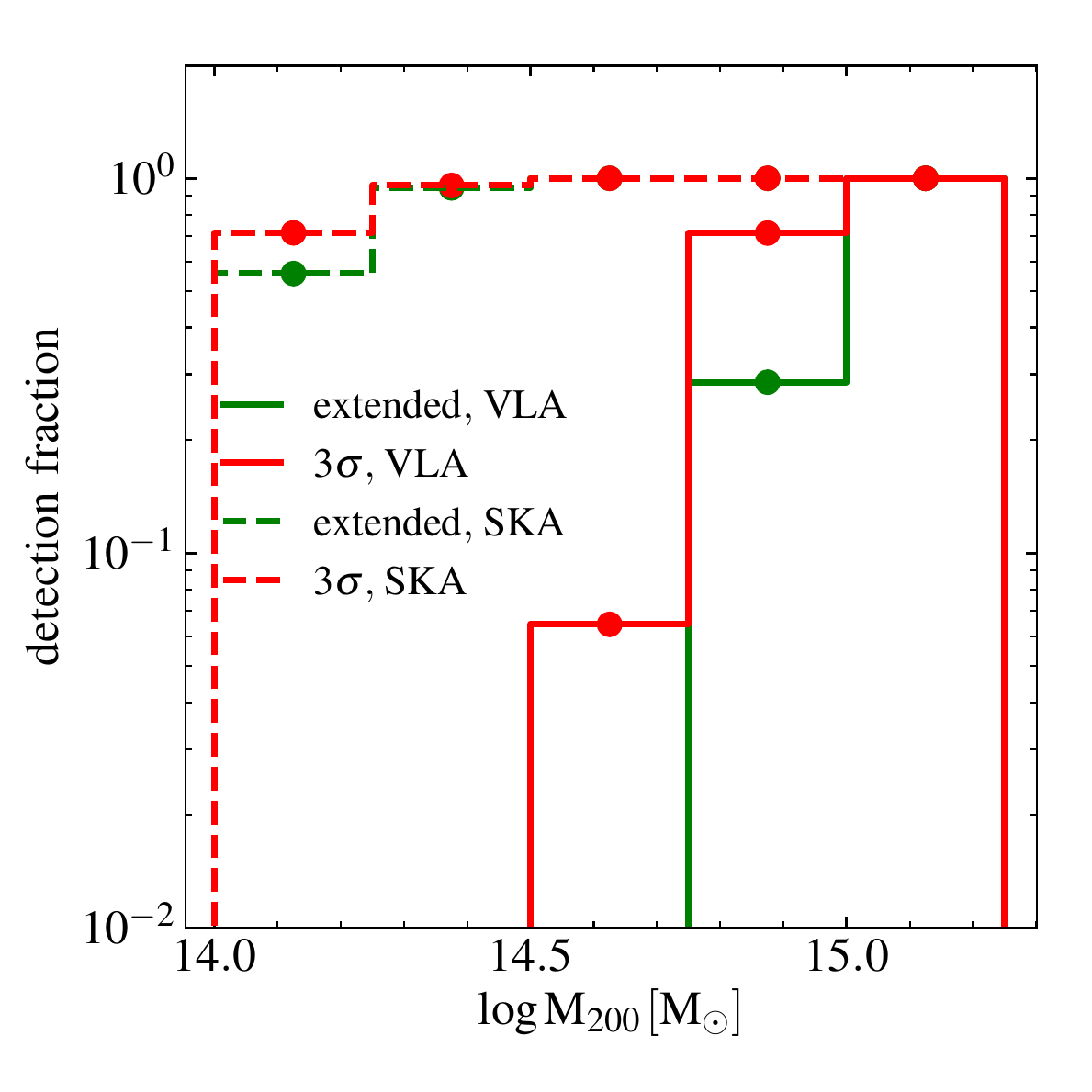}
\includegraphics[width=0.49\textwidth]{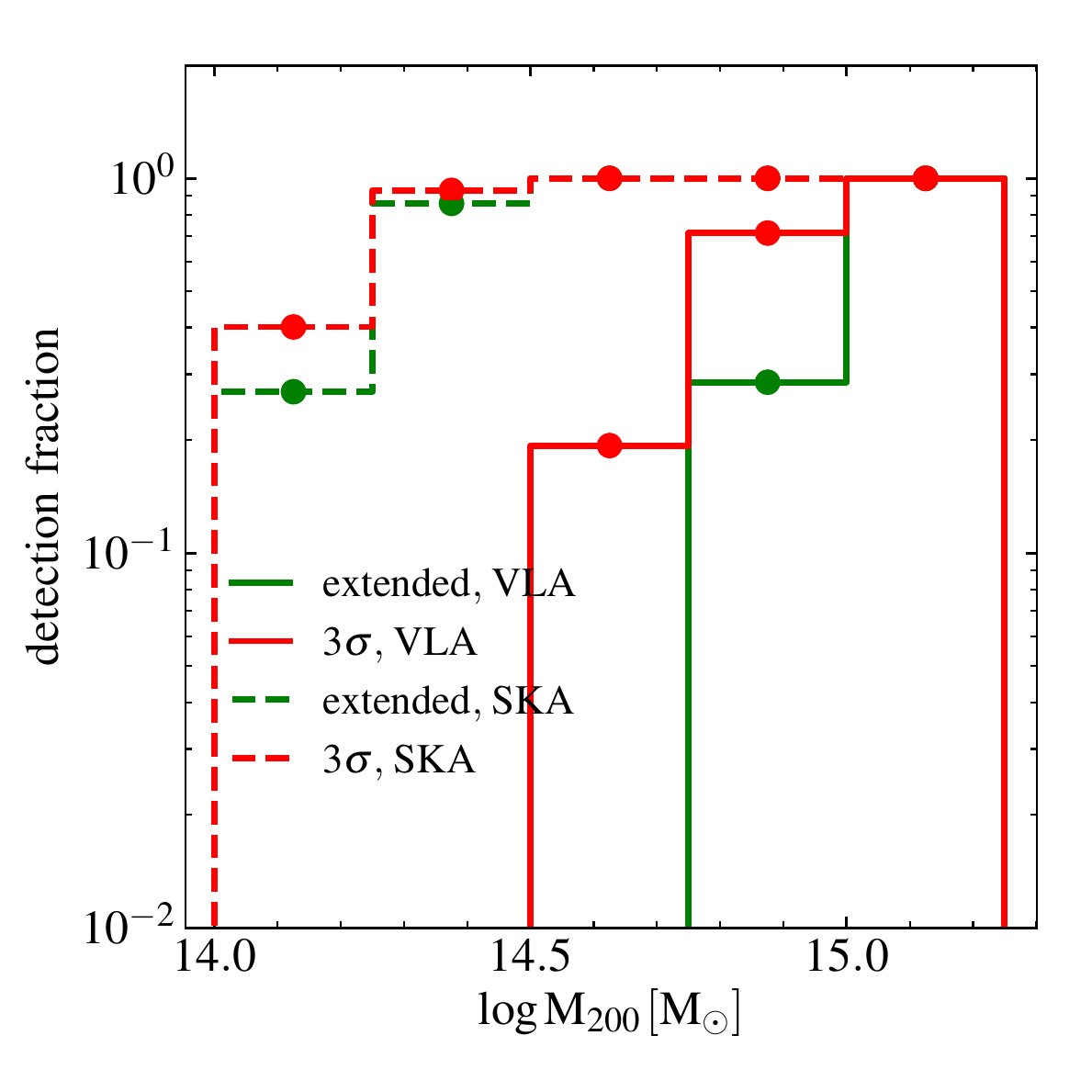}
\caption{Fraction of detected extended radio haloes as a function of virial mass 
in objects more massive than $10^{14}\,{\rm M_{\odot}}$ for the TNG300 
simulation. Solid lines show the detection fraction for mock VLA observations 
while dashed lines give the corresponding prediction for mock SKA observations. 
See Table~\ref{tab:mapsproperties} for details on mock observation properties. 
The different line colours in both panels represent the two different criteria 
for detection adopted in this work: a central surface brightness cut 
\citep[][green]{Murgia2009} and fixed threshold in surface brightness of three 
times the rms noise level of the mock observation (red; at least three points must 
be above the threshold for detection). All mock observations have been created 
at a fiducial redshift of 0.2. \FMR{Top and bottom panels show the predictions 
of synchrotron models 1 and 2, respectively (see Appendix~\ref{sec:app} for details).}} 
\label{fig:Radioprob} 
\end{figure}

Observationally, galaxy clusters hosting extended radio emission populate two 
distinct regions. Mini haloes are found at large central brightness values 
($\gsim 20\,\mu{\rm Jy\,arcsec^{-2}}$) and small spatial extent ($r_{\rm e} 
\lsim 50\,{\rm kpc}$) while giant radio haloes have relatively low central 
surface brightness ($\lsim 3\,\mu{\rm Jy\,arcsec^{-2}}$) but a much larger 
spatial extent ($r_{\rm e}$ up to $\sim 400\,{\rm kpc}$). While there is roughly 
the same division in our simulated haloes \FMR{for radio model 1, radio model 
2 features a more continuous distribution with no clear gap at a central surface
brightness of $\gsim 10\,\mu{\rm Jy\,arcsec^{-2}}$}. \FMR{Moreover, for both radio models, }the group of haloes with high central 
brightness is in general more extended ($r_{\rm e} \sim 100\,\kpc$) than 
observed mini haloes. Simulated giant radio haloes match the observed properties 
quite well. It is apparent that most of the objects hosting giant radio haloes 
would have a central surface brightness too low for detection with the typical 
VLA parameters that we have used, and more targeted and sensitive observations 
would be needed to unveil their radio emission. These more sensitive 
observations have been performed for all the observed objects present in the 
figure which are below our adopted detection limit. In general, there appears to 
be a lack of very extended radio haloes ($r_{\rm e} \gsim 150\,{\rm kpc}$) in 
our simulated sample. \FMR{The giant radio hales simulated with model 2 appear to be 
in general more centrally concentrated and less extended than the ones obtained with radio 
model 1.}

The bottom panel of Fig.~\ref{fig:Radiofits} shows the relation between the 
radio power at 1.4 GHz and the e-folding radius of the exponential fit of the 
haloes with a virial mass above $10^{14}{\, \rm M_{\odot}}$ in the TNG300 
simulation. This relation is compared to the observational results of 
\citet{Murgia2009}, who find a $P_{\rm 1.4\,GHz} \propto r_{\rm e}^3$ relation 
for giant radio haloes \FMR{(the scaling relation that they derive by assuming 
an average emissivity of $10^{-42} {\rm erg\,s^{-1}\,cm^{-2}\,Hz^{-1}}$ for 
giant radio haloes \textit{only} is shown by the grey dashed line)}. As in the 
observations, radio luminosities are computed within $3\,r_{\rm e}$. Please note 
that all the gas cells which are cooling and not star-forming within $3\,r_{\rm 
e}$ have been considered and no surface brightness cuts have been imposed. We 
have estimated the error bars in the radio luminosities of the 
\citet{Murgia2009} data set by taking the upper and lower bound in the average 
volumetric synchrotron emissivity and computing the associated luminosities 
within $3\,r_{\rm e}$. This yields luminosity errors that are larger than those 
presented in \citet{Murgia2009}, but this overestimation is not relevant for the 
analysis carried out in this work. \FMR{Our simulated results broadly reproduce 
the division in two groups between giant and mini haloes found observationally for radio model 1.
If we divide the sample of our simulated radio haloes in giant haloes and 
mini haloes by taking a cut in  central brightness at $10\,\mu{\rm Jy\,arcsec^{-2}}$ 
as suggested by a visual inspection of the top panel (giant radio haloes are below this 
cut whereas mini haloes are above), the simulation 
predictions match the observational trends for giant radio haloes. With 
this we mean that their average synchrotron emissivity is comparable with the 
one ($\simeq 10^{-42} {\rm erg\,s^{-1}\,cm^{-2}\,Hz^{-1}}$) determined 
observationally. We caution, however, that the majority of the systems would 
have not been detected for the standard VLA mock observation parameters given in 
Table~\ref{tab:mapsproperties}, and only 1 of 5 objects would have been 
classified as an extended source according to the criterion presented in 
equation~(\ref{eq:extemission})}. In the case of mini haloes the agreement is not 
as good, especially at high luminosities. Specifically, the high luminosity 
simulated objects are too extended to be part of the mini halo class, but not 
extended enough to be classified as giant radio haloes. \FMR{These discrepancies 
become more pronounced for radio model 2. This is due to the higher central surface brightness
and smaller $r_{\rm e}$ compared to the values obtained with radio model 1. However,
also in this case, the least bright simulated radio haloes have average emissivities in line with those 
observed in giant radio haloes.}

Summarizing, our simulated radio haloes broadly capture the observed dichotomy 
in surface brightness between mini and giant radio haloes. However, there are 
noticeable discrepancies with observations, which signal shortcomings of our 
rather crude \FMR{modelling} for synchrotron emission and might be alleviated by a more 
sophisticated treatment, for example in the spatial and energy distribution of 
relativistic electrons.

\subsection{Observational scaling relations}\label{sec:scaling}

In Fig.~\ref{fig:Radioprob} we show the fraction of detected radio haloes as a 
function of the virial mass of the hosting cluster for the TNG300 simulation. We 
show this quantity for two radio telescopes observing at different frequencies: 
VLA (solid lines) and SKA (dashed lines). To estimate the detection fraction we 
grouped all haloes with virial mass above $10^{14}\,{\rm M}_{\odot}$ in $0.25$ 
dex width mass bins, and computed the fraction of haloes showing detectable 
radio emission. We used two criteria for detection. The first one, shown with 
green lines, uses equation (\ref{eq:extemission}). The second criterion is less 
restrictive and considers as detected objects with at least three points in the 
radial brightness profile above the sensitivity limit ($3\sigma$ rms noise 
level, red lines). Radio profiles used to create the figure have been calculated 
at the fiducial redshift $z=0.2$. \FMR{Top and bottom panels show the results for synchrotron
models 1 and 2, respectively.}

\begin{figure}
\centering
\includegraphics[width=0.453\textwidth]{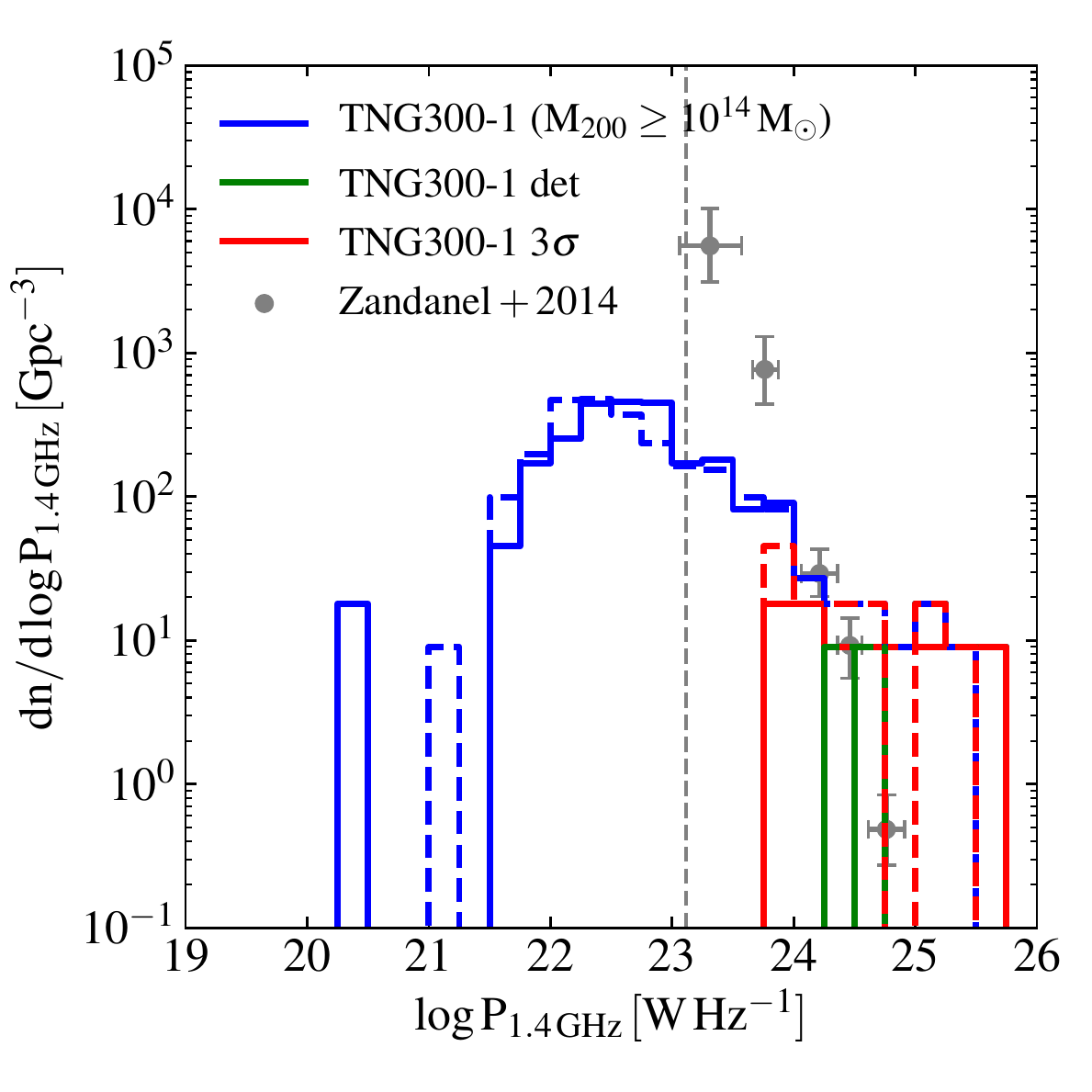}
\includegraphics[width=0.453\textwidth]{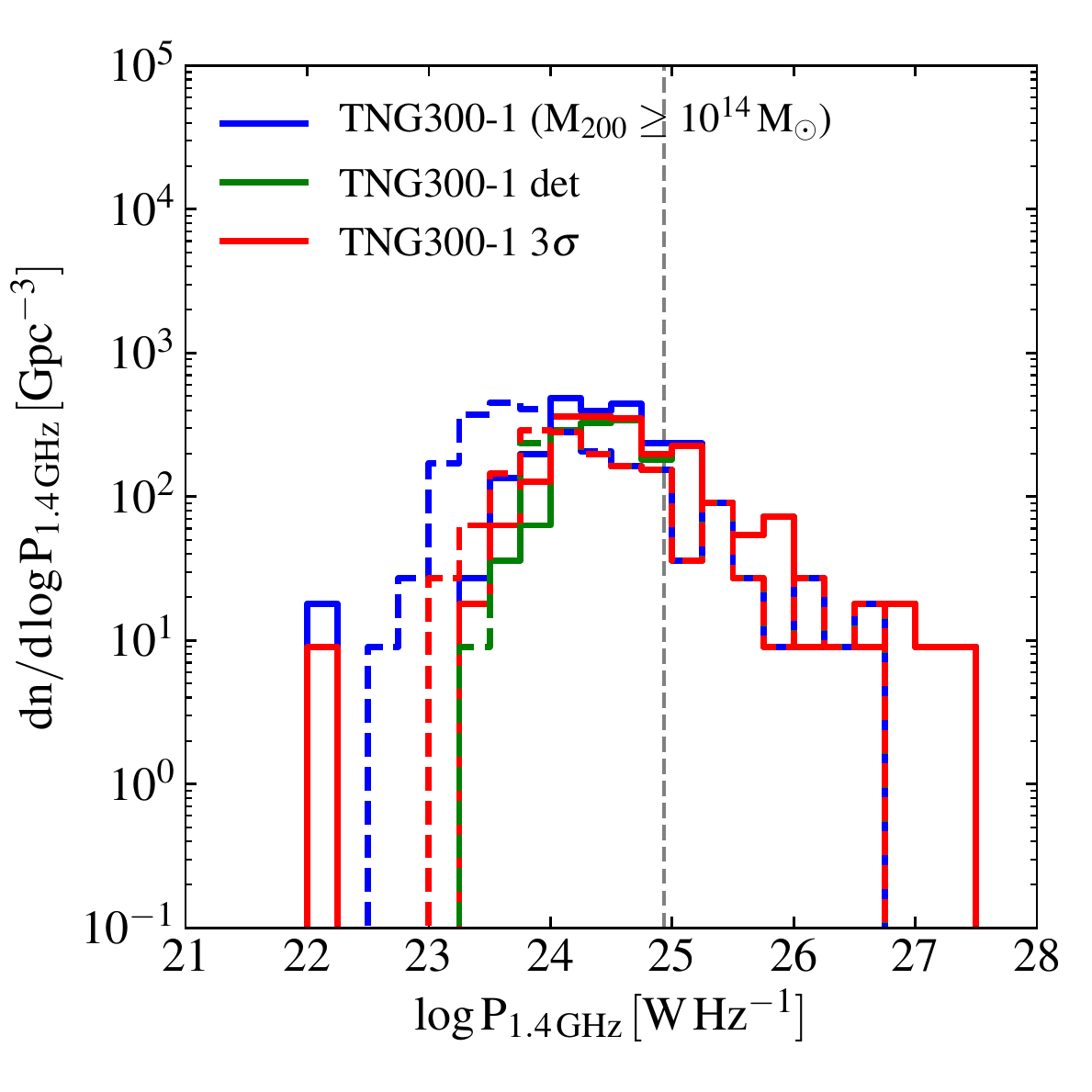}
\caption{Source count per unit volume as a function of the total synchrotron 
power in the TNG300 simulation for VLA (top) and SKA (bottom) mock observations. 
Only objects with a virial mass larger than $10^{14}\,{\rm M}_{\odot}$ have been 
considered. The plots show the resulting source count taking into account all 
sources above the virial mass cut (blue) or only detectable source according to 
two different criteria applied to their surface brightness profiles: a central 
surface brightness cut \citep[][green]{Murgia2009} and fixed threshold in 
surface brightness of three times the rms noise level of the mock observation 
(red; at least three points must be above the threshold for detection). For both 
panels a fiducial redshift of 0.2 has been used \FMR{Solid and dashed lines 
show the results for synchrotron models 1 and 2, respectively (more details in 
Appendix~\ref{sec:app})}. The dashed vertical lines 
indicate a completeness level of 95 per cent given our virial mass cut. VLA mock 
observations are compared to the NVSS data set obtained in \citet[][see also 
\citealt{Giovannini1999}]{Zandanel2014}.} 
\label{fig:Radiohist} 
\end{figure}

\begin{figure}
\centering
\includegraphics[width=0.48\textwidth]{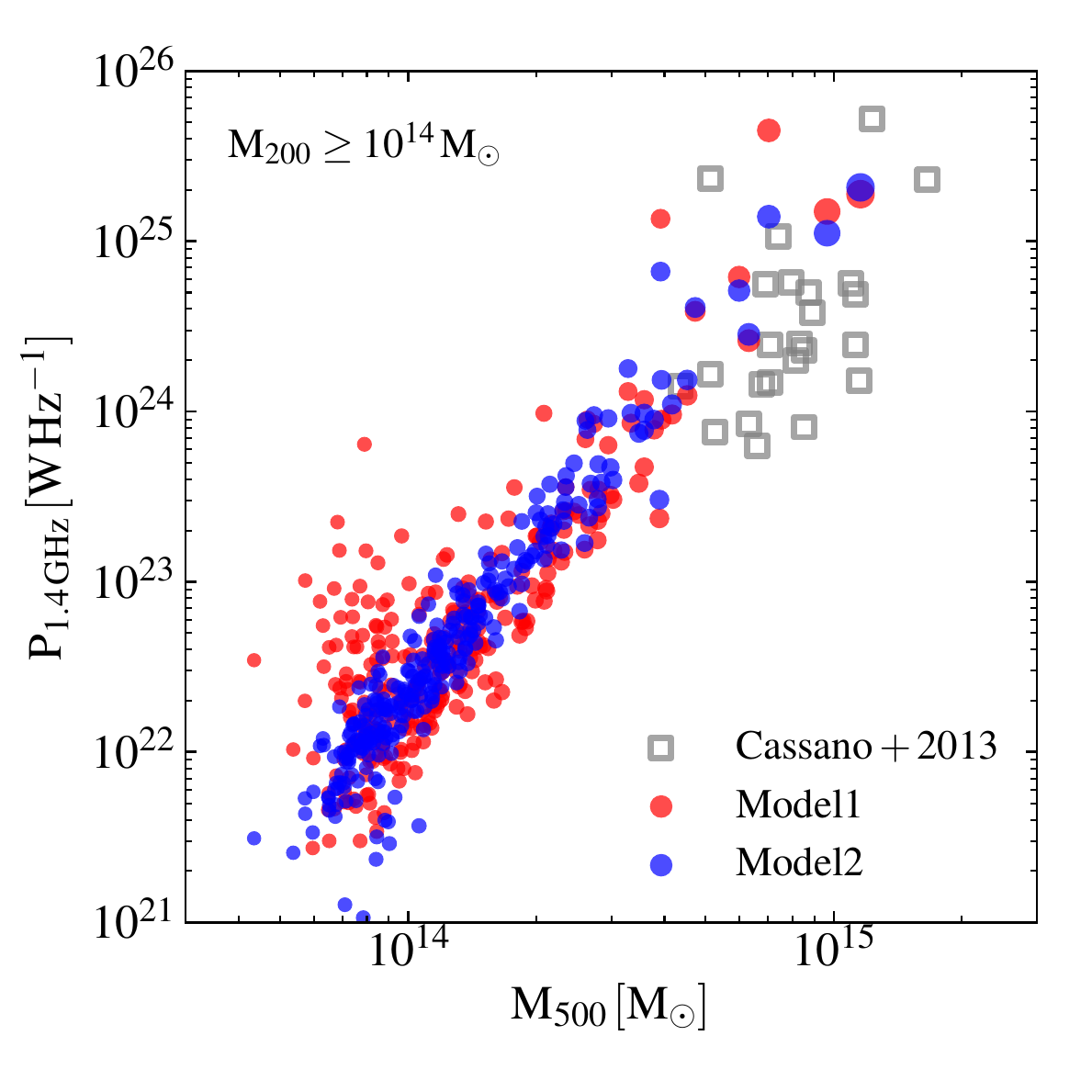}
\includegraphics[width=0.48\textwidth]{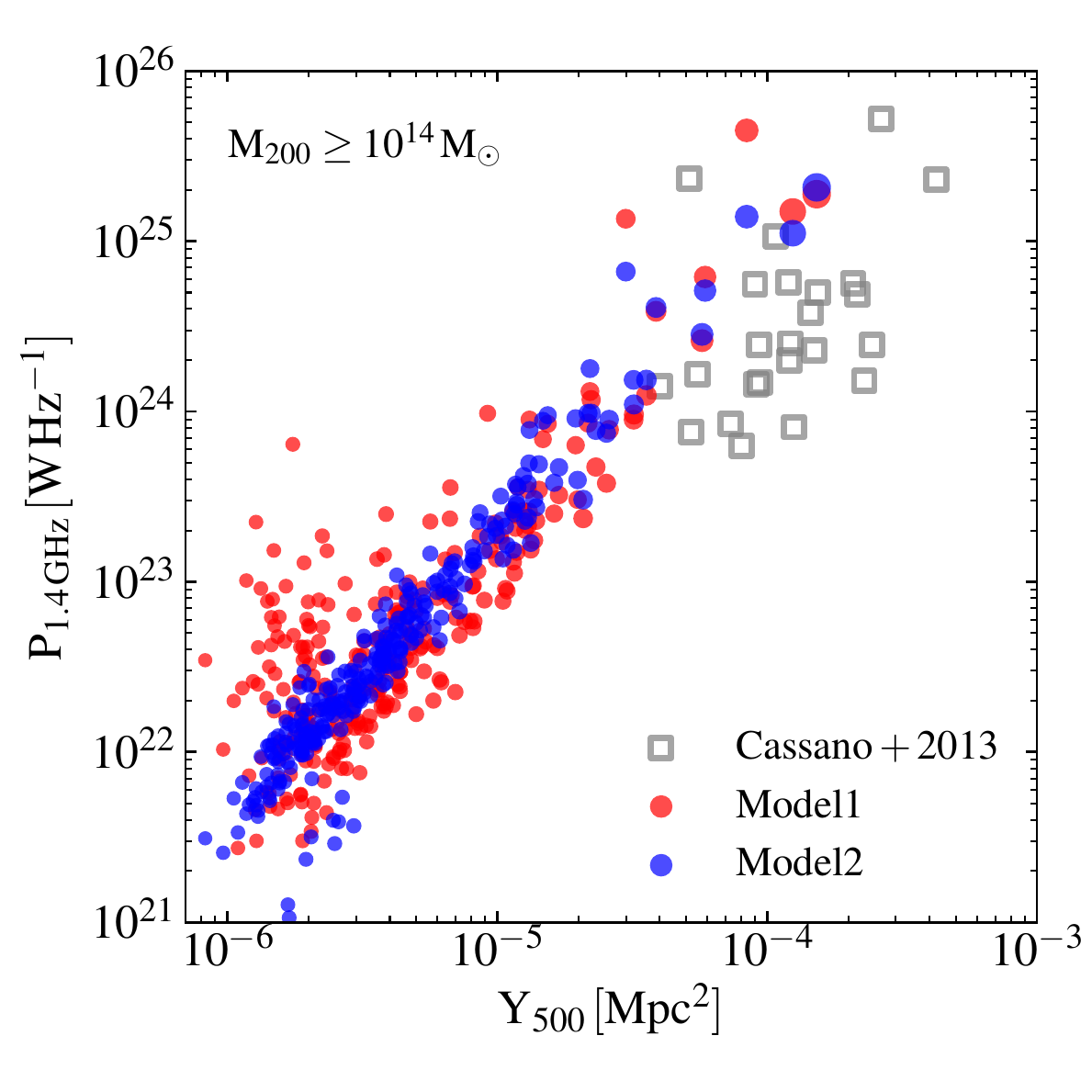}
\caption{Top panel: Total synchrotron power versus $M_{500}$ for the TNG300 
simulation at redshift zero. The radio emissivity has been computed within 
$r_{\rm 500}$ of each halo and with typical VLA parameters (see 
Table~\ref{tab:mapsproperties}). Only objects with a virial mass larger than 
$10^{14}\,{\rm M}_{\odot}$ are shown in this plot. The size of each circle is 
scaled linearly with the halo virial mass. Bottom panel: Total synchrotron power 
versus the Sunyaev-Zel'dovich parameter $Y_{500}$ for the TNG300 simulation at 
redshift zero. To make this plot the same set up of the top panel has been used. 
Simulations are compared to data points presented in \citet{Cassano2013}.
\FMR{Red points show the predictions for synchrotron model 1, whereas blue points
the one obtained for synchrotron model 2 (more details in Appendix~\ref{sec:app}).}} 
\label{fig:RvsM500}
\end{figure}

Given the comparatively low sensitivity of VLA observations compared to SKA, 
only haloes with virial masses above $\sim 3\times 10^{14} {\rm M_\odot}$ show a 
significant probability \FMR{($\gsim$ a few per cent for model 1 and $\gsim 20$ per 
cent for model 2)} of hosting detectable radio 
emission. The detection fraction is monotonically increasing with mass, arriving 
at 100 per cent for the most massive bin. Moreover, there is a difference in the 
detection fraction trends with respect to the detection criterion adopted. In 
particular, the less restrictive one, dubbed $3\sigma$ in the figure caption, 
yields substantially higher detection fractions at small masses. This 
discrepancy reduces at the high-mass end, where the trends given by the two 
criteria come into agreement. For the SKA observations, the increasing trend of 
the detection fraction with virial mass is also present. Thanks to the superior 
sensitivity -- and also of a larger radio flux at 120 MHz -- the detection 
fractions are substantially larger in this case \FMR{(above 50 per cent for model 1 
and 30 per cent for model 2)}. The less 
restrictive detection criterion yields larger detection fractions at small 
masses. \FMR{The differences between the two radio models are within a factor 
of $\sim 3$, but are more pronounced at lower virial masses. Interestingly, radio model 2 
predicts a fraction of detectable radio haloes that is smaller than radio model 1
in the SKA case, whereas the trend reverses for VLA observations. This is 
largely due to the fact that the predicted radio luminosities of model 2 in the VLA 
case are in general somewhat larger at fixed virial mass (see also Fig.~\ref{fig:RvsM500}) 
than the ones that are obtained in model 1. However, due to the flatter spectral index $\alpha$
the trend is the opposite at lower frequencies.}

In Fig.~\ref{fig:Radiohist} we show the luminosity function of radio haloes, 
i.e. source count per comoving unit volume and total radio power, for the TNG300 
simulation. All haloes with virial mass above $10^{14}\,{\rm M_\odot}$ have been 
considered. The luminosity function is presented for VLA (top panel) and SKA 
(bottom panel) mock observations for a fiducial redshift of $z=0.2$ 
\FMR{and for both radio models 1 (solid lines) and 2 (dashed lines)}. Both the 
luminosity function of all the haloes above the adopted mass cut and those of 
the objects hosting detectable radio emission -- according to the criteria 
presented above -- are  displayed in the figure. For the VLA case, we have also 
added observations at 1.4 GHz taken from \citet{Zandanel2014} and based on the 
NVSS survey \citep{Giovannini1999}. The observed points are derived from an 
X-ray flux limited sample with 13 detected objects. For the simulated luminosity 
function, radio luminosities are computed within the virial radius of each 
object without imposing any cut on surface brightness. The dashed vertical lines 
present in both panels indicate the luminosity at which the radio luminosity 
function is complete at the 95 per cent level given our virial mass cut.

Both panels show the same shape of the luminosity function. At low luminosities 
the number of radio haloes per unit comoving volume increases with radio power 
until it reaches a maximum at around $\sim 10^{22.5}\,{\rm W\,Hz^{-1}}$ for VLA 
and $\sim 10^{24}\,{\rm W\,Hz^{-1}}$ for SKA mock observations. After the peak 
there is a steady decline of the number of haloes for increasing radio power. 
Note that this peak is an artefact of the virial mass selection criterion that 
we have applied to select the cluster sample, which causes the luminosity 
function to turn over at low luminosities \citep[see also][]{Zandanel2014}. Both 
luminosity functions cover a span in radio power of about 4 to 5 dex, with SKA 
mock observations shifted towards higher radio power. The higher radio power for 
the SKA mock observations is due to the different observing frequency from the 
VLA in conjunction with the assumed spectral shape for the synchrotron radiation 
($P_\nu \propto \nu^{-\alpha}$, with $\alpha = 1.7$ \FMR{for radio model 1 and 
$\alpha = 1.3$ for radio model 2}). 

When considering only detected haloes, we find some differences. It can be seen 
that only radio haloes at the most luminous end can be detected. For the VLA 
case this translates into a minimum power of $\approx 10^{24}\,{\rm 
W\,Hz^{-1}}$. The shape of the luminosity function of detected haloes above this 
minimum power is practically indistinguishable from the total luminosity 
function, i.e. the red and green lines, which show the luminosity function for 
detected haloes, cover the blue line, which shows the luminosity function of all 
the 280 clusters of the sample. There exist small differences in the form of the 
luminosity function of detected haloes depending on the detection criterion 
adopted, but in general the agreement is good at high luminosity, in line with 
the results presented in Fig.~\ref{fig:Radioprob}. Compared to the observations, 
it can be readily seen that our simulated radio haloes are overly luminous, and 
the steep drop in the source count is not reproduced by our model. This 
discrepancy may just be a reflection of our simplified model for relativistic 
electrons, which are not directly tracked in the simulation. In particular, 
their energy density follows directly from the magnetic field distribution, and 
to reproduce the observations a different and more physically motivated 
distribution of relativistic particles (derived, for instance, as in 
\citealt{Zandanel2014}) might be required. For mock SKA observations, the 
differences in the shape of the luminosity function of detected radio haloes for 
the two detection criteria that we have adopted in this work is not very 
pronounced and disappears at high radio power. Differently from the VLA case, 
the luminosity function of detected haloes extends down to lower luminosities, 
spanning a larger range in power. This is again due to the superior sensitivity 
of SKA and the larger synchrotron emission at lower frequencies. \FMR{The trends
of the two different radio models are very similar for VLA mock observation, while there 
is a shift towards lower radio luminosity for SKA mocks in the case of synchrotron model 2
due to the flatter ($1.3$ versus $1.7$) spectral index adopted.}

In Fig.~\ref{fig:RvsM500}, we present the relationship between the total radio 
power (computed with VLA parameters) within $r_{500}$ with $M_{500}$ 
(top panel) and with Sunyaev-Zel'dovich parameter $Y_{500}$ (bottom 
panel) for the haloes more massive than $10^{14}\,{\rm M_\odot}$ in the 
TNG300 simulation. Observational data (grey squares) are taken from 
\citet{Cassano2013}\FMR{, whereas red and blue points represent the predictions
of radio models 1 and 2, respectively}. 

\begin{figure*}
\centering
\includegraphics[width=0.49\textwidth]{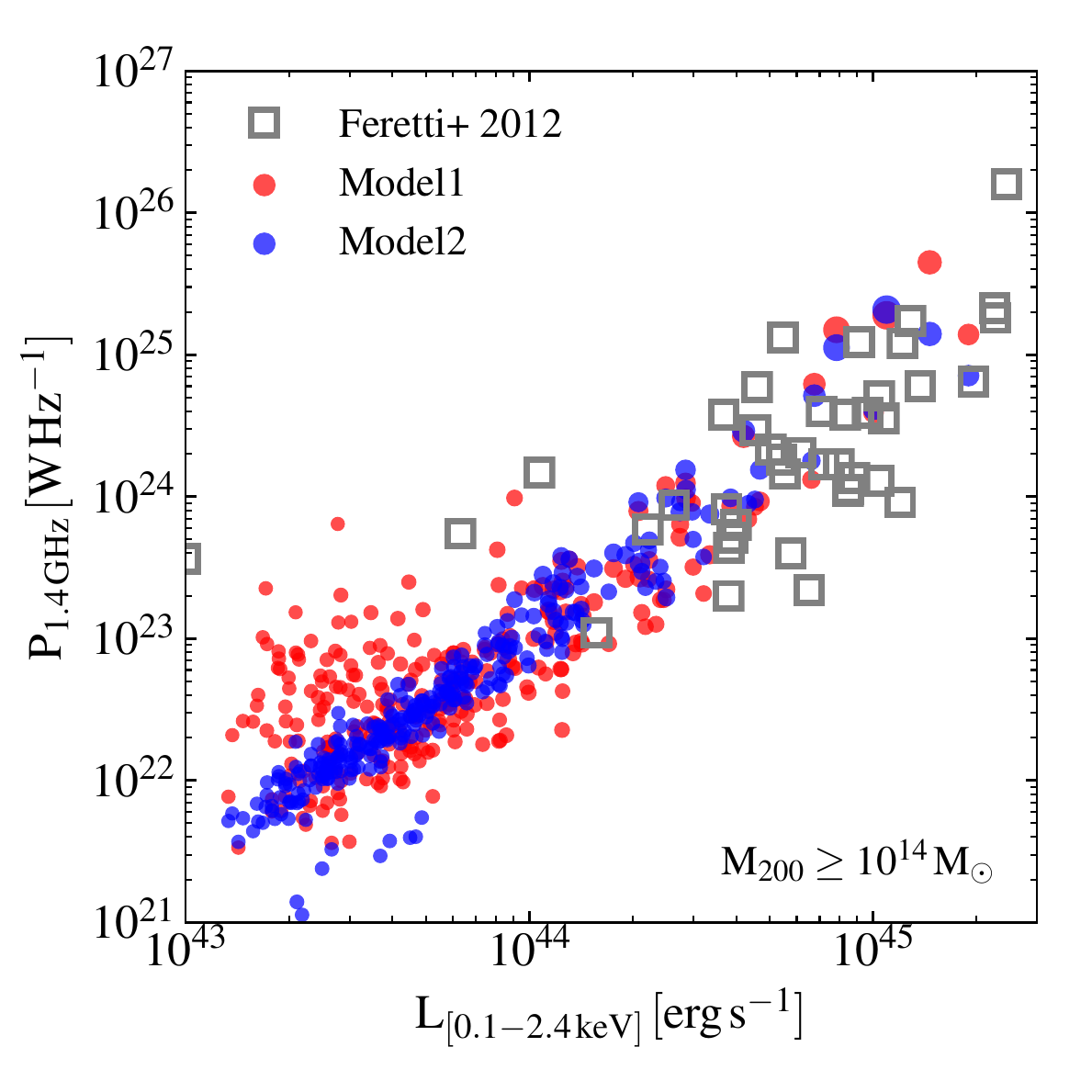}
\includegraphics[width=0.49\textwidth]{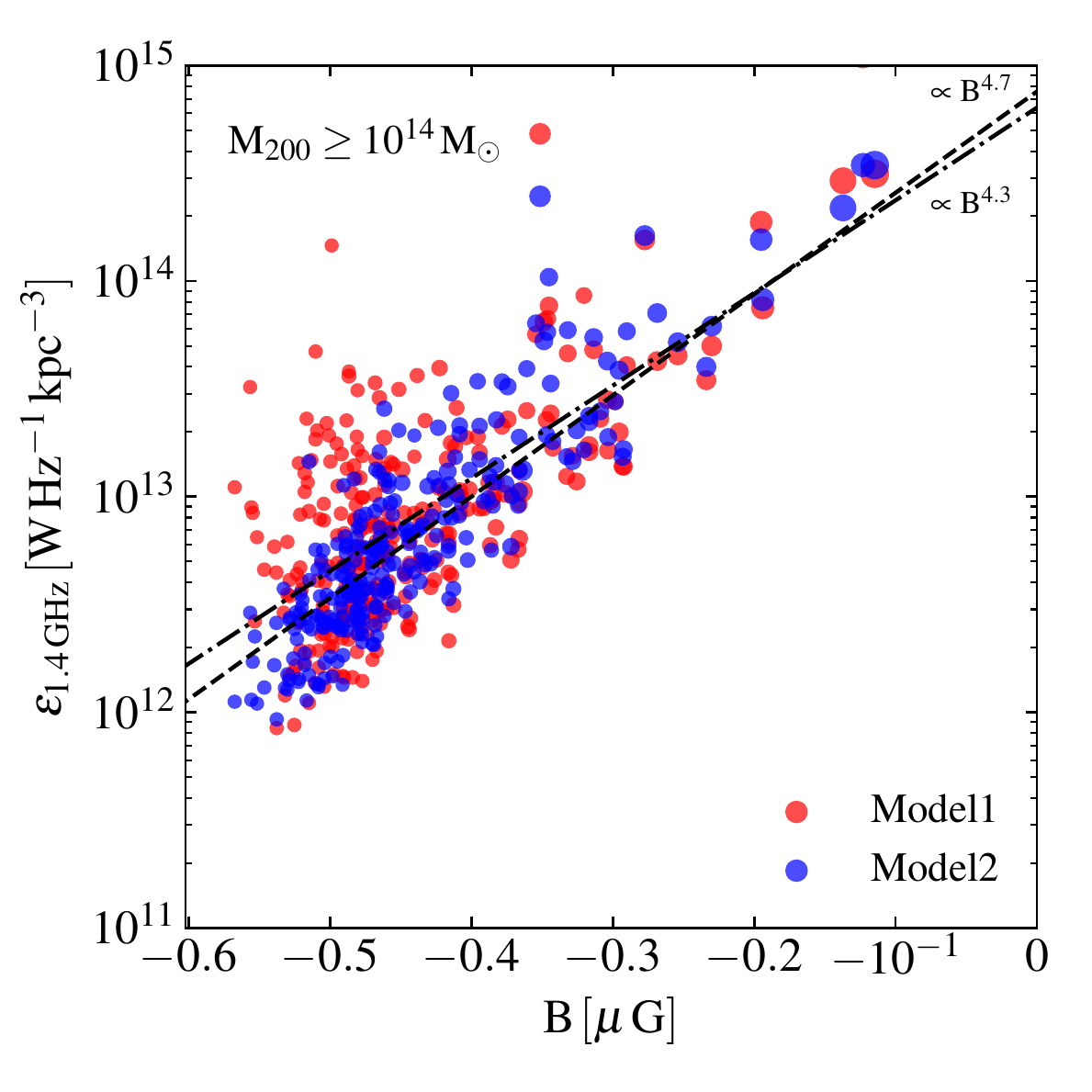}
\caption{Left-hand panel: Total synchrotron power versus X-ray luminosity for the TNG300 simulation 
at redshift zero. Luminosities for both emission mechanisms have been computed 
within the virial radius of each halo and only objects with a virial mass larger 
than $10^{14}\,{\rm M}_{\odot}$ have been considered. The size of each circle is 
scaled linearly with the halo virial mass. Simulation results are compared with 
the observed relation for galaxy clusters (grey squares). Data are taken from 
the list compiled in Table 1 of \citet[][see references therein for studies of 
individual objects]{Feretti2012}. Right-hand panel: Total synchrotron emissivity versus 
volume-weighted B field intensity for the TNG300 simulation at redshift zero. 
The radio emissivity and B field intensity have been computed within the virial 
radius of each halo and only objects with a virial mass larger than 
$10^{14}\,{\rm M}_{\odot}$ are shown in this plot. The size of each circle is 
scaled linearly with the halo virial mass. The black dashed lines show the 
expected scaling of the radio emissivity $(\epsilon\propto B^{3+\alpha})$ 
for the synchrotron emission \FMR{models} used in this work \FMR{($\alpha=1.7$ for 
model 1 and $\alpha = 1.3$ for model 2, respectively)}. Radio power and 
emissivities have been computed with typical VLA parameters (see 
Table~\ref{tab:mapsproperties}).
\FMR{In both panels red points show the relations for synchrotron model 1, whereas blue points
the one obtained for synchrotron model 2 (see Appendix~\ref{sec:app}).}} 
\label{fig:RvsX} 
\end{figure*}

\begin{figure}
\centering
\includegraphics[width=0.49\textwidth]{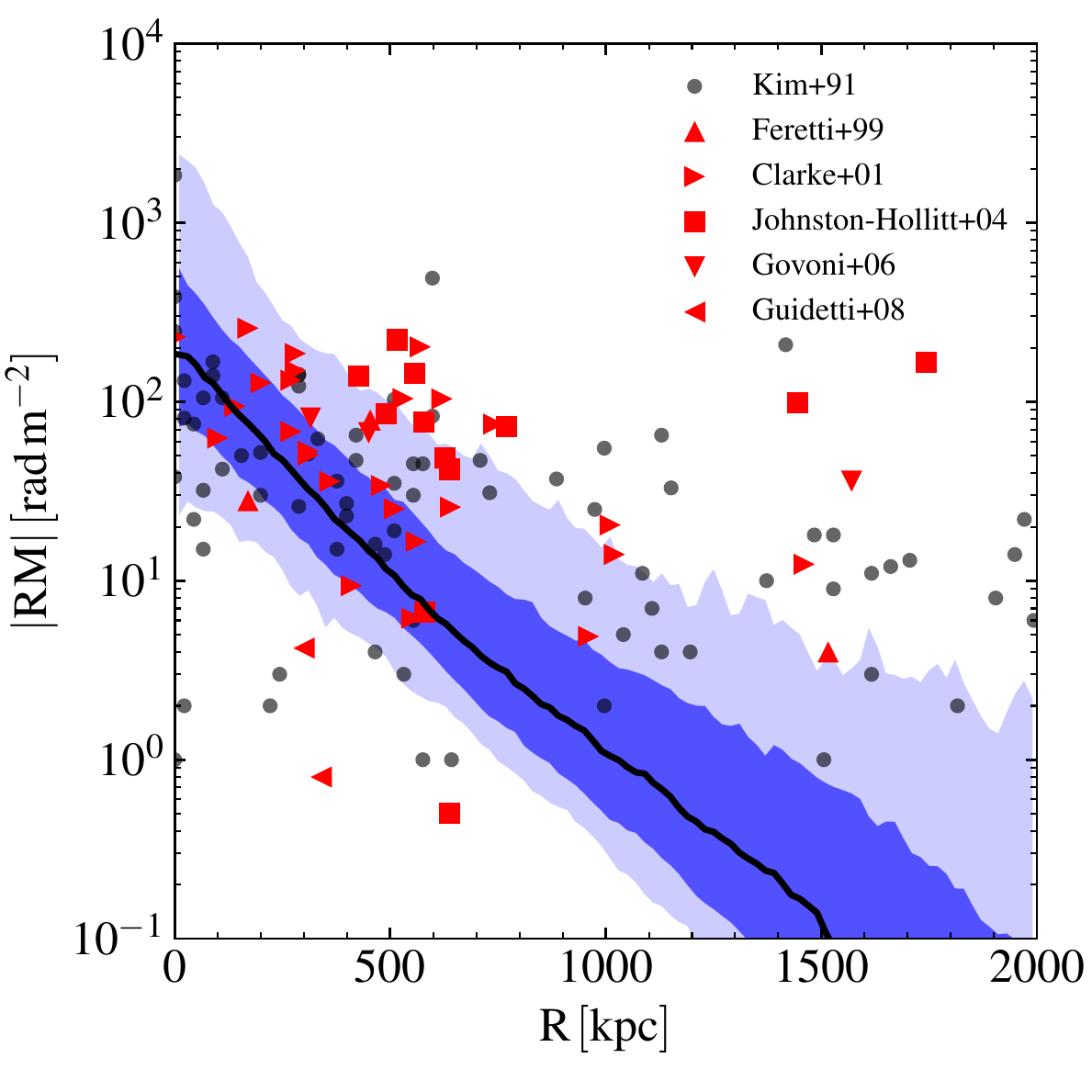}
\caption{
\FM{Radial profile of the absolute value of the Faraday rotation measure for  
haloes with $M_{200\,} > 10^{14}\,{\rm M}_{\odot}$ in the TNG300 simulation. The 
solid black line represents the median profile of the sample while the dark blue 
and the light blue regions enclose the $16{\rm th}-84{\rm th}$ and the $2{\rm nd}-98{\rm th}$ percentiles, 
respectively. Simulation results are compared to actual measurements of rotation 
measure in galaxy clusters (coloured symbols) taken from the references 
indicated in the legend.} } 
\label{fig:RMvsr} 
\end{figure}

There exists a correlation between $M_{500}$ and radio power, with more massive 
haloes having larger radio power. The scatter progressively increases towards 
lower masses \FMR{in the case of radio model 1, whereas radio model 2 gives rise 
to a tighter relation across the full mass spectrum}.  Note that if we had applied the detection criteria discussed 
previously, only the most massive objects $M_{200} \gsim 10^{14.5}\,{\rm 
M_\odot}$ in this plot would have been detected (see Fig.~\ref{fig:Radioprob} on 
the detection probability). Compared to the observations, the scaling \FMR{relations}
examined here at the high-mass end is tighter and less steep. Moreover, there is 
no clear indication of the observed bimodal behaviour in the radio emission -- 
only about 30 per cent of the observed massive/luminous clusters posses 
detectable radio emission -- for clusters at high masses (or equivalently high 
X-ray luminosity, e.g.~\citealt{Brunetti2007}). The tightness of the 
relationship, and the absence of bi-modality, are a consequence of the 
assumption that the energy density of relativistic electrons is proportional to 
the energy content of the B field in each cell. \FMR{This occurs also for 
radio model 2, although in this case the relativistic electron energy density is 
normalized to the cluster thermal energy content. However, the ratio between magnetic and thermal
energies in our simulated cluster is approximately constant at about $1$\% level 
(see Fig.~\ref{fig:BetaandBstackedfullphys}, bottom right-hand panel).}

The steepness of the relation can be varied by changing the assumed spectral 
index $\alpha$ of the synchrotron radiation, which determines the dependence of 
the radiated radio power on the B field strength in our model (see Appendix 
\ref{sec:app}). However, we caution that changing $\alpha$ has also an effect on 
the normalization of this scaling relation other than on its slope, and that 
there exist different, but still physically motivated, parametrizations 
controlling synchrotron emissivity in our model that can reproduce the 
observations equally well, \FMR{as it is explicitly demonstrated by 
the choices of parameters adopted in our radio models 1 and 2}. There is no obvious way\FM{, with our present models 
for synchrotron-emitting electrons,} to increase the scatter of the correlation 
at high virial masses \FM{nor to reproduce the observed bimodality in the 
population of radio emitting clusters \citep[e.g.][]{Kale2015}, which therefore 
still remains an unsolved issue in our simulations. Finally, we would also like 
to mention that in our current modelling there is no evidence of a dependence of 
radio emission with the host halo dynamical state \citep[e.g.][]{Cassano2010}. 
These shortcomings call for a more sophisticated treatment of relativistic 
particles in our analysis.} 

The same considerations also apply to the $P_{\rm 1.4\,GHz}$ versus $Y_{500}$ 
relation. We would like to note that the same set of parameters controlling the 
radio emissivity can reproduce both scaling relations with a comparable level of 
accuracy. This is expected, since the $Y_{500}$ Compton parameter is also used 
in observations to estimate total cluster masses, being proportional to the gas 
mass enclosed within $r_{500}$ and therefore to $M_{500}$ \citep[see 
e.g.][equation~2]{Cassano2013}. This is another indirect confirmation that our 
galaxy cluster sample features realistic gas fractions, therefore addressing one 
of the major Illustris \FM{issues} \citep{Genel2014}. Cluster gas fractions and 
scaling relations will be analysed in more detail in future work.

In Fig.~\ref{fig:RvsX}, we present scaling relations for galaxy clusters, 
relating magnetic fields, radio emission properties and cluster X-ray emission 
for all haloes more massive than $10^{14}\,{\rm M_\odot}$ in the TNG300 
simulation. \FM{The figure} relates the total radio power at 1.4 GHz to the 
total X-ray luminosity in the $[0.1-2.4]\,{\rm keV}$ energy band (left-hand panel), and the 
synchrotron emissivity at 1.4 GHz as a function of the volume-weighted average 
magnetic field strength within the virial radius of each object (right-hand panel). 
\FMR{Each panel presents the results for both radio model 1 (red symbols) and 
2 (blue symbols)}. For the 
left-hand panel, we compare our result to observational findings collected in 
\citet{Feretti2012}. We use this comparison to calibrate the free parameters of 
our radio emission \FMR{models} in such a way that it broadly matches this scaling 
relation at the high-mass end, where radio emission is detected for galaxy 
clusters. The parameters that we have chosen are compatible with theoretical and 
observational estimates (see Appendix \ref{sec:app} for their values). The 
interpretation of these results largely follows the discussion already presented 
for Fig.~\ref{fig:RvsM500}, given the relation between the cluster $M_{500}$ and 
its X-ray emission. For the right-hand panel, we compare the results to the expected 
scaling for the synchrotron emissivity of our \FMR{models (dashed and dash-dotted black lines)}, which 
\FMR{are  proportional to $B^{4.7}$ and $B^{4.3}$, for models 1 and 2, respectively\footnote{\FMR{The scaling for model 
2 implicitly assumes that the ratio between magnetic and thermal energy densities within each 
cluster is constant. This appears to be a good approximation in our simulated clusters
(see Fig.~\ref{fig:BetaandBstackedfullphys}, bottom right-hand panel)}.}} 
(see again Appendix \ref{sec:app}). At low average 
magnetic fields, which in general correspond to low virial masses, the 
emissivity values present a large scatter \FMR{(more pronounced in the case of radio model 1)}, while the relation progressively 
tightens at larger average magnetic field intensities (large virial masses). The 
expected scaling is reproduced quite well, but globally the relation appears to 
be slightly steeper than the theoretical expectation. However, it is difficult 
to establish a trend given the significant dispersion in the radio emissivity at 
low virial masses.

\FM{In Fig.~\ref{fig:RMvsr} we present the median radial profile of the Faraday 
rotation measure (RM) for all haloes with $M_{200\,} > 10^{14}\,{\rm M}_{\odot}$ 
in the TNG300 simulation. RM is defined as
\begin{equation}
 \mathrm{RM} = \frac{e^3}{2\pi m_{\rm e}^2 c^4}\int_0^L n_{\rm 
e}(s)B_{\parallel}(s)\,{\rm d}s,
 \label{eq:rotmeasure}
\end{equation}
where $e$, $m_{\rm e}$, and $c$ are the electron charge, electron mass, and speed 
of light, respectively; $n_{\rm e}$ is the electron density and $B_{\parallel}$ 
is the component of the magnetic field along the line-of-sight element ${\rm 
d}s$, and  gives information about the strength and coherence length of the 
magnetic field, being therefore dependent on its amplification and topology. We 
compute the RM using a similar set up as for the X-ray and radio maps of  
Fig.~\ref{fig:XrayRadio}. For each pixel of the map -- which extends two times 
the virial radius of each object across and uses a pixel size of $2\,\kpc$ -- 
equation (\ref{eq:rotmeasure}) is computed. Then pixels are radially binned in 
100 equally spaced bins in the radial range $0-2000\,{\rm kpc}$ and the RM 
profile for each halo is extracted. The median (solid black line) and the 
associated percentiles of the distribution showing the spread of the values 
(dark and light blue, see caption for details) are finally determined and 
plotted. Simulation results are compared to actual measurement of RM in galaxy 
clusters (coloured symbols) taken from \citet{Kim1991, Feretti1999, Clarke2001, 
J-Hollitt2004, Govoni2006, Guidetti2008}, as indicated in the legend. }

\FM{It can be seen that the median RM profile is a declining function of the 
radius. Overall there is a good agreement between the RM values of the simulated 
clusters and the observations especially in the innermost regions ($R \lsim 
1000\,\kpc$). At larger distances our median profile is declining too fast 
compared to the observations \citep[see also][]{Marinacci2015}. 
\FMRR{This faster decline is likely due to insufficient resolution at large distances.
This in turn is caused by the pseudo-Lagrangian nature of the default refinement scheme adopted in \arepo,
which tries to keep the gas cell mass within a factor of two from a predefined target mass,
effectively making spatial resolution dependent on gas density \citep{Nelson2016}. However, 
higher resolution simulations targeted at individual galaxy clusters 
\citep[see e.g.][]{Vazza2018} are needed to fully investigate this aspect.} There is a large 
spread in the values of the RM in both panels, which is becoming more pronounced 
at larger distances. The large spread is a consequence of either (i) the 
turbulent amplification of the B field within haloes, (ii) an indication of 
general lack of resolution, or (iii) missing physics within our model that can 
reorient the magnetic field \citep[e.g. thermal conduction, ][]{Kannan2016b}. We 
also note that the computation of the RM signal is sensitive to the pixel size 
used to produce the map. Therefore the results can also depend on this choice, 
although increasing or decreasing the pixel size by a factor of a few (up to 
$10\,\kpc$)  does not lead to a dramatic change of the median RM profile. }

\FMR{Finally, we would like to point out that in our analysis we did not attempt 
to distinguish between different types of galaxy clusters. This is of course a simplification because 
some of the scaling relations that we have investigated (e.g. rotation measure 
profiles and the relation between radio and X-ray luminosities) have been 
determined for objects falling into disjoint classes (e.g. relaxed clusters with a 
cool core and merging clusters). The fact that we are able to reproduce to some 
extent these observational relations in our cluster sample stems from the 
observation that our simulated clusters show a continuum of properties that at 
the extreme ends encompass cool-core and non-cool core systems \citep[with no 
strong dichotomy present unlike suggested by observations, see][]{Barnes2017} 
combined with the large scatter in the relations that we have investigated, 
in particular in the values of the rotation measure.}

\section{Summary and conclusions}\label{sec:conclusions}
We have introduced the IllustrisTNG project, the successor of Illustris, and
presented the general properties of radio haloes and magnetic fields in the
TNG100 and TNG300 simulations, which are the most novel physics aspects of 
IllustrisTNG compared to Illustris. 

We first quantified the magnetic field strengths as a function of environment,
and then specifically explored the scaling relations between the total radio
power and various cluster properties (mass, X-ray emission, and SZ signal), the
structural properties of radio haloes (central surface brightness, spatial
extent, and radial profiles), and the detection fraction and luminosity function
of such objects. We have performed a detailed comparison to current and
upcoming observational data by producing mock radio observations for VLA,
LOFAR, ASKAP, and SKA. Our cluster sample represents the largest numerically
studied radio halo sample to date including $280$ haloes above a virial mass of
$10^{14}\,{\rm M_\odot}$. Our main results can be summarized as follows: 

\begin{itemize}
\item The magnetic field intensity traces density peaks in the matter 
distribution closely. At low overdensities (i.e. in filaments and voids, 
$\rho/\langle\rho_{\rm b}\rangle \lsim 10^{2}$), the intensity of magnetic 
fields scales according to  flux conservation ($B \propto \rho^{2/3}$). Closer 
to and within haloes ($\rho/\langle\rho_{\rm b}\rangle \gsim 10^{2}$) additional 
processes, such as gas turbulent motions and shear flows, amplify magnetic 
fields up to $\sim 10\,\muG$, a value that is $4$ to $5$ orders of magnitude 
larger than the value expected from `flux freezing'. By $z=2$, most of the 
magnetic field amplification at large overdensities has already occurred. We do 
not find large differences in these trends in the two simulation boxes (TNG100 
and TNG300) analysed in this work. Only in the transition region between 
low-density and high-density gas ($\rho/\langle\rho_{\rm b}\rangle \sim 10^{2}$) 
and at high redshift ($z=2$), the magnetic field amplification in the TNG300 
simulation is less pronounced than in TNG100 because of the lower resolution of 
the former run.
\vspace{0.25cm}

\item The magnetic field topology within galaxies depends on their morphology 
and in particular on the existence of a rotationally supported gaseous disc. In 
early-type galaxies, in which a gaseous disc is largely absent due to the 
disruptive effects of the AGN feedback, the field orientation is not well 
defined although magnetic fields tend to align with gas filaments if any is 
present. Disc galaxies generally feature an ordered, large-scale magnetic field 
in the disc plane, which is also (anti-)aligned with the gas velocity field. 
\vspace{0.25cm}

\item Within haloes, magnetic field intensities are declining with radius. The 
field strength tends to decline faster once a radial distance of $\sim 
0.3\,r_{\rm 200}$ is reached. In the halo centres the field intensity reaches 
values of $\sim 10\,\muG$, in line with observational determinations, to then 
drop by about three orders of magnitude in the external regions. This value is 
however larger than the one expected from `flux freezing', signalling again the 
efficient magnetic field amplification occurring within haloes. There is good 
agreement between the results obtained from TNG100 and TNG300 notwithstanding 
the coarser resolution of the latter. 
\vspace{0.25cm}

\item The ratio between magnetic and thermal pressure $\beta^{-1}$ reaches 
maximum values of about $3$ in halo centres ($r \sim 0.1\,r_{200}$) for virial 
masses $\lsim 10^{12.5}\,{\rm M_{\odot}}$. This value declines with virial mass 
and for increasing distances from the halo centres. In the most massive haloes 
($M_{200} \gsim 10^{13.5}\, {\rm M_{\odot}}$) $\beta^{-1} \simeq 10^{-2}$, and 
this value remains basically constant with radius. In general, magnetic fields 
are dynamically unimportant on the scales resolved in IllustrisTNG. Again, the 
agreement between the results obtained from TNG100 and TNG300 is very good. 
\vspace{0.25cm}

\item Our simulated galaxy clusters show extended radio emission with morphology 
that matches the extension and shape of bremsstrahlung X-ray radiation 
originating from the hot gas. The detectability of this radio emission depends 
on the mass of the cluster -- more X-ray luminous/massive clusters are easier to 
detect in radio -- and on the sensitivity and the observing frequency of the 
radio telescope. In particular, we find that in current radio surveys (VLA mock 
observations) only clusters with a virial mass larger than $\sim 3\times 
10^{14}\, {\rm M_{\odot}}$ have a significant probability ($\gsim$ few per cent) 
of hosting detectable radio emission, which increases steadily reaching 100 per 
cent for the most massive ($M_{200} \sim 10^{15}\,{\rm M_{\odot}}$) objects. The 
minimum virial mass for detection decreases down to $10^{14}\, {\rm M_{\odot}}$ 
(the lowest virial mass analysed in this paper) for SKA mock observations. 
Moreover, the detection probability for SKA mock observations is substantially 
larger -- above \FMR{30} per cent at all masses -- and shows the same increasing trend 
with virial mass. 
\vspace{0.25cm}

\item Emission from radio haloes exhibits a radially declining profile, which is 
described quite well by an exponential. At the fiducial redshift $z=0.2$ used to 
generate the mock radio maps, central surface brightnesses, averaged over the 20 
most massive objects of our simulated sample, of $\simeq 3\times10^{-3}\,{\rm 
mJy\,beam^{-1}}$ for instruments observing at 1.4 GHz (such as VLA and ASKAP) or 
of $\simeq 0.2\,{\rm mJy\,beam^{-1}}$ for instruments observing at a 140 MHz 
(such as LOFAR and SKA) are reached, respectively. On average, radio emission is 
detectable up to $0.1-0.3\,r_{200}$ after which it goes below the 3$\sigma$ 
instrument sensitivity. The simulated VLA profiles are broadly consistent with 
observations although discrepancies from the observed trends are noticeable. In 
particular, observed clusters featuring high radio surface brightness are more 
compact than the simulated objects, while observed radio haloes with lower 
surface brightness appear to be more radially extended than our findings. 
\vspace{0.25cm}

\item The average trends are in agreement with the results of fitting 
exponential surface brightness profiles to our simulated radio haloes on an 
object by object basis. The fitting procedure reveals that, while our 
simulations broadly capture the observational division between mini and giant 
radio haloes, the simulated mini haloes are in general more extended ($r_{\rm e} 
\sim 100\,\kpc$) than their observed counterpart ($r_{\rm e} \sim 50\,\kpc$) and 
there is a general lack of very extended ($r_{\rm e} \gsim 150\,\kpc$) radio 
haloes in our simulated sample. The relationship between the total radio 
luminosity with $r_{\rm e}$ ($P_{1.4\,{\rm GHz}} \propto r_{e}^3$) is \FMR{roughly} recovered 
in our simulations for giant haloes \FMR{for radio model 1 and the least 
bright haloes of model 2}, while our simulated mini haloes appear to 
be too extended for their radio luminosity.
\vspace{0.25cm}

\item We find a good agreement with scaling relations observed between the total 
radio power and galaxy cluster properties -- such as mass, X-ray luminosity, SZ 
Compton parameter. Calibrating the free parameters of our radio emission 
modelling on one of these relations (we have adopted the radio power versus 
X-ray luminosity relation in this work) reproduces the other scaling relations 
with the same degree of accuracy. This presents further evidence that the 
magnetic field amplification (which controls the total radio power in our model) 
and the assembly of the cluster and its ICM are tightly linked. 
\end{itemize}

Our results indicate that the magnetic field properties predicted by our
simulations are in line with various observational constraints. There are,
however, some tensions. Those are inevitable due to the rather simple \FMR{models}
that we have adopted to describe relativistic particles, whose evolution is not
followed self-consistently in the IllustrisTNG simulations. These shortcomings
of our treatment of relativistic particles can likely be addressed with a more
complete description of their spatial and energy distributions
\citep[e.g.][]{Pinzke2017}, or by directly modelling their evolution in a
self-consistent way in the simulations \citep[][]{Pfrommer2017}. We intend to
explore both possibilities, and thus present a more detailed characterization
of radio emission in simulated galaxy clusters, in future work adopting a
galaxy formation model going beyond this limitation of IllustrisTNG.

\section*{Acknowledgements}
\FM{We thank an anonymous referee for insightful comments that helped improving 
the paper}. FM also thanks Christoph Pfrommer, Irina Zhuravleva, and Rahul Kannan 
for helpful discussions on radio haloes in galaxy clusters. VS, RW, and RP 
acknowledge support through the European Research Council under the ERC-StG 
grant EXAGAL-308037 and would like to thank the Klaus Tschira Foundation. MV 
acknowledges support through an MIT RSC award, the support of the Alfred P. 
Sloan Foundation, and by the NASA ATP grant NNX17AG29G. SG and PT are supported 
by NASA through Hubble Fellowship grants HST-HF2-51341.001-A and 
HST-HF2-51384.001-A, respectively, awarded by the STScI, which is operated by 
the Association of Universities for Research in Astronomy, Inc., for NASA, under 
contract NAS5-26555. JN acknowledges the support of the NSF AARF award 
AST-1402480. The Flatiron Institute is supported by the Simons Foundation. 
Simulations were run on the Hazel-Hen Cray XC40-system at the High Performance 
Computing Centre Stuttgart as part of project GCS-ILLU of the Gauss Centre for 
Supercomputing (GCS). Additional simulations were carried out on the Hydra and 
Draco supercomputers at the Max Planck Computing and Data Facility, on the 
Stampede supercomputer at the Texas Advanced Computing Center through the XSEDE 
project AST140063, and on the joint MIT-Harvard computing cluster supported by 
MKI and FAS. All the figures in this work were produced by using the {\sc 
matplotlib} graphics environment \citep{Matplotlib}.

\bibliographystyle{mnras}
\bibliography{paper}

\newpage
\appendix

\section{Synchrotron radiation}\label{sec:app}

\subsection{Synchrotron emissivity}

\FMR{To compute the total emitted synchrotron power we assume a relativistic 
population of electrons in each gas cell whose number density per energy 
interval is distributed as a power law of the form} 
\begin{align}
 \frac{{\rm d}n(\gamma)}{{\rm d}\gamma} = n_0 \gamma^{-p} 
\qquad\qquad\qquad\qquad
 {\rm for}\quad \gamma > \gamma_{\rm min},
\label{eq:elecdens}
\end{align}
\FMR{where $\gamma$ is the Lorentz factor, $n_0$ a normalizing factor (to be 
determined), $p$ is the electron spectral index and $\gamma_{\rm min}$ is a 
lower energy cut-off that takes into account the fact that only very energetic 
electrons ($\gamma\sim10^{3}$) contribute significantly to synchrotron emission. 
We make further assumption that each electron emits at the synchrotron 
critical frequency $\nu_s \equiv \gamma^2\nu_{\rm L}$. $\nu_{\rm L}$ is the 
so-called Larmor frequency given by} 
\begin{equation}
 \nu_{\rm L} = \frac{eB}{2\pi m_{\rm e} c},
\end{equation}
\FMR{with $e$ and $m_{\rm e}$ being the electron charge and mass, respectively, 
$B$ the intensity of the B field in the cell and $c$ the speed of light.
Under these assumptions we can write the power emitted per unit frequency 
by a single electron as \citep[see e.g.][]{Longair2011}}
\begin{equation}
 P(\nu) = \frac{4}{3}\sigma_{\rm T} c U_{B} \beta^2\gamma^2\delta(\nu - \nu_{\rm 
s}) \approx \frac{4}{3}\sigma_{\rm T} c U_{B}\gamma^2\delta(\nu - \nu_{\rm s}),
\end{equation}
\FMR{in which $\sigma_{\rm T}$ is the Thomson cross section, $U_{B} = B^2 / 
(8\pi)$ is the magnetic energy density and the last approximate equality holds 
in the ultrarelativistic limit $\gamma \gg 1$, $\beta = v/c \approx 1$.}

\FMR{The total power radiated per unit frequency and volume by the electron 
population distributed as in equation (\ref{eq:elecdens}) is obtained as} 
\begin{equation}
 \epsilon_{\rm tot}(\nu) = \int_{\gamma_{\rm min}}^{\gamma_{\rm 
max}}\frac{4}{3}\sigma_{\rm T} c U_{B}\gamma^{2 - p} n_0 \delta(\nu - \nu_{\rm 
s}) {\rm d}\gamma,
 \label{eq:Ptot}
\end{equation}
\FMR{where $\gamma_{\rm min}$ and $\gamma_{\rm max}$ are the minimum and maximum 
Lorentz factors of the electron population, respectively. With the change of 
variable $\nu^{'} = \gamma^{2} \nu_{\rm L}$ equation (\ref{eq:Ptot}) yields} 
\begin{equation}
 \epsilon_{\rm tot}(\nu) = \frac{2}{3}\frac{\sigma_{\rm T} c U_{B}n_0}{\nu_{\rm 
L}} \left(\frac{\nu}{\nu_{\rm L}}\right)^{-\alpha},
 \label{eq:spect}
\end{equation}
\FMR{with $\alpha = (p-1)/2$.}

\FMR{We compute the total synchrotron power per unit frequency and volume at 
$\nu_0$ over a bandwidth of $\Delta\nu$  by integrating the power given by 
equation (\ref{eq:spect}) over the bandwidth and averaging the resulting total 
power over $\Delta\nu$ \citep[see also][]{Xu2012}. This results in} 
\begin{equation}
 \epsilon_{\nu_0} = \frac{2}{3}\frac{\sigma_{\rm T} c U_{B} n_0}{\Delta\nu 
(1-\alpha)}
 \left[\left(\frac{\nu_+}{\nu_{\rm L}}\right)^{1-\alpha} - 
\left(\frac{\nu_-}{\nu_{\rm L}}\right)^{1-\alpha}\right],
 \label{eq:GHzpower}
\end{equation}
\FMR{where $\nu_\pm = \nu_0 \pm \Delta\nu / 2$. The total synchrotron emission 
in the target objects is then obtained by multiplying the previous equation by 
the cell volume and summing over all gas cell contributions. The parameters that 
we have considered for different radio telescopes are listed in 
Table~\ref{tab:mapsproperties}. We assume that the integration times of the mock 
observations are the ones necessary to reach the sensitivity reported in 
Table~\ref{tab:mapsproperties}. These can vary between $\approx 2$ hours per 
pointing in the case of VLA \citep{Govoni2001} and up to $\approx 8$ hours per 
pointing in the case of LOFAR \citep{Shimwell2017}. The remaining quantities 
that need to be determined in order to compute the synchrotron emissivity are 
the spectral index $\alpha$, which can be readily fixed from observations 
(usually $\alpha > 1$), and the electron density normalization 
$n_0$. In the subsections below we present two different parametrizations to 
determine the latter quantity.}

\subsection{Model 1}\label{sec:model1}

\FMR{In our first model we only consider relativistic particles above the low 
energy cut-off $\gamma_{\rm min}$, which are the only ones that contribute to 
the emitted synchrotron radiation.} To find an expression for $n_0$ we assume 
that in each gas cell the energy density in relativistic particles \FM{(protons 
\textit{and} electrons)} is proportional to  the magnetic energy density 
according to
\begin{equation}
 (1 + k) m_e c^2n_0\int_{\gamma_{\rm min}}^{\gamma_{\rm max}}\gamma^{-2\alpha}\, 
{\rm d}\gamma = \eta U_{B},
 \label{eq:ergdens}
\end{equation}
where $k$ is the energy density ratio between (relativistic) protons and 
electrons and $\eta$ the ratio between the energy density in relativistic 
particles and the magnetic energy density. \FM{We note that 
eq.~(\ref{eq:ergdens}) makes the assumption that protons and electrons are 
distributed in energy according to the same power law (\ref{eq:elecdens}). 
Equation~(\ref{eq:ergdens}) yields} 
\begin{multline}
 n_0 = \frac{\eta}{(1+k)}\frac{B^2}{8\pi m_{\rm e} c^2} \left(\int_{\gamma_{\rm 
min}}^{\gamma_{\rm max}}\gamma^{-2\alpha}\, {\rm d}\gamma\right)^{-1} = \\
 \frac{\eta}{(1+k)}\frac{B^2}{8\pi m_{\rm e} c^2} \frac{(2\alpha - 
1)}{\gamma_{\rm min}^{1 - 2\alpha} - \gamma_{\rm max}^{1 - 2\alpha}}.
\end{multline}
\begin{table}
\centering
\begin{tabular}{cccccc}
\hline
    Telescope & $\nu_0$ & $\Delta\nu$ & beam & rms noise \\
              & $[\rm{MHz}]$  & $[\rm{MHz}]$  & $[{\rm arcsec}]$ & $[{\rm 
mJy\,beam^{-1}}]$ \\
\hline
      VLA       &  1400  & 25  & 35 & 0.1  \\
      LOFAR-HBA &   120  & 32  & 25 & 0.25 \\
      ASKAP     &  1400  & 300 & 10 & 0.01 \\
      SKA-LOW   &   120  & 32  & 10 & 0.02 \\
\hline
\end{tabular}
\caption{Telescope configurations for radio maps. The columns list (from left to 
right): instrument name, observation frequency, frequency bandwidth, beam size 
(FWHM), and rms noise. Values for these parameters have been taken from 
\citet[][VLA; Fig.1 right panel]{Govoni2001}, \citet[][LOFAR]{Rottgering2011}, 
\citet[][ASKAP]{Norris2013}, and \citet[][SKA; see also \citealt{Vazza2015a}]{Cassano2015}.}
\label{tab:mapsproperties}
\end{table}
Common choices for the $\eta$ parameter are those known as minimum energy 
\citep[$\eta=4/3$,][]{Longair2011}, in which the total energy content in 
relativistic particles and magnetic fields is minimized at fixed total 
synchrotron power, and equipartition ($\eta = 1$), in which magnetic fields and 
relativistic particles have the same energy density. In what follows, we  assume 
that B fields and relativistic particles \FM{(protons \textit{plus} electrons)} 
have the same energy density (i.e.~$\eta=1$), which is justified in galaxy 
clusters \citep[e.g.][]{Pinzke2010, Arlen2012}. Furthermore, we fix $k = 10$, 
consistent with determinations of cosmic rays energy fluxes in the Milky Way and 
other theoretical work \citep{Ensslin2011, Vazza2014c, Persic2014}. 

\FM{The choice of these parameters implies that, for our cluster sample, about 
0.6 per cent of  thermal energy in any given cluster is present in the form of 
relativistic particles. Indeed for our simulated objects the magnetic energy 
density is only about 1 per cent of the gas thermal energy (see 
Fig.~\ref{fig:BetaandBstackedfullphys}, bottom right-hand panel). Breaking the 
energetics up in relativistic electron and proton contributions, we find that 
electrons carry about 0.05 per cent of the cluster thermal energy while protons 
approximately 0.5 per cent. The proton energy content is in agreement with what 
has been determined by recent Fermi upper limits. For instance, in the Coma 
cluster \citet{Brunetti2017} find that the ratio between the energy density of 
cosmic rays proton to the gas thermal energy density, albeit dependent on the 
distance to the centre, is about $\lsim 1$ per cent (reaching up to 10 per cent 
in the outskirts), which is consistent with our modelling.} \FMR{However, in that 
paper, the spectrum of relativistic particles extends down to $\gamma = 1$. This 
is the major limitation of the present model, since in principle, even though 
electrons below $\gamma_{\rm min}$ do not contribute substantially to the total 
synchrotron emission, they can nevertheless carry the bulk of the energy of the 
whole population. We address this limitation of the model in the 
next section.} 

\begin{figure}
\centering
\includegraphics[width=0.49\textwidth]{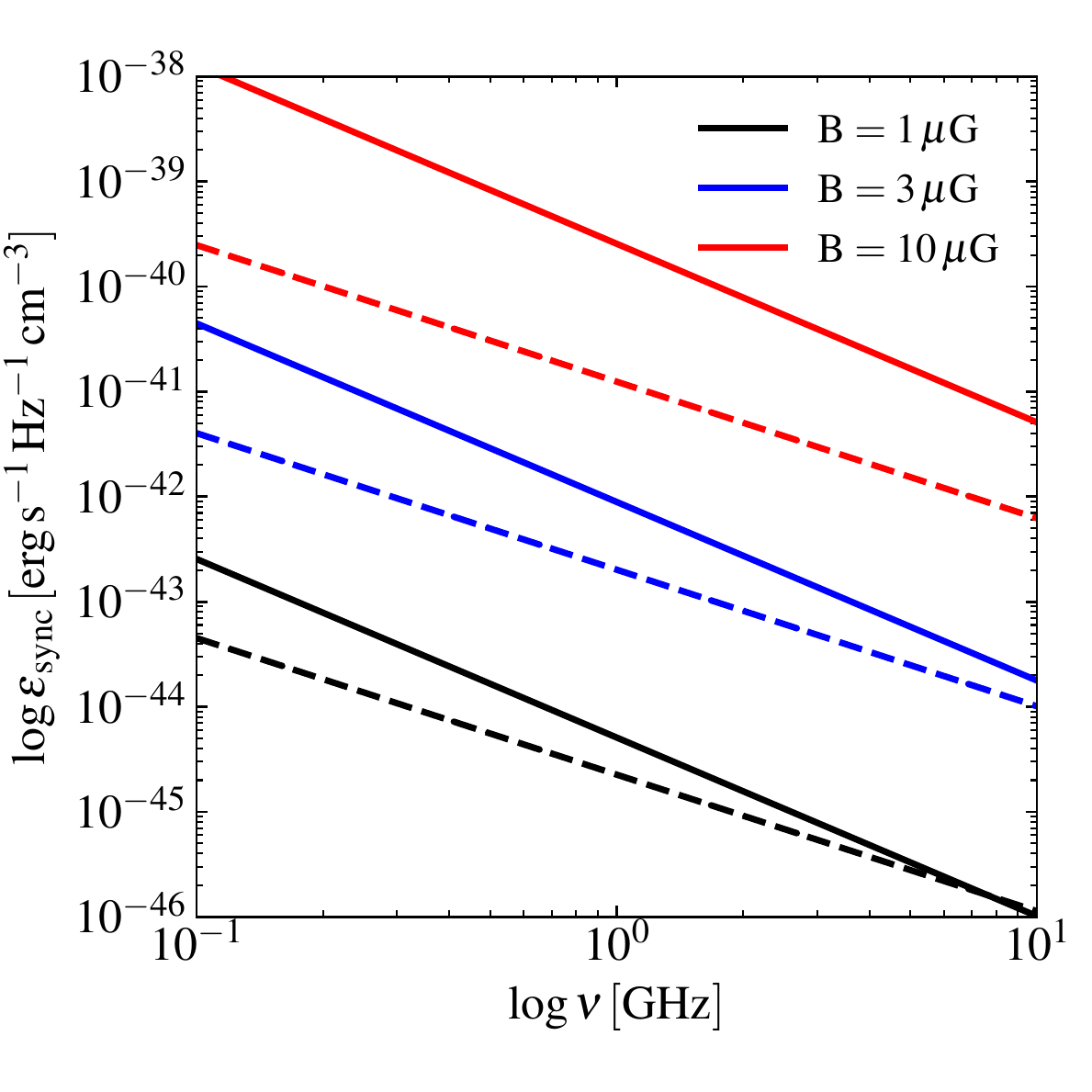}
\includegraphics[width=0.49\textwidth]{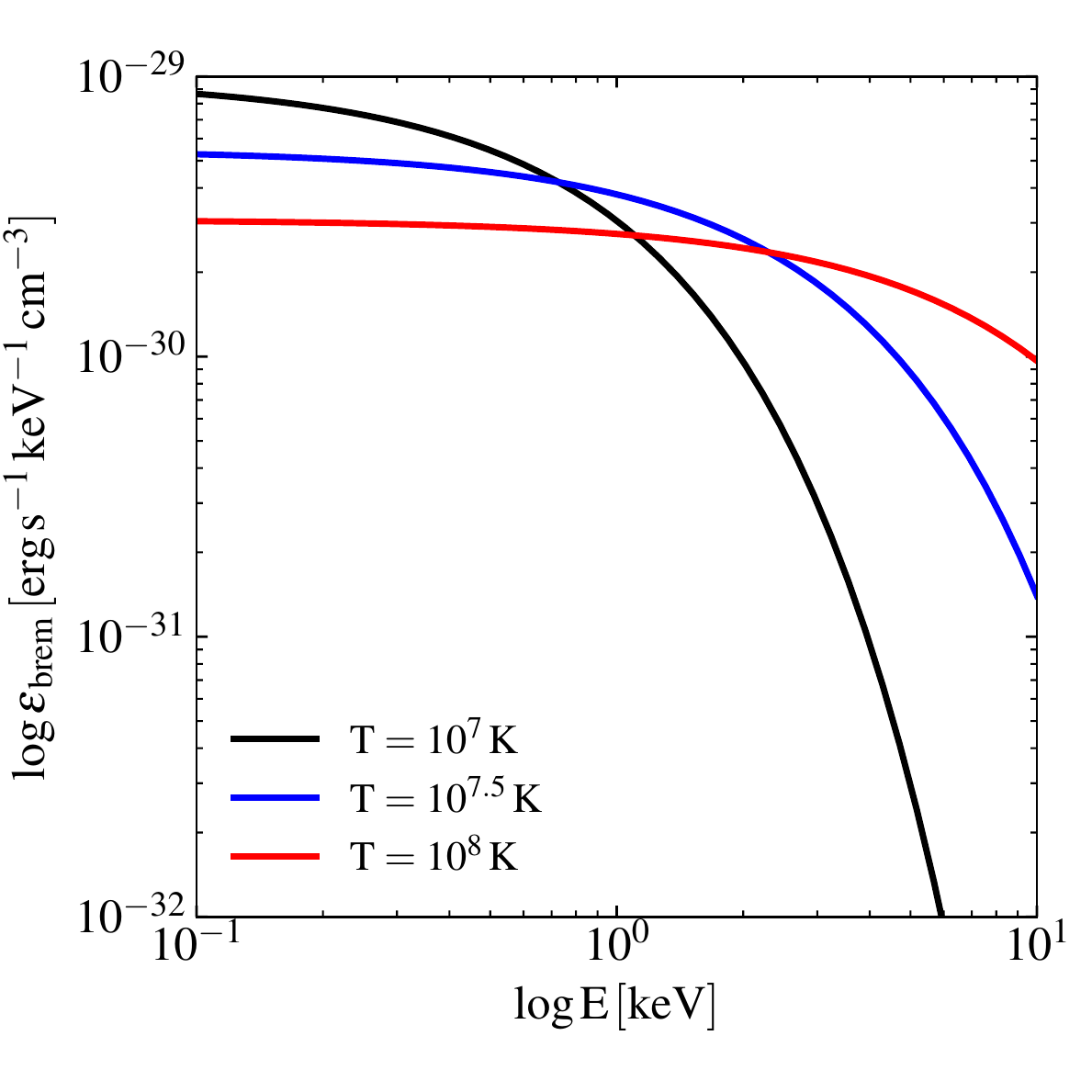}
\caption{Top panel: Volumetric synchrotron emissivity as a function of frequency 
for \FMR{synchrotron model 1 (solid lines) and 2 (dashed lines)}. The emissivity is computed 
for constant magnetic field intensities as indicated in the legend \FMR{and with the typical parameters 
in Section~\ref{sec:model1} for model 1, whereas the emission for model 2 has been estimated 
by assuming a value for the gas thermal energy density equal to the magnetic energy contained in the
fields listed in the legend and a constant (thermal) electron density $n_{\rm e} = 10^{-3}\,\cm^{-3}$}. Note the 
power-law slope of the resulting spectrum (index $-\alpha$) and its strong 
dependence on the field strength ($\propto B^{3+\alpha}$). Bottom panel: 
Volumetric bremsstrahlung emissivity as a function of energy for the typical 
parameters that we have adopted in this work (see Section~\ref{sec:ff}). The 
resulting spectrum is computed for different gas temperature values, as 
indicated in the legend, and for $n_{\rm H} = 10^{-3}\,{\rm cm^{-3}}$. Note how 
the characteristic exponential cut-off at high energies moves towards higher 
values for increasing gas temperatures.} 
\label{fig:emissivity}
\end{figure}

Finally, we use as lower and upper limits for the electron Lorentz factor 
$\gamma_{\rm min} = 300$ and $\gamma_{\rm max} = \infty$, respectively. \FMR{For 
a justification of the cut-off at low $\gamma$ see Section~\ref{sec:Sarazin}}. The 
spectral index $\alpha$ is fixed to 1.7, appropriate for the cluster temperature 
range \FM{($T \lsim 8\,{\rm keV}$)} analysed here \citep[\FM{see discussion 
in}][\FM{page 15}]{Feretti2012}. The parameters thus chosen reproduce fairly 
well the scaling relation between the total radio power and the X-ray luminosity 
on galaxy clusters (see Fig.~\ref{fig:RvsX} left-hand panel). Indeed, we have used 
this observational relation, as presented in \citet{Feretti2012}, to calibrate 
\FMR{the synchrotron emission in our simulations}. Figure~\ref{fig:emissivity} (top panel\FMR{, solid lines}) shows 
the main characteristic features of the emitted radiation (described by 
equation~\ref{eq:spect}) as a function of frequency for three different 
(constant) magnetic field intensities typical of the ICM \FMR{for the present model}. 

\subsection{Model 2}\label{sec:Sarazin}

\FMR{In our second model we address the major shortcoming of the model presented 
in Section~\ref{sec:model1} by extending the spectrum of relativistic electrons 
down to $\gamma = 1$. To do so we follow \citet{Sarazin1999} and assume that 
relativistic electrons are injected continuously in the ICM 
according to} 
\begin{equation}
 Q(\gamma) = Q_0 \gamma^{-2\alpha},
\end{equation}
\FMR{where $\alpha$ is the spectral index of the synchrotron radiation, $\gamma$ 
is the Lorentz factor, and $Q_0$ is the electron injection rate per unit volume 
(which is related to $n_0$ that needs to be determined). The electron population 
evolves subject to Coulomb losses (at low energies) and inverse Compton and 
synchrotron losses (at high energies). If the time-scale of these losses is less 
than the cluster age (as is usually the case), the electron population 
reaches a steady state distribution given by \citep[see][equation 38]{Sarazin1999}} 
\begin{equation}
 n(\gamma) = \frac{1}{b(\gamma)}\int_\gamma^{\infty} Q(\gamma')\,{\rm d}\gamma',
 \label{eq:steadyst}
\end{equation}
\FMR{where $b(\gamma)$ is the so-called loss function that parametrizes Coulomb 
and radiation losses from the ageing electrons. The loss function can be written 
as \citep[see again][equations $6-9$]{Sarazin1999}} 
\begin{equation}
 b(\gamma) = b_{\rm Coulomb} + b_{\rm IC + synch}(\gamma),
\end{equation}
\FMR{with} 
\begin{multline}
 b_{\rm Coulomb} \simeq 1.2\times 10^{-15} \left(\frac{n_{\rm e}}{10^{-3}\,{\rm 
cm^{-3}}}\right)\,\,\, s^{-1} = \\A \left(\frac{n_{\rm e}}{10^{-3}\,{\rm 
cm^{-3}}}\right)\,\,\, s^{-1},
\end{multline}
\begin{multline}
 b_{\rm IC + synch}(\gamma) \simeq 1.37\times10^{-20} \gamma^2 \\ \left[(1 + 
z)^4 + 0.095 \times \left(\frac{B}{1\,\mu{\rm G}}\right)^2\right]\,\,\, s^{-1},
\end{multline}
\FMR{which represent the expression for the Coulomb and radiation losses due to 
synchrotron and inverse Compton on the cosmic microwave background photons, 
respectively. Substituting these expression in eq.~(\ref{eq:steadyst}), the steady 
state electron distribution becomes} 
\begin{equation}
 n(\gamma) = \frac{Q_0}{A(2\alpha - 1)}\frac{\gamma^{1-2\alpha}}{1 + 
(\gamma/\gamma_{\rm br})^2} = n_0\frac{\gamma^{1-2\alpha}}{1 + 
(\gamma/\gamma_{\rm br})^2},
\end{equation}
\FMR{with}
\begin{equation}
 \gamma_{\rm br} = \left(\gamma^2\frac{b_{\rm Coulomb}}{b_{\rm IC + 
synch}(\gamma)}\right)^{1/2}.
\end{equation}
\FMR{Please note that $\gamma_{\rm br}$ is independent of $\gamma$. Please also 
note that in the asymptotic limit $\gamma \gg \gamma_{\rm br}$, this 
distribution can be approximated by} 
\begin{equation} n(\gamma) \sim
 \displaystyle n_0\gamma_{\rm br}^2\gamma^{-(1+2\alpha)}
 \label{eq:asymp}
\end{equation}
\FMR{which, modulo the constant factor $\gamma_{\rm br}^2$ that can be 
reabsorbed in the electron normalization, recovers the electron distribution 
adopted in Section~\ref{sec:model1}. Moreover, for typical cluster conditions at 
$z=0$ ($n_{\rm e} = 10^{-3}\,{\rm cm^{-3}}$, $B = 1\,{\rm \mu{\rm G}}$) 
$\gamma_{\rm br} \simeq 282$, which justifies our previous value for 
$\gamma_{\rm min}$.}

\FMR{To fully characterize the model $n_0$ or alternatively $Q_0$ must be 
determined. We do so by imposing that the total energy content in the electrons 
is a fraction $\eta$ of the cluster thermal energy per unit volume $U_{\rm 
therm}$ for any gas cell contained in the cluster. This leads to the condition} 
\begin{multline}
 E_{\rm el} = m_{\rm e} c^2 n_0\int_1^{\infty}\frac{\gamma^{2-2\alpha}}{1 + 
(\gamma/\gamma_{\rm br})^2} {\rm d}\gamma = \\
 m_{\rm e} c^2 n_0 G(\gamma_{\rm br}, \alpha) = \eta U_{\rm therm},
\end{multline}
\FMR{which yields}
\begin{equation}
  n_0 = \frac{\eta U_{\rm therm}}{m_{\rm e} c^2 G(\gamma_{\rm br}, \alpha)}, 
\,\,\,\,\, Q_0 = A (2\alpha -1)\frac{\eta U_{\rm therm}}{m_{\rm e} c^2 
G(\gamma_{\rm br}, \alpha)}.
  \label{eq:n0norm}
\end{equation}

\FMR{From the previous expressions it is also possible to determine the total 
energy injection rate to sustain the cluster radio emission. The injection rate 
is given by} 
\begin{equation}
 \dot{E}_{\rm inj} = m_{\rm e}c^2 Q_0 \int_1^\infty\gamma^{1-2\alpha}{\rm 
d}\gamma = \frac{m_{\rm e}c^2 Q_0}{(2\alpha - 2)},
\end{equation}
\FMR{and upon substituting into eq.~(\ref{eq:n0norm}) one obtains}
\begin{equation}
 \dot{E}_{\rm inj} = A\frac{(2\alpha - 1)}{(2\alpha-2)}\frac{\eta U_{\rm 
therm}}{G(\gamma_{\rm br}, \alpha)}.
\end{equation}
\FMR{The total energy (in terms of the cluster thermal energy) needed to sustain 
the cluster radio emission can be found by multiplying the previous expression 
by $t_{\rm Hubble} \sim 10\,{\rm Gyr}$. This yields} 
\begin{equation}
 \frac{E_{\rm inj}}{U_{\rm therm}} = At_{\rm Hubble}\frac{(2\alpha - 
1)}{(2\alpha-2)}\frac{\eta}{G(\gamma_{\rm br}, \alpha)},
\end{equation}
\FMR{which for typical galaxy cluster parameters at $z=0$ ($n_{\rm e} = 
10^{-3}\,{\rm cm^{-3}}$, $B = 1\,{\rm \mu{\rm G}}$) and $\alpha = 1.3$  becomes} 
\begin{equation}
 \frac{E_{\rm inj}}{U_{\rm therm}} = 0.04\left(\frac{\eta}{10^{-3}}\right).
\end{equation}
\FMR{This is only a small fraction of the total cluster energy and thus the 
model is energetically feasible. For the results presented in the paper we have 
adopted $z=0.2$, $\alpha = 1.3$, $\eta = 6\times10^{-3}$, whereas the (thermal) 
electron density $n_{\rm e}$ and $B$ field entering the loss equations are the 
volume-weighted average within the virial radius of each cluster. In this way 
the total energy requirements of the model for a Hubble time are about $\sim 6$\%
of the cluster thermal energy, and relativistic electrons with Lorentz factors 
in the range $\gamma \in[1,\infty]$ contain presently as much energy as the magnetic 
field in each cluster. The main features of the emitted radiation  as a function of 
frequency are shown in Figure~\ref{fig:emissivity} (top panel\FMR{, dashed lines}) 
by assuming a value for the gas thermal energy density equal to the magnetic energy 
contained in the field strengths listed in the legend and a constant
(thermal) electron density $n_{\rm e} = 10^{-3}\,\cm^{-3}$.
Finally, we would like to note that, with the choice of the fiducial parameters, 
relativistic particles carry approximately the same fraction of the total cluster energy in 
models 1 and 2. However, the major difference is that, while in model 1 this is the 
total energy of ultrarelativistic ($\gamma > 300$) protons \textit{and} electrons,
in model 2 only relativistic electrons, with a spectrum extending down to $\gamma = 1$,
are present. The total energy carried by ultrarelativistic electrons (the only ones that 
contribute in a significant way to synchrotron emission) is about the same in both
models, which implies similar levels of radio emissivity.}

\section{Bremsstrahlung radiation}\label{sec:ff}
To compute the X-ray emission from the galaxy cluster sample selected in this 
work, we make the assumption that the X-ray radiation is produced by 
bremsstrahlung from thermal electrons in the hot gas. We note that this model is 
somewhat a simplification because, for instance, it does not attempt to model 
line emission, which can be important in lower mass (colder) systems. However, 
it is accurate enough to investigate the scaling relations that we have 
presented in the paper.

We compute the volumetric bremsstrahlung emissivity as \citep[see e.g.][]{Kannan2016}
\begin{equation}
 \epsilon_\nu = 6.8\times10^{-38} T^{-1/2} \exp\left(-\frac{h\nu}{kT}\right) Z^2 
g_{\rm ff}\,n_{\rm e} n_{\rm i},
 \label{eq:freefree}
\end{equation}
where $T$ is the gas temperature, $h$ is the Planck constant, $\nu$ the 
frequency of the radiation, $k$ the Boltzmann constant, $Z$ the mean ionic 
charge, $g_{\rm ff}$ the mean gaunt factor, and $n_{\rm e}$ and $n_{\rm i}$ the 
number density of electrons and ions, respectively. Figure~\ref{fig:emissivity} 
(bottom panel), show the characteristic resulting spectrum for three different 
values of the gas temperature, as indicated in the legend, in terms of $E = 
h\nu$. A constant hydrogen number density of $10^{-3}\,{\rm cm^{-3}}$ is assumed 
(see  below for the values of the remaining parameters).

To determine the X-ray luminosity of any given gas cell $j$ we integrate 
equation (\ref{eq:freefree}) over a prescribed energy range $[E_{\rm low}, 
E_{\rm high}]$ and multiply by the cell volume, obtaining 
\begin{multline}
 L_{j, [E_{\rm low}, E_{\rm high}]} = 
6.8\times10^{-38}\,\frac{kT_{j}^{1/2}}{h}\\
 \left[\exp\left(-\frac{E_{\rm low}}{kT_j}\right) - \exp\left(-\frac{E_{\rm 
high}}{kT_j}\right)\right] \\
 Z^2 g_{\rm ff}\frac{X_{\rm e}}{(X_{\rm i} + X_{\rm e})^2} 
\left(\frac{\rho_j}{\mu_j\,m_{\rm p}}\right)^2 V_j.
 \label{eq:freefree2}
\end{multline}
In equation (\ref{eq:freefree2}), the subscript $j$ indicates quantities 
associated to a single gas cells, $V_j$ is the cell volume, $\rho_j$ the gas 
density, $\mu_j$ the gas mean molecular weight, and $X_{\rm i}$ and $X_{\rm e}$ 
are the ratios $n_{\rm i} / n_{\rm H}$ and $n_{\rm e} / n_{\rm H}$, with $n_{\rm 
H}$ being the hydrogen particle density. We fixed the free parameters in 
equation (\ref{eq:freefree2}) to $Z=\sqrt{1.15}$, $g_{\rm ff}=1.3$, $X_{\rm 
i}=1.079$, $X_{\rm e}=1.16$, which are appropriate for a fully ionized gas with 
primordial composition. We checked, with the {\sc pyxsim} package 
\citep{Zuhone2014}, that metal lines do not contribute significantly to the 
total X-ray emission in the virial mass range selected for our analysis.

\section{Generation of surface brightness maps}\label{sec:appc}
We now briefly describe the procedure that we have used to create radio and 
X-ray emission maps presented in this work. In practice,  we have adopted the 
following steps: 
\begin{enumerate}
 \item given the radio or the X-ray emissivity of gas cells, we have projected 
this quantity on a uniform grid, obtaining the radio or X-ray luminosity per 
unit area; 
 \item by assuming a fiducial redshift (we have adopted a value of 0.2, 
 which is the average redshift of detected radio haloes,
 \citealt{Feretti2012}), we have computed the luminosity and angular diameter 
distance for the cosmology adopted for IllustrisTNG; in this way surface 
brightness in the appropriate observational units can be determined;
 \item radio maps are smoothed via a convolution with a Gaussian kernel 
 to the nominal resolution of the radio  telescope given by its FWHM beam size;
 we recall that the relation between the dispersion of the Gaussian and the beam
 size is given by
 \begin{equation}
  \sigma_{b} = \frac{{b_{\rm FWHM}}}{2(\ln2)^{1/2}}.
 \end{equation}
\end{enumerate}

Sensitivity limits of the different instruments are presented  as rms noise in 
Table~\ref{tab:mapsproperties} for radio emission \citep[see][and references 
therein]{Vazza2015a}. For X-rays, we have adopted the typical {\it Chandra} ACIS 
background $1.3\times10^{-12}\,{\rm erg\,s^{-1}\,arcmin^{-2}}$ 
\citep{Anderson2010}. We convert it to a background surface brightness by 
dividing by the effective telescope area, which we assume to be $200\,\cm^2$ 
\citep[see again][]{Anderson2010}. The X-ray emission is detected if the surface 
brightness of a given pixel is above 0.3 times the background value thus 
determined. Please note that these brightness cuts are only used to produce the 
maps and are not imposed to compute the total X-ray luminosities.

\bsp
\label{lastpage}

\end{document}